\documentclass[12pt,english]{elsarticle}
\pdfoutput=1
\usepackage[T1]{fontenc}
\usepackage[latin9]{inputenc}
\usepackage[a4paper]{geometry}
\usepackage{color}
\usepackage{textcomp}
\usepackage{amsbsy}
\usepackage{graphicx}
\usepackage{subscript}

\makeatletter

\providecommand{\tabularnewline}{\\}

\definecolor{background-color}{gray}{0.98}

\@ifundefined{showcaptionsetup}{}{%
 \PassOptionsToPackage{caption=false}{subfig}}
\usepackage{subfig}
\makeatother

\usepackage{babel}
\begin{document}
\begin{frontmatter} 

\title{First Principles Study of Structural and Optical Properties of B$_{12}$
Isomers}

\author{Pritam Bhattacharyya}

\address{Department of Physics, Indian Institute of Technology Bombay, Powai,
Mumbai 400076, India}

\ead{pritambhattacharyya01@gmail.com}

\author{Ihsan Boustani }

\address{Theoretical and Physical Chemistry, Faculty of Mathematics and Natural
Sciences, Bergische Universit\"at, Wuppertal, Gauss Strasse 20, D-42097
Wuppertal, Germany }

\ead{boustani@uni-wuppertal.de}

\author{Alok Shukla}

\address{Department of Physics, Indian Institute of Technology Bombay, Powai,
Mumbai 400076, India}

\ead{shukla@phy.iitb.ac.in}
\begin{abstract}
In this work we undertake a comprehensive numerical study of the ground
state structures and optical absorption spectra of isomers of B$_{12}$
cluster. Geometry optimization was performed at the coupled-cluster-singles-doubles
(CCSD) level of theory, employing cc-pVDZ extended basis sets. Once
the geometry of a given isomer was optimized, its ground state energy
was calculated more accurately at the coupled-cluster-singles-doubles
along with perturbative treatment of triples (CCSD(T)) level of theory,
employing larger cc-pVTZ basis sets. Thus, our computed values of
binding energies of various isomers are expected to be quite accurate.
Our geometry optimization reveals eleven distinct isomers, along with
their point group, and electronic ground state symmetries. We also
performed vibrational frequency analysis on the three lowest energy
isomers, and found them to be stable. Therefore, we computed the linear
optical absorption spectra of these isomers of B\textsubscript{12},
employing large-scale multi-reference singles-doubles configuration-interaction
(MRSDCI) approach, and found a strong structure-property relationship.
This implies that the spectral fingerprints of the geometries can
be utilized for optical detection, and characterization, of various
isomers of B$_{12}$. We also explored the stability of the isomer
with with the structure of a perfect icosahedron, with $I_{h}$ symmetry.
In bulk boron icosahedron is the basic structural unit, but, our vibrational
frequency analysis reveals that it is unstable in the isolated form.
We speculate that this instability could be due to Jahn-Teller distortion
because five-fold degenerate HOMO orbitals in $I_{h}$ structure are
unfilled. 
\end{abstract}
\end{frontmatter}

\section{Introduction}

Since the eighties, and until now, the structures and energetics of
small atomic clusters have been of great interest, both experimentally,
and theoretically.\citep{deheer} Initially, the clusters were considered
as a bridge, or an accumulation at nanoscale to solids, but over the
years, due to sustained research effort, they have become well established
as a separate research discipline. Since then, the exploration and
synthesis of structures and energetics of pure atomic clusters has
acquired both academic and practical importance. Thereupon, the mass
spectra of alkali-metal, non-metal, carbon and boron clusters were
investigated as well as the related magic numbers were determined.\citep{deheer}
For example, among the allotropes of carbon, all sp\textsuperscript{2}-types
are closely related and have been extensively studied: graphene (flat
monoatomic sheet of graphite), spherical fullerenes, and nanotubes.\citep{full-graph}
Also for boron, carbon's left neighbor in the periodic table, the
landscape of fullerene-like possibilities is just beginning to emerge:
from small quasi-planar clusters, spherical cages and nanotubes \citep{bous-RSC},
then to borophene (single atom-thin monolayer sheet of boron atoms)
\citep{boro-mannix}.

However, the development of non crystalline boron beyond the icosahedral
arrangements began in the late eighties with theoretical and experimental
studies on small boron clusters. Anderson group \citep{anderson-JPC88}
carried out the first experimental and theoretical study of bonding
and structures in boron cluster ions B\textsubscript{n}\textsuperscript{+}
for (n $\le$ 13) in comparison with the well known closo boron hydrides.
Further important study on boron clusters was carried out by Kawai
and Weare \citep{kawai-JCP91} using Car-Parrinello \emph{ab initio}
molecular dynamics simulation. They found that an open 3D structure
is more stable than the icosahedral boron. However, Kato $et\,al.$\citep{kato-CPL92}
investigated boron clusters B\textsubscript{n} for (n=8-11) using
\emph{ab initio} molecular orbital theory, and concluded that the
most stable clusters have planar or pseudo-planar cyclic structures.
The breakthrough in boron cluster research happened in 1997 due to
the work of one of us,\citep{PRB97} in which it was discovered that
the most stable boron clusters have quasi-planar structures of dovetailed
hexagons including the B\textsubscript{12} cluster. In this work,
the so-called Aufbau principle was proposed, according to which highly
stable novel structures in form of boron sheets (nowadays called borophene),
nanotubes, and spheres can be constructed from only two basic units:
pentagonal and hexagonal pyramids B\textsubscript{6}, and B\textsubscript{7},
respectively.\citep{PRB97} These novel structures are different from
the conventional allotropes of solid boron.

It is well known that the bulk boron exists in several crystalline
phases.\citep{albert-angw09} The most famous boron solids are the
rhombohedral $\alpha$-B\textsubscript{12} and $\beta$-B\textsubscript{106}
phases with 12 and 106 atoms per unit cells, respectively, The $\alpha$-rhombohedral
phase transforms at 1200 °C into the more stable $\beta$-rhombohedral
one. Besides the $\alpha$-tetragonal, and $\beta$-tetragonal phases
with 190-192 atoms per unit cell, there is a new phase of crystalline
boron called orthorhombic $\gamma$-B\textsubscript{28} boron \citep{oganov},
with two B$_{12}$ clusters, and two B$_{2}$ pairs in the unit cell,
accompanied by a charge transfer as boron boride (B$_{2}$)$^{\delta+}$(B$_{12}$)$^{\delta-}$.
Remarkable is the common basic unit cell of all these structures:
the B$_{12}$ icosahedron. This regular B$_{12}$ icosahedron operates
as a building block for all phases mentioned above. Due to the multi-center
covalent inter-atomic bonds between the icosahedra, the electron deficiency
is reduced, and the electron distribution saturates the twelve boron
atoms in each icosahedron, leading to high stability. However, as
soon as this B$_{12}$ icosahedron is separated from the bulk, losing
its neighbors, and becoming a free standing cluster, it loses its
stability, and flattens to transform into a quasi-planar structure.\citep{PRB97}
Based upon molecular orbital occupancies, we argue in this work that
this flattening of the B$_{12}$ - I$_{h}$ icosahedron into the quasi-planar
B$_{12}$ - C$_{3v}$ structure could be a consequence of Jahn-Teller
distortion.

Given the importance of the B$_{12}$ cluster, in this work, we study
all its isomers including the icosahedron, and compute the optical
absorption spectra of its most stable isomers. For the purpose, we
have utilized standard wave function based first principles quantum
chemical methodologies, employing Cartesian Gaussian basis functions.
Geometry optimization was carried out using the coupled-cluster-singles-doubles
(CCSD) approach, while the optical absorption spectra of various isomers
were computed using the multi-reference singles-doubles configuration
interaction (MRSDCI) approach, which has been extensively utilized
in the group of one of us to study optical properties of conjugated
polymers,\citep{PhysRevB.65.125204Shukla65,PhysRevB.69.165218Shukla69,sony-acene-lo,doi:10.1021/jp408535u,himanshu-triplet}
graphene quantum dots,\citep{:/content/aip/journal/jcp/140/10/10.1063/1.4867363Aryanpour,Tista1}
along with atomic clusters such as those of boron,\citep{Shinde_nano_life}
aluminium,\citep{Shinde_PCCP} sodium,\citep{epjd-pradip} and magnesium.\citep{epjd-shinde-mg}

\section{Theoretical approach and Computational details}

\label{sec:theory}

\subsection{Geometry Optimization}

\label{subsec:geometry}

\begin{figure}
\begin{centering}
\subfloat[Quasi-planar ($C_{3v}$)]{\begin{centering}
\includegraphics[scale=0.15]{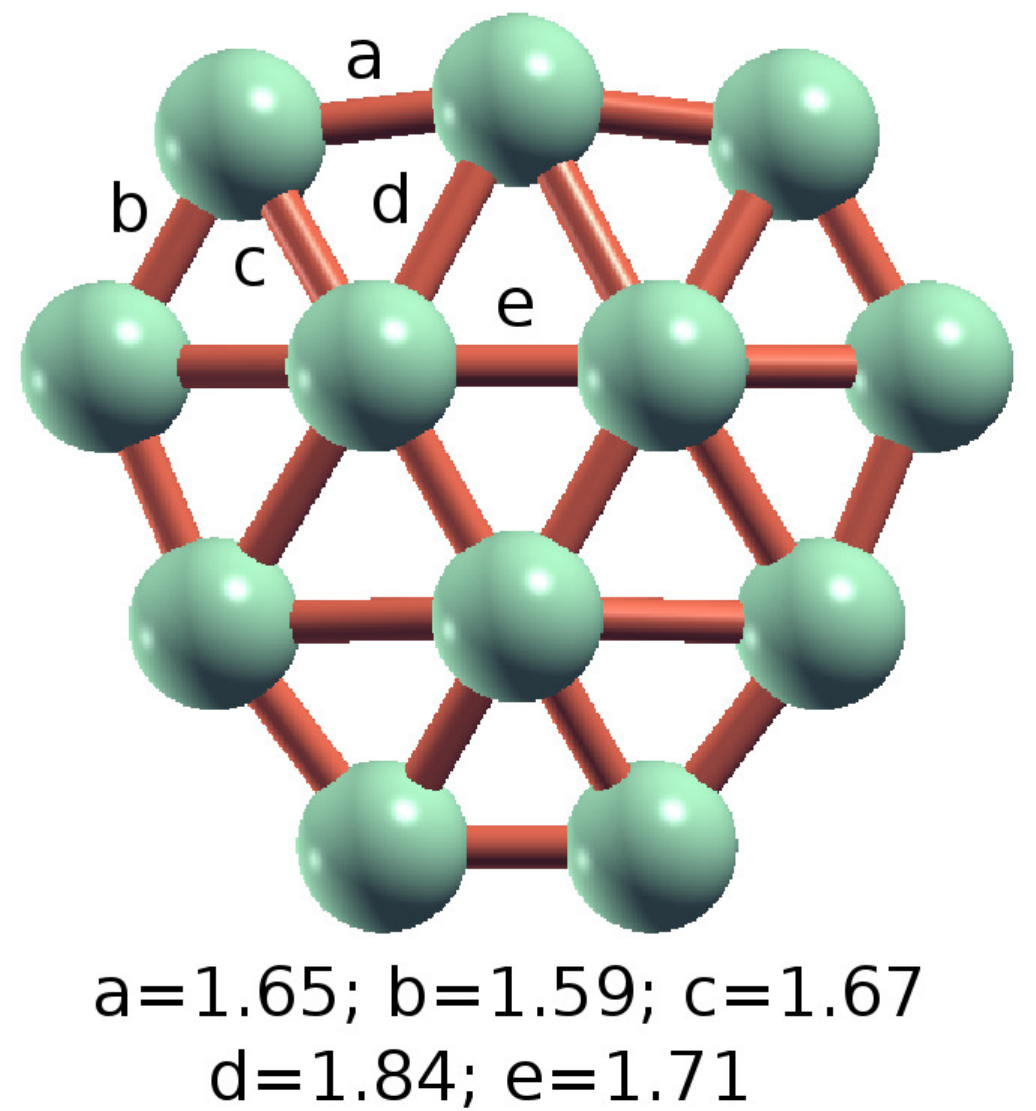} 
\par\end{centering}
}\subfloat[Double ring ($D_{6d}$)]{\begin{centering}
\includegraphics[scale=0.15]{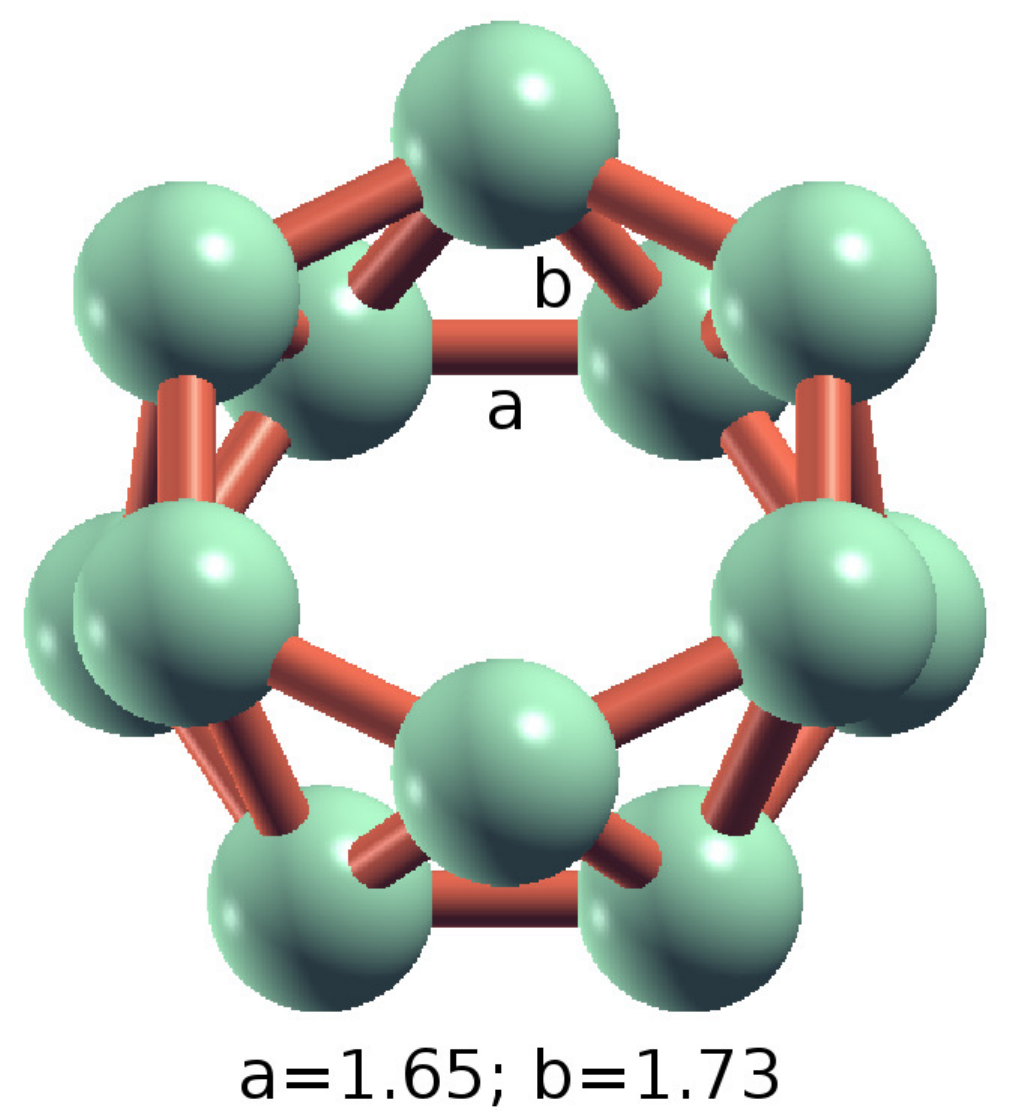} 
\par\end{centering}
}\subfloat[Chain ($C_{2h}$)]{\begin{centering}
\includegraphics[scale=0.19]{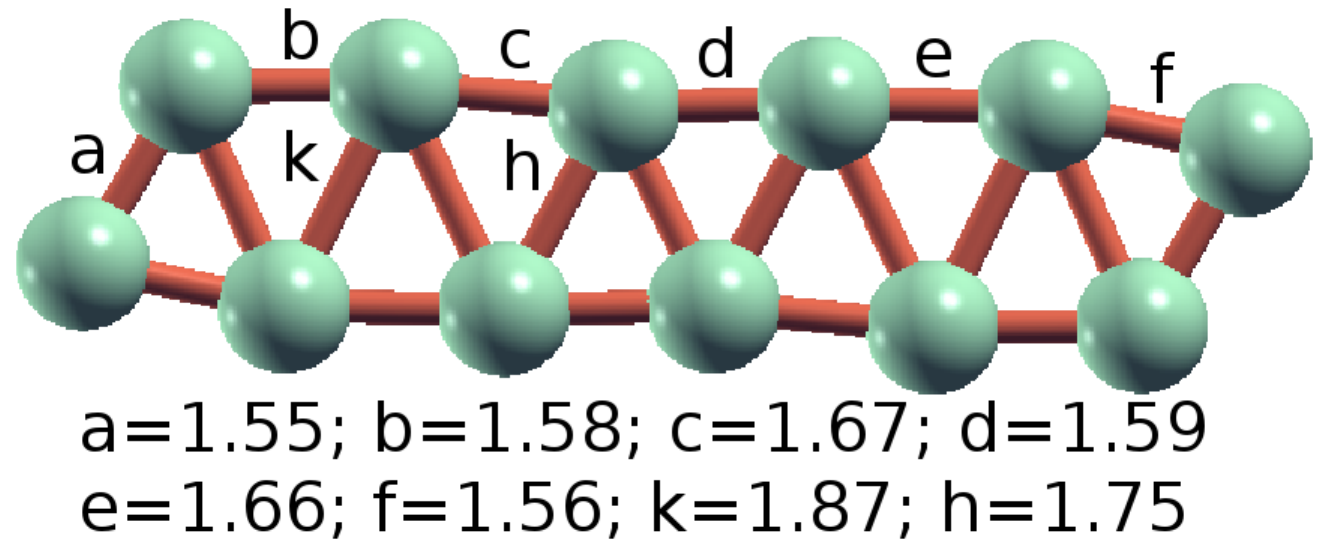} 
\par\end{centering}
}
\par\end{centering}
\begin{centering}
\subfloat[Hexagon ($D_{2h}$)]{\begin{centering}
\includegraphics[scale=0.16]{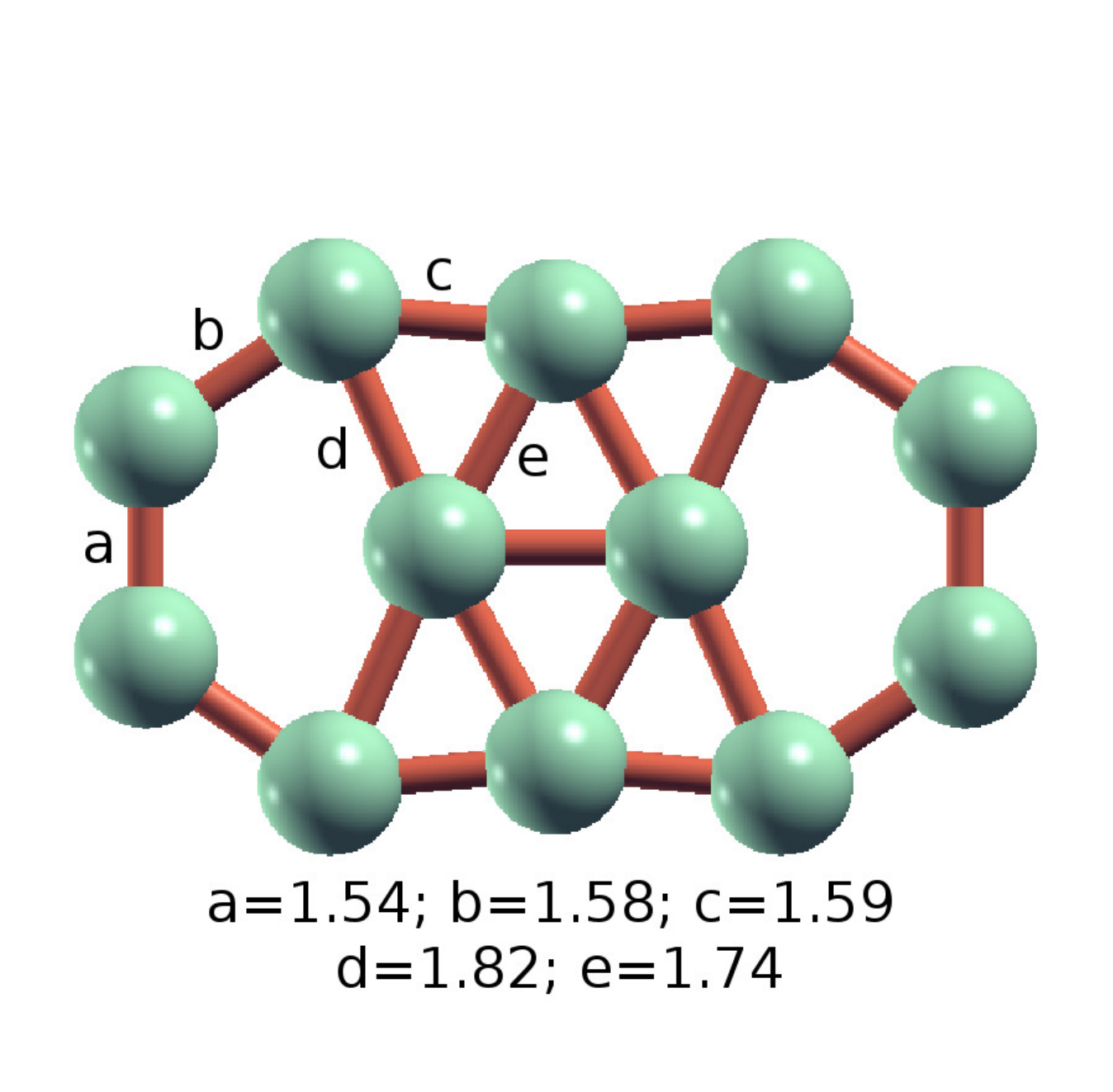} 
\par\end{centering}
}\subfloat[Perfect Icosahedron ($I_{h}$)]{\begin{centering}
\includegraphics[scale=0.17]{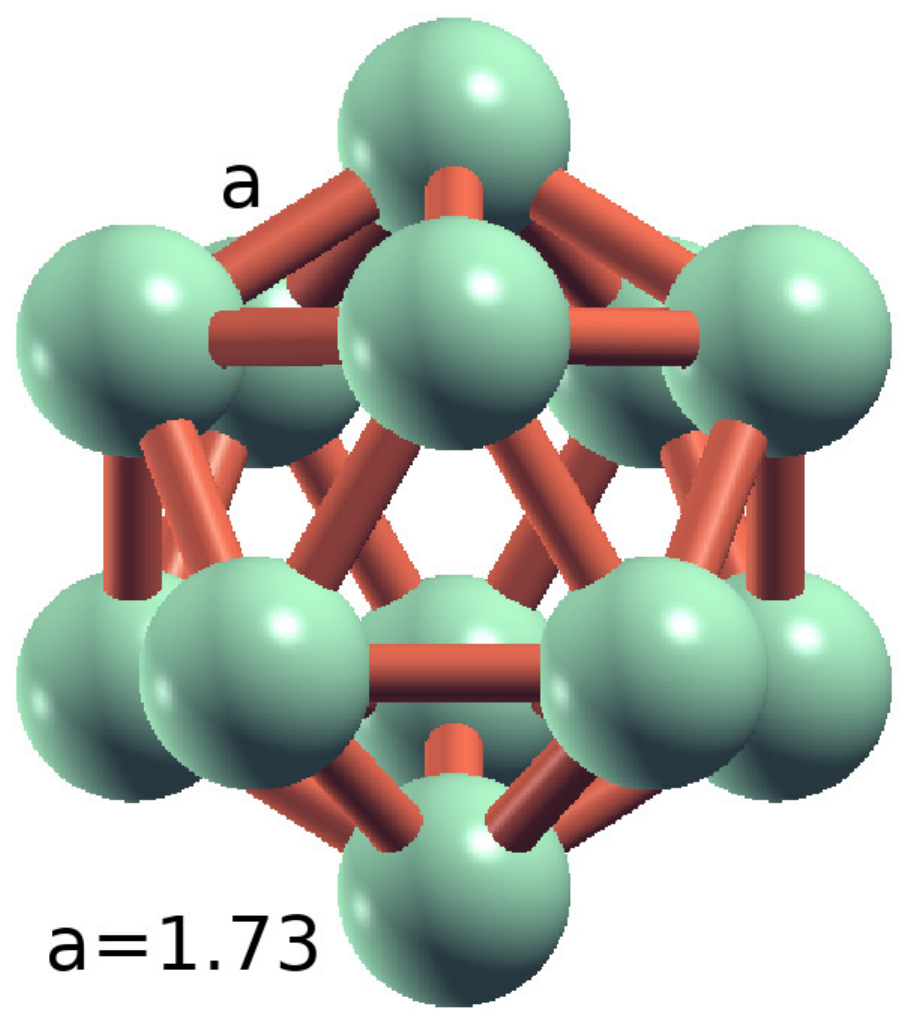} 
\par\end{centering}
}\subfloat[Distorted Icosahedron ($D_{2h}$)]{\begin{centering}
\includegraphics[scale=0.16]{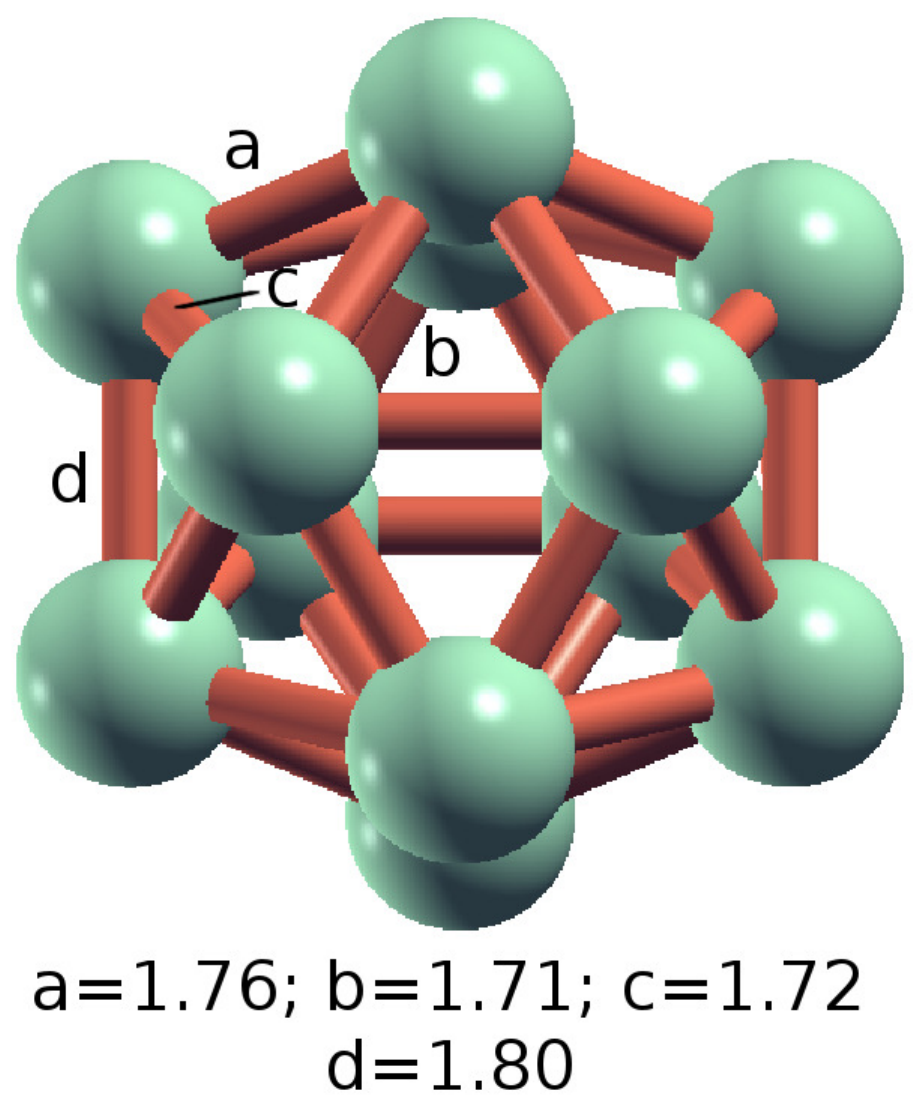} 
\par\end{centering}
}\subfloat[Quad ($D_{2h}$)]{\begin{centering}
\includegraphics[scale=0.16]{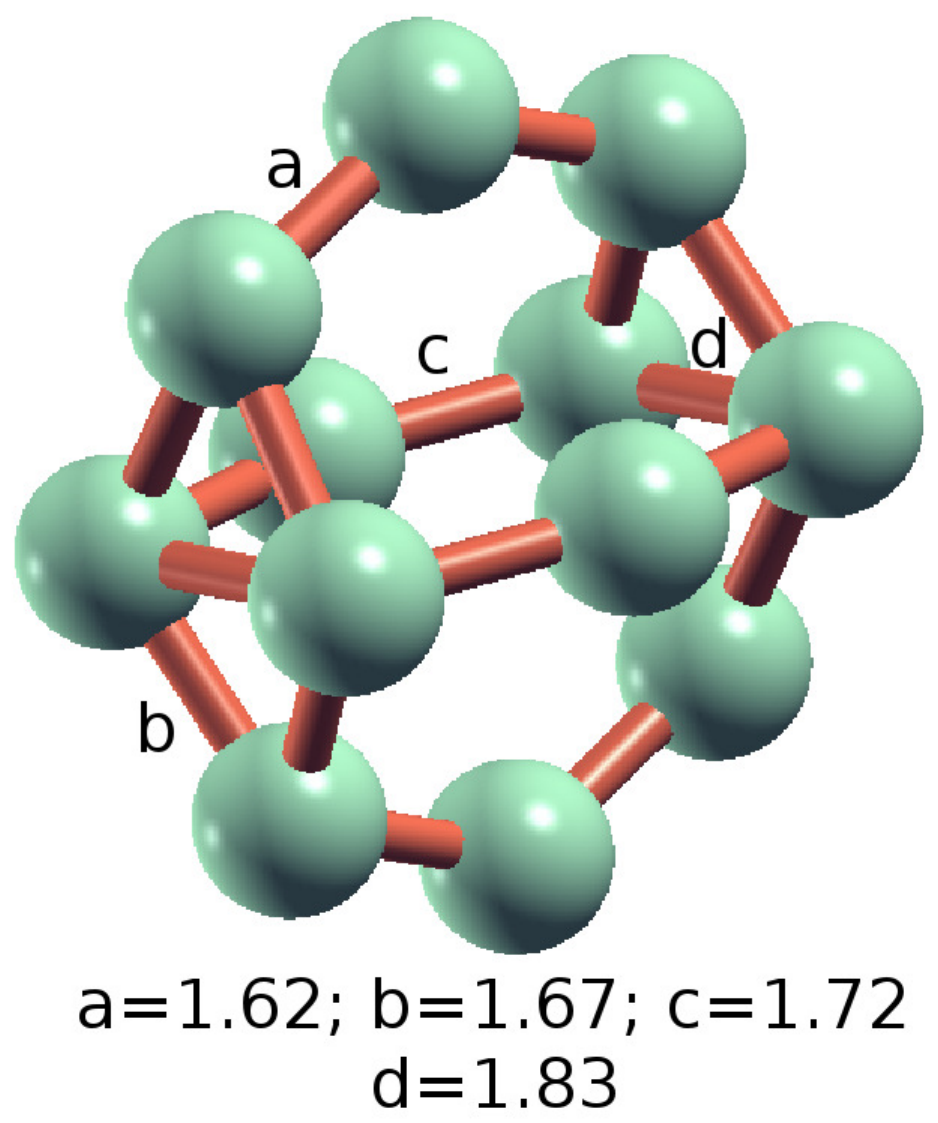} 
\par\end{centering}
}
\par\end{centering}
\begin{centering}
\subfloat[Pris-3-square-1 ($D_{4h}$)]{\begin{centering}
\includegraphics[scale=0.18]{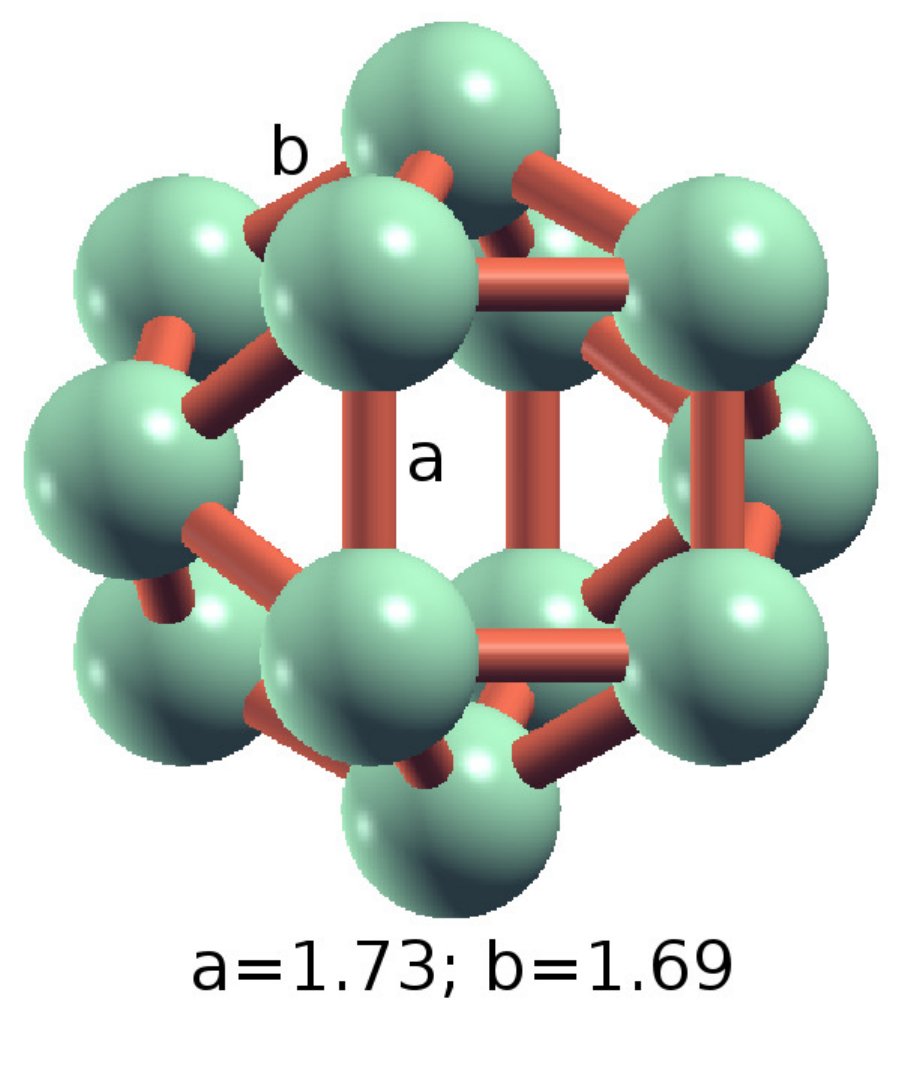} 
\par\end{centering}
}\subfloat[Pris-3 square-2 ($D_{4h}$)]{\begin{centering}
\includegraphics[scale=0.2]{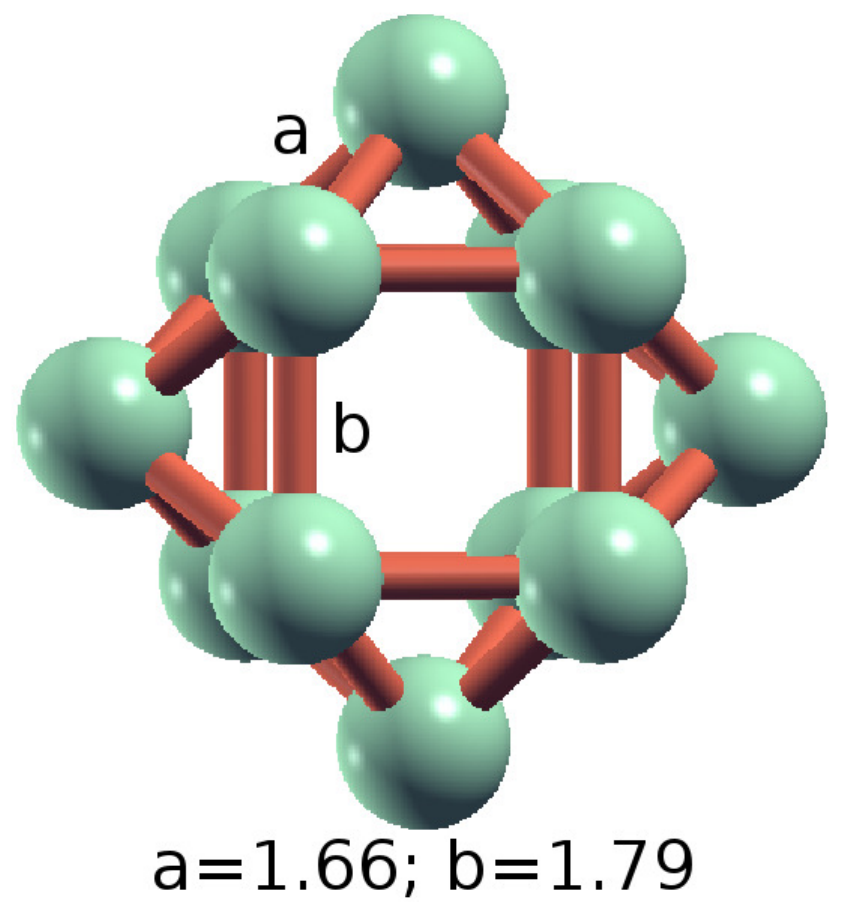} 
\par\end{centering}
}\subfloat[Parallel triangular ($D_{3h}$)]{\begin{centering}
\includegraphics[scale=0.2]{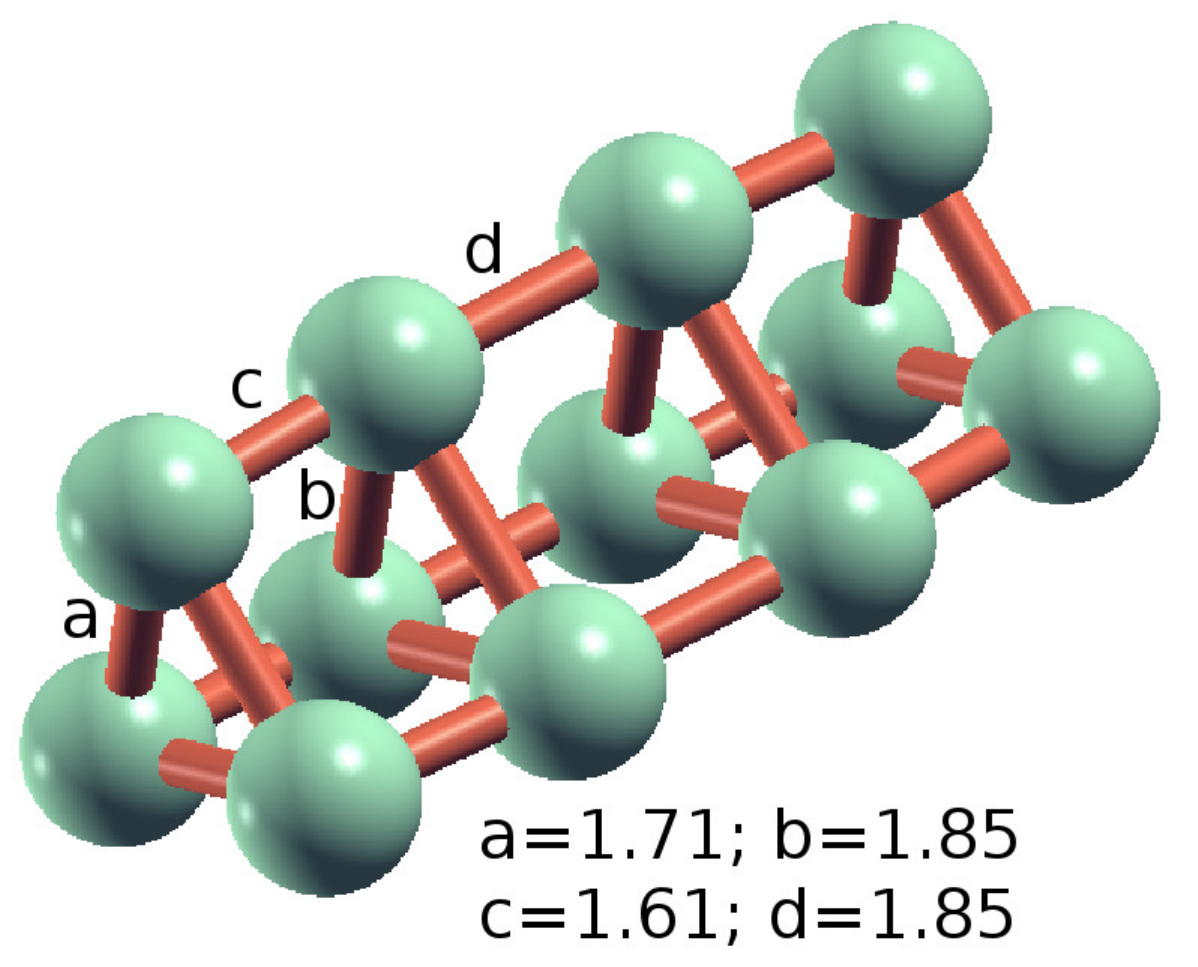} 
\par\end{centering}
}\subfloat[Planar square ($D_{4h}$)]{\begin{centering}
\includegraphics[scale=0.18]{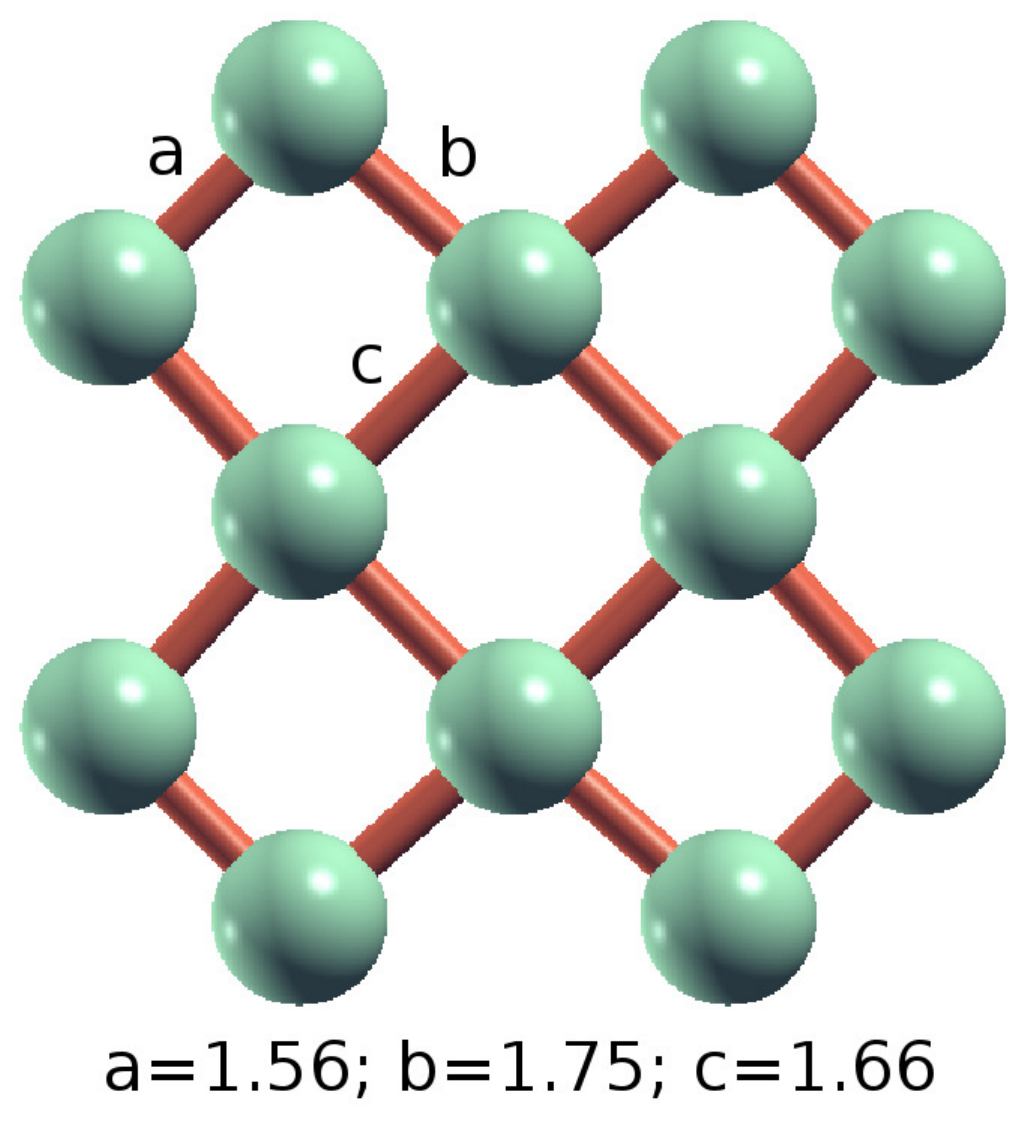} 
\par\end{centering}
}
\par\end{centering}
\caption{Optimized geometries of various isomers of B$_{12}$ cluster, along
with their point group symmetries. Optimizations were performed using
CCSD level of theory, and cc-pVDZ basis set, as implemented in the
Gaussian 09 package.\citep{g09} All isomers have singlet multiplicity,
and the bond lengths are in Å\ unit. \label{fig:optimized-geometries}}
\end{figure}

The geometries of various isomers of B$_{12}$ cluster were optimized
using the coupled-cluster singles doubles (CCSD) method, and the cc-pVDZ
basis set, as implemented in the Gaussian 09 package.\citep{g09}
The optimized geometries of all the isomers of B\textsubscript{12}
cluster are shown in the Fig. \ref{fig:optimized-geometries}. The
final total ground state energies of various isomers reported in Table
\ref{tab:total-energies}, were obtained by performing single-point
energy calculations at those optimized geometries, using the coupled-cluster
singles doubles triples (CCSD(T)) method, and the cc-pVTZ basis set.
Thus, the final ground state total energies of various isomers were
computed using a higher level of theory, and a larger basis set. One
important criteria, i.e., the binding energy per atom which is directly
related to the stability of the cluster has been computed using the
formula

\begin{equation}
\frac{E_{b}}{n}=E_{1}-\frac{E_{n}}{n},
\end{equation}

where $E_{1}$ is the energy of a single boron atom, $E_{n}$ is the
total energy of the cluster, and $n$ is the number of atoms present
in the cluster. For boron atom $E_{1}=-24.598101$ Hartree was used,
obtained using CCSD(T) approach as implemented in Gaussian09 program,
employing the cc-pVTZ basis set. Computed binding energies per atom
of all the eleven isomers are listed in Table \ref{tab:total-energies}.

We have also computed the optical absorption spectra of the three
lowest-energy structures, which not only have the highest binding
energies, but were also found to be stable from the vibrational frequency
analysis. 

\begin{table}
\caption{All the isomers of B$_{12}$ cluster have been listed below along
with their point group, ground state symmetry, and the ground state
energy (in Hartree). Also their relative energy (in eV) with respect
to the lowest isomer, i.e., quasi-planar structure is presented in
the same table. The correlation energy is the difference of the total
energies of the system at the CCSD(T) and the Hartree-Fock levels.
In the last column, the binding energy per atom (in eV) is shown to
illustrate the stability of the structure. \label{tab:total-energies}}

\begin{tabular}{ccccccc}
\hline 
Isomer  & Point  & Ground state & Ground state & Relative  & Correlation  & Binding \tabularnewline
 & group & symmetry  & energy$^{\dagger}$(Ha)  & energy & energy per & energy per\tabularnewline
 &  &  &  & (eV)  & atom (eV)  & atom (eV)\tabularnewline
\hline 
\hline 
 &  &  &  &  &  & \tabularnewline
Quasi-planar  & C\textsubscript{3v}  & \textsuperscript{1}A\textsubscript{1}  & -297.208187  & 0.0  & 2.92  & 4.60\tabularnewline
 &  &  &  &  &  & \tabularnewline
Double ring  & D\textsubscript{6d}  & \textsuperscript{1}A\textsubscript{1}  & -297.1399191  & 1.858  & 3.00  & 4.45\tabularnewline
 &  &  &  &  &  & \tabularnewline
Chain  & C\textsubscript{2h}  & \textsuperscript{1}A\textsubscript{g}  & -297.1029824  & 2.863  & 2.92  & 4.36\tabularnewline
 &  &  &  &  &  & \tabularnewline
Hexagons  & D\textsubscript{2h}  & \textsuperscript{1}A\textsubscript{g}  & -297.0853492  & 3.342  & 2.97  & 4.33\tabularnewline
 &  &  &  &  &  & \tabularnewline
Perfect Icosahedron & I\textsubscript{h} & \textsuperscript{1}A\textsubscript{g}  & -297.0662229 & 3.863 & 3.36 & 4.28\tabularnewline
 &  &  &  &  &  & \tabularnewline
Distorted Icosahedron  & D\textsubscript{2h}  & \textsuperscript{1}A\textsubscript{g}  & -297.077724  & 3.550  & 3.35  & 4.31\tabularnewline
 &  &  &  &  &  & \tabularnewline
Quad  & D\textsubscript{2h}  & \textsuperscript{1}A\textsubscript{g}  & -297.0180888  & 5.173  & 3.08  & 4.17\tabularnewline
 &  &  &  &  &  & \tabularnewline
Pris-3-square-1 & D\textsubscript{4h}  & \textsuperscript{1}A\textsubscript{1g}  & -296.8945639  & 8.534  & 3.44  & 3.89\tabularnewline
 &  &  &  &  &  & \tabularnewline
Pris-3-square-2 & D\textsubscript{4h}  & \textsuperscript{1}A\textsubscript{1g}  & -296.8192288  & 10.583  & 3.29  & 3.72\tabularnewline
 &  &  &  &  &  & \tabularnewline
Parallel triangular  & D\textsubscript{3h}  & \textsuperscript{1}A$^{\prime}$\textsubscript{1}  & -296.7691979  & 11.945  & 3.19  & 3.61\tabularnewline
 &  &  &  &  &  & \tabularnewline
Planar square  & D\textsubscript{4h}  & \textsuperscript{1}A\textsubscript{1g}  & -296.7132189  & 13.468  & 3.36  & 3.48\tabularnewline
 &  &  &  &  &  & \tabularnewline
\hline 
\end{tabular}

$\dagger$ Ground state energy was computed using CCSD(T) level of
theory and cc-pVTZ basis set. 
\end{table}

\subsection{Optical Absorption Spectrum}

\label{subsec:optics}

With the aim of understanding structure property relationship, we
also computed the optical absorption spectra of three of the lowest
energy isomers: (a) B\textsubscript{12} - quasi-planar , (b) B\textsubscript{12}
- double ring, and (c) B\textsubscript{12} - chain. In order to compute
the optical absorption spectra, we utilized the \emph{ab initio} MRSDCI
approach, as implemented in the computer program MELD,\citep{MELD}
and described in recent works on the optical properties of clusters.\citep{Shinde_nano_life,Shinde_PCCP,epjd-pradip,epjd-shinde-mg}
For the purpose, we utilized the geometries of the isomers (quasi-planar,
double ring, and chain) presented in Fig. \ref{fig:optimized-geometries},
and the cc-pVDZ basis set, which was also used for geometry optimization.
Calculations were initiated by performing the restricted Hartree-Fock
calculations, and the resultant orbital set (occupied and virtual),
served as the single-particle basis for the CI calculations. Because
the orbital basis set was fairly large, the following approximations
were employed to truncate it: (a) frozen-core approximation, and (b)
removal of high-energy virtual orbitals. Both these approximations
are justified on the physical grounds in that the orbitals which are
far away from the Fermi level are unlikely to contribute to optical
excitations. Nevertheless, we critically examine the influence of
these approximations on our results in the following section. After
fixing the single-particle basis, the next step involves singles-doubles
configuration interaction calculations (SDCI), employing single reference
wave functions. Ground and the excited states obtained from these
calculations are used to calculate the optical absorption spectrum
$\sigma(\omega)$, according to the formula 
\begin{equation}
\sigma(\omega)=4\pi\alpha\sum_{i}\frac{\omega_{i0}|\langle i|\boldsymbol{\hat{e}.r}|0\rangle|^{2}\gamma^{2}}{(\omega_{i0}-\omega)^{2}+\gamma^{2}}.\label{eq:sigma}
\end{equation}
Above, $\omega$ denotes frequency of the incident light, $\boldsymbol{\hat{e}}$
denotes its polarization direction, $\boldsymbol{r}$ is the position
operator, $\alpha$ is the fine structure constant, $0$ and $i$
denote, respectively, the ground and excited states, $\omega_{i0}$
is the frequency difference between those states , and $\gamma$ is
the line width, taken as 0.1 eV, in these calculations. The summation
over $i$ in Eq. \ref{eq:sigma} involves an infinite number of excited
states which are dipole connected to the ground state, while in practice
it is restricted to states with excitation energies up to 10 eV. We
carefully examine the excited states contributing to the peaks in
the absorption spectrum, and the configurations contributing to those
are included in the reference set for MRSDCI calculations. A new set
of configuration space is generated by considering singles and doubles
excitations from the reference set, and the entire procedure is repeated.
This process is continued until the calculated absorption spectrum
has converged within reasonable limits.\citep{Shinde_nano_life,Shinde_PCCP,epjd-pradip,epjd-shinde-mg}

We make a few brief comments about the line width $\gamma$ occurring
in the expression for the optical absorption spectrum (see Eq. \ref{eq:sigma}
above). The line width incorporates all kinds of uncertainties associated
with the values of energy levels due to reasons such as: (a) natural
line widths, (b) vibrational broadening, because of finite temperature
effects, (c) collisional broadening in the gas or liquid phase, (d)
impurities and disorder in the samples, and (e) experimental uncertainties.
The value $\gamma=0.1$ eV chosen by us is fairly reasonable, and
has been employed in all our past calculations of optical absorption
spectra. From the structure of Eq. \ref{eq:sigma} it is obvious that
a smaller value of $\gamma$ leads to sharper and higher peaks, while
a larger value causes the peaks to become broader, and lower in height.
Therefore, as per the requirement, one can use any reasonable value
of $\gamma$.

\subsection{Orbital Truncation Scheme}

\begin{figure}
\subfloat[quasi-planar]{\begin{centering}
\includegraphics[scale=0.35]{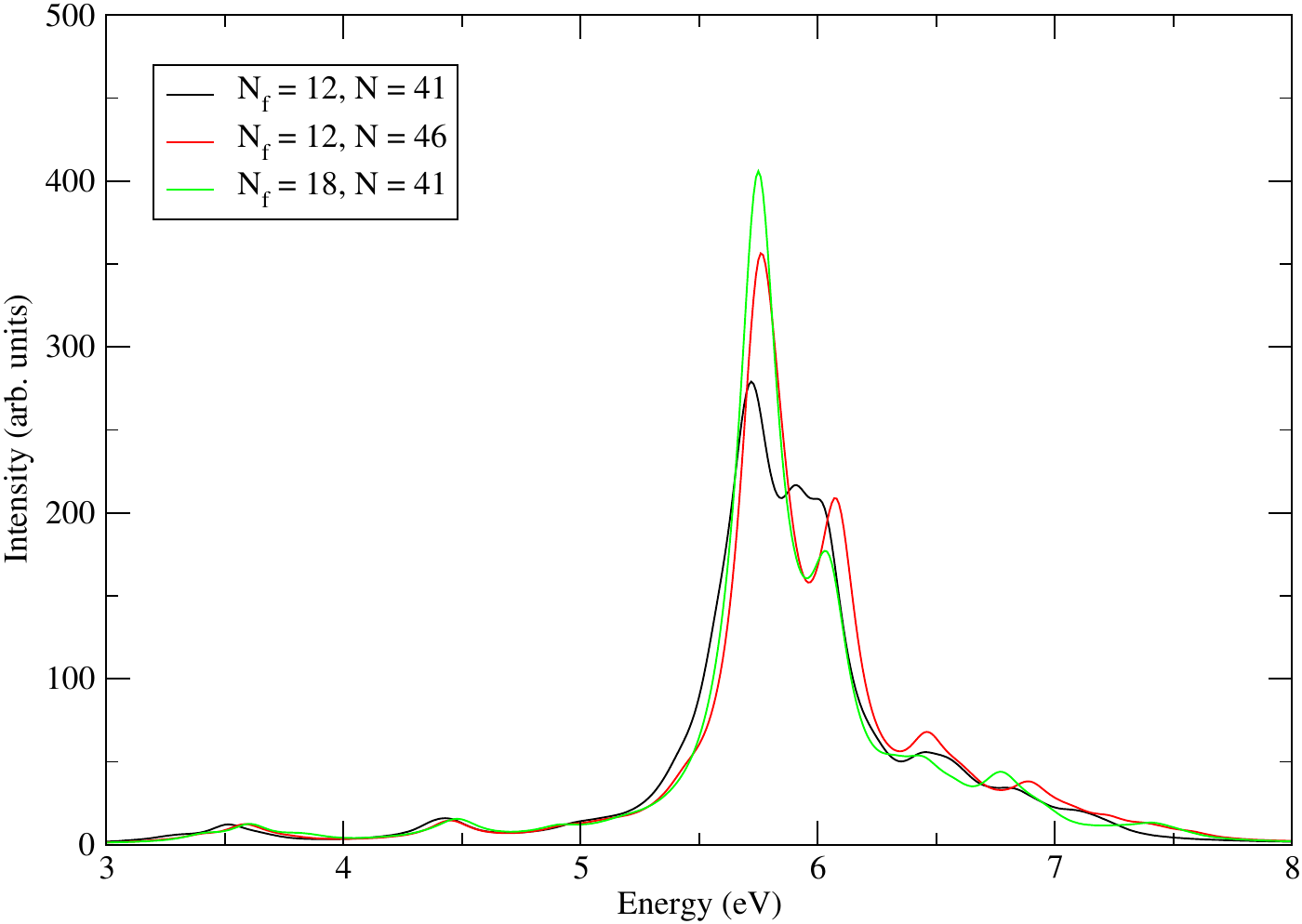} 
\par\end{centering}
}~~~~\subfloat[Double ring]{\begin{centering}
\includegraphics[scale=0.35]{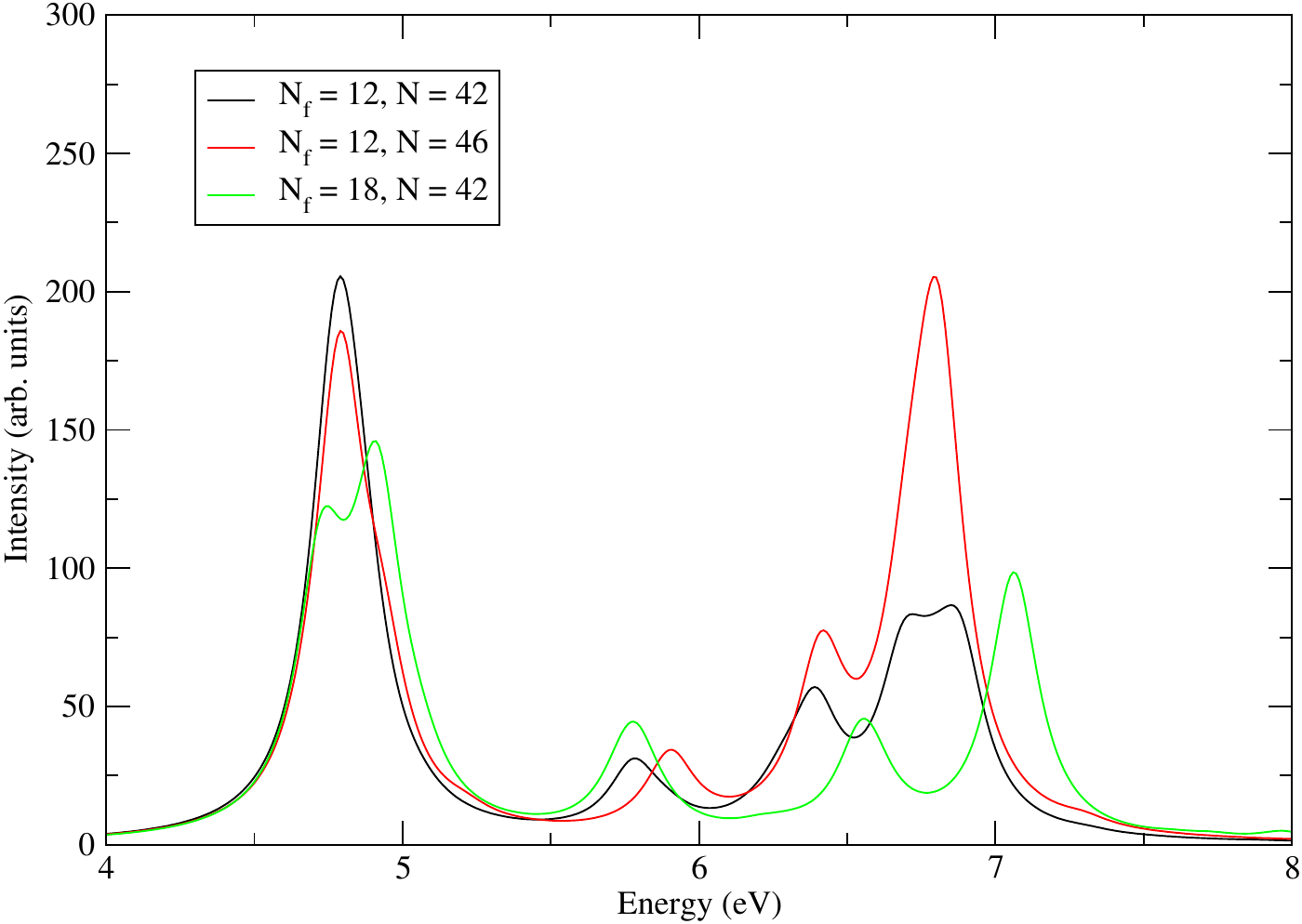} 
\par\end{centering}
}

\vspace{1cm}

~~~~~~~~~~~~~~~~~~~~~~~~~~~~~~~~~~~~~~\subfloat[Chain]{\begin{centering}
\includegraphics[scale=0.35]{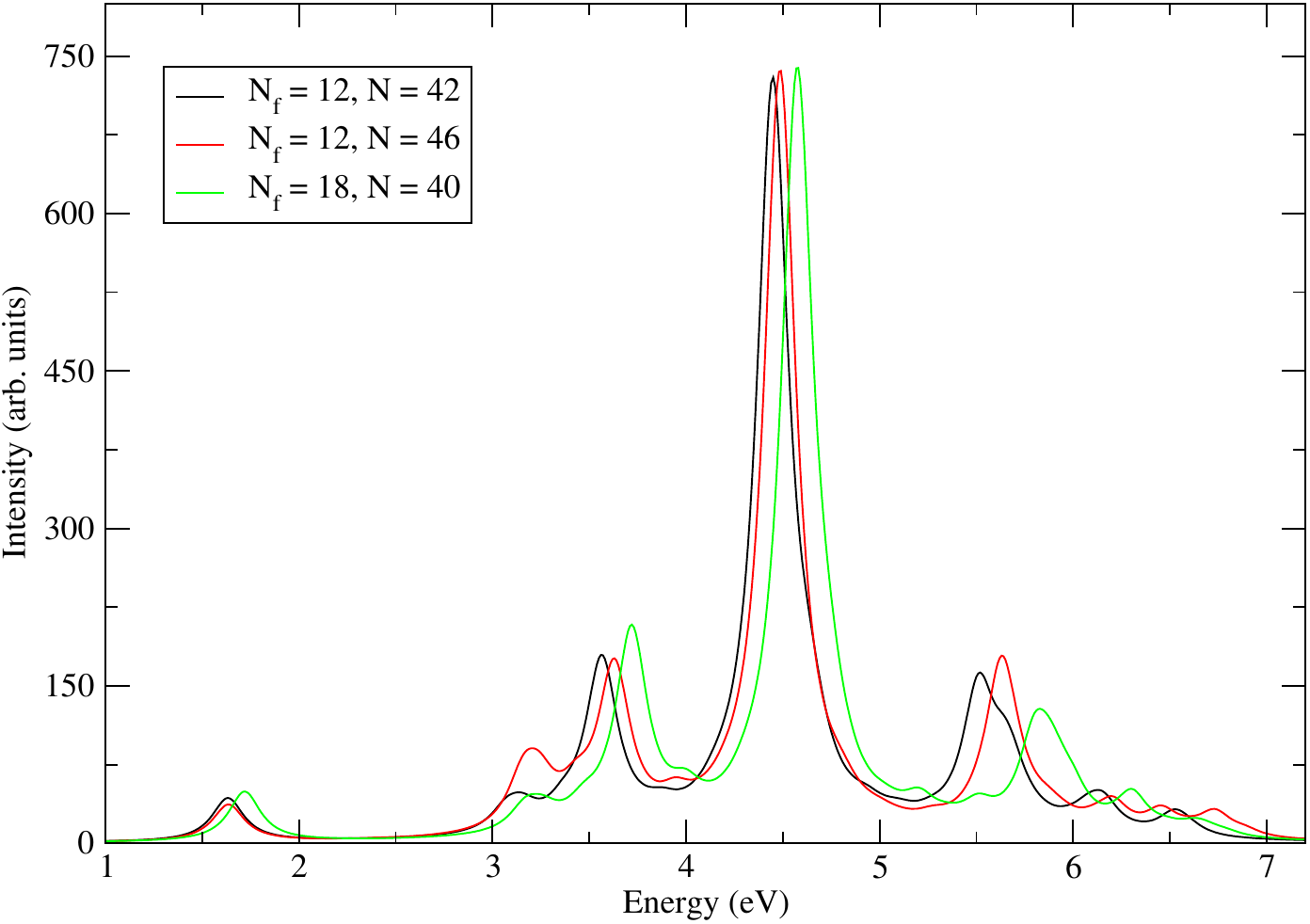} 
\par\end{centering}
}

\caption{Optical absorption spectra of three most stable isomers of B\protect\protect\protect\protect\textsubscript{12},
calculated using varying numbers of frozen and active orbitals.\label{fig:ci-orb-converg}}
\end{figure}

The computational effort in CI calculations grows \textsuperscript{}$\approx N^{6}$,
where $N$ is the total number of active orbitals. Thus, the size
of the CI expansion, and hence the computational effort involved proliferates
very rapidly with the increasing $N$. Therefore, the choice of the
active orbitals becomes crucial, and determines the feasibility and
the quality of the calculations. As discussed earlier, there are two
ways to control the value of $N$: (a) by the choosing how many low-lying
orbitals will be frozen during the CI calculations, and (b) how many
virtual orbitals will be included in the CI calculations. In Fig.
\ref{fig:ci-orb-converg}, we examine the convergence of the computed
optical absorption spectrum of the three isomers of B$_{12}$, with
respect to the choice of the active orbitals. Assuming that $N_{f}$
implies the total number of frozen orbitals in the calculations, we
performed three sets of calculations for each isomer: (i) $N_{f}=12$,
$N=40/41/42$, (ii) $N_{f}=12$, $N=46$, and (iii) $N_{f}=18$, $N=40/41/42$.
Here $N_{f}=12$ implies that 1s core orbital of each boron atom of
the cluster was frozen during the CI calculations, while for $N_{f}=18$
implies that in addition to the atomic core orbitals, six additional
orbitals were frozen during the calculations. Furthermore, for $N_{f}=12$,
total number of valence electrons considered in the CI calculations
was $N_{val}=36$ , while for $N_{f}=18$, this reduces to $N_{val}=24$.
When we examine the computed spectra, it is obvious that: (a) for
the quasi-planar and the chain structures, three sets of calculations
are in good qualitative and quantitative agreement with each other,
(b) for the case of the double ring structure, three sets of calculations
are in good agreement with each other for photon energy $E<5$ eV,
but at $E\approx7$ eV, results of three sets of calculations differ
from each other. Nevertheless, for the double ring, even in that energy
region $N_{f}=12$, $N=42$ and $N_{f}=12$, $N=46$ calculations
agree with each other as far as peak locations are concerned, but
disagree on peak intensities. Therefore, we conclude that the calculations
corresponding to $N_{f}=12$, $N=46$, are well converged, and analyze
those results in the next section.

\subsection{Size of the CI expansion.}

As mentioned previously, the MRSDCI procedure adopted in these calculations
is iterative in nature. Thus, once the number of active/frozen orbitals
$N/N_{f}$ is fixed, MRSDCI procedure is iterated until the absorption
spectrum converges within reasonable tolerance levels. In Table \ref{tab:config-data},
we present the data regarding the final MRSDCI calculations performed
on the three isomers, which includes: (a) point group symmetry used
in the calculations, and (b) total number of configurations $N_{total}$
in the MRSDCI expansion for various irreducible representations. We
note that $N_{total}$ ranges roughly from five millions to ten millions,
implying that these were large-scale calculations, and, therefore,
electron-correlation effects have been included to a high-level both
for the ground state, and the excited states.

\begin{table}
\caption{The average number of total configurations involved in MRSDCI calculations
(N\protect\textsubscript{total}) of the optical absorption spectra
of the most stable isomers of B\protect\textsubscript{12} cluster.\label{tab:config-data}}

\begin{tabular}{cccccc}
\hline 
Isomer  & \hspace{0.5cm}  & Point  & ~~  & Symmetry  & N\textsubscript{total}\tabularnewline
 &  & group  &  &  & \tabularnewline
\hline 
\hline 
Quasi-planar $^{\dagger}$  &  & C\textsubscript{3v}  &  & A$^{\prime}$  & 6903907\tabularnewline
 &  &  &  & A$^{\prime\prime}$  & 6207478\tabularnewline
 &  &  &  &  & \tabularnewline
 &  &  &  &  & \tabularnewline
Double ring$^{\dagger}$  &  & D\textsubscript{6d}  &  & A$^{\prime}$  & 5210210\tabularnewline
 &  &  &  & A$^{\prime\prime}$  & 6741895\tabularnewline
 &  &  &  &  & \tabularnewline
 &  &  &  &  & \tabularnewline
Chain  &  & C\textsubscript{2h}  &  & A\textsubscript{g}  & 556980\tabularnewline
 &  &  &  & A\textsubscript{u}  & 9353654\tabularnewline
 &  &  &  & B\textsubscript{u}  & 9588626\tabularnewline
\hline 
\end{tabular}

$\dagger$ C\textsubscript{s} point group symmetry was used during
the calculations 
\end{table}

\section{Results and Discussion}

\label{sec:results} Next, we discuss the geometry and the electronic
structure of isomers of B$_{12}$ considered in this work. Furthermore,
we also discuss the optical absorption spectra of three of the most
stable isomers.

\subsection{Icosahedron}

This is one of the most studied isomers of the B$_{12}$ cluster,
perhaps, because of its high-level of symmetry, i.e., a perfect icosahedral
structure, and $I_{h}$ symmetry group.\citep{b12-icos-bamba-wagner,BOUSTANI-cpl-1995,PRB97}
Bambakidis and Wagner used the Slater's SCF-Xa-SW approach to study
the electronic structure and cohesive properties of B$_{12}$ icosahedral
cluster, while Boustani employed first-principles DFT and Hartree-Fock
approaches to study it.\citep{BOUSTANI-cpl-1995,PRB97} Furthermore,
B$_{12}$ icosahedron forms the fundamental units in bulk boron, and
other boron-rich solids, therefore, it has always made one curious
whether or not it exists in isolated form.\citep{kawai-JCP91,nature-b12,prl-b12}
Hayami\citep{b12-icos-hayami} studied the encapsulation properties
of the B$_{12}$ icosahedral cluster using first-principles DFT based
methodology. Kawai and Weare\citep{kawai-JCP91} based upon ab initio
molecular dynamics simulations, and Boustani using first-principles
DFT and Hartree-Fock methods,\citep{BOUSTANI-cpl-1995,PRB97} concluded
that the isolated B$_{12}$ icosahedral cluster is unstable, and actually
stabilizes to a lower-energy open structure. 

\begin{figure}
\begin{centering}
\includegraphics[scale=0.75]{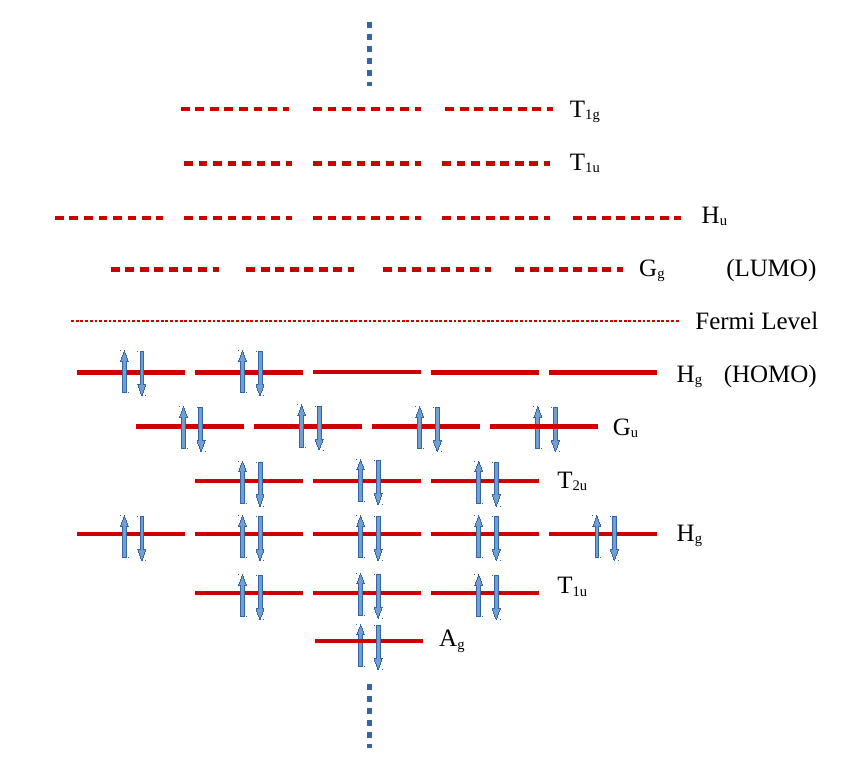} 
\par\end{centering}
\caption{A rough sketch of orbital energies of B$_{12}$ - icosahedron ($I_{h}$
symmetry group), close to the Fermi level. Irreducible representations
of the corresponding orbitals are indicated next to them. HOMO, with
its five-fold degeneracy, is partially occupied, thereby making the
system a candidate for Jahn-Teller distortion, to a lower symmetry
structure. \label{fig:orbitals-b12-ih}}
\end{figure}

In order to examine the fundamental reasons behind the instability
of the singlet $I_{h}$ structure against structural distortions,
we consider the possibility of Jahn-Teller effect, and to that end
we examine the single-particle energy levels of B$_{12}$ perfect
icosahedron, obtained from calculations without any electrons. These
calculations were performed using a STO-3G basis set, and the energy
levels obtained, along with their symmetries, are shown in Fig. \ref{fig:orbitals-b12-ih}.
If we start filling the energy levels using auf-bau principle, we
note that the five-fold degenerate highest-occupied molecular orbital
(HOMO) of H\textsubscript{g} symmetry is partially filled, thus making
the isomer a candidate for Jahn-Teller distortion to a structure of
lower symmetry.

Another manifestation of this instability is that automated geometry
optimization procedures, such as the one based on Berny algorithm
implemented in Gaussian 09 package,\citep{g09} inevitably lead to
a lower symmetry structure, if the starting geomertry is a perfect
icosahedron, with $I_{h}$ symmetry. Since we are interested in computing
the total energy of the perfect icosahedral structure, we optimized
corresponding geometry manually, by varying the nearest-neighbor bond
length, and looking for the minimum of energies computed using the
CCSD approach, and cc-PVDZ basis functions. At this point, an objection
can be raised against the use of single-reference approaches such
as CCSD to describe the ground state of the $I_{h}$ structure, given
the orbital degeneracy at the Fermi level as depicted in Fig. \ref{fig:orbitals-b12-ih}.
However, this is not a problem if one uses the Hartree-Fock (HF) molecular
orbitals (MOs) for performing the correlated-electron calculations,
because the electron-electron repulsion lifts the orbital degeneracies
to a great extent. Furthermore, we verified this by performing MRSDCI
calculations on the $I_{h}$ structure at its minimum energy geometry
using the HF MOs, and found that the coefficient of the HF reference
state is 0.81 in the lowest singlet-state wave function (see Table
S16 of Supporting Information, and the related discussion), thereby,
justifying the use of single-reference approaches to compute the ground
state energy. The minimum energy $I_{h}$ structure obtained from
these CCSD/cc-PVDZ calculations is depicted in Fig. \ref{fig:optimized-geometries}(e),
with the corresponding total energy listed in Table \ref{tab:total-energies}.
Next, we checked the stability of the perfect icosahedral structure
by performing vibrational frequency analysis at the CCSD level, using
the cc-PVDZ basis set, and found it to be unstable with five imaginary
frequencies. 

Next, we performed automated geometry optimization\citep{g09} on
the icosahedral structure, with the starting atomic coordinates corresponding
to the $I_{h}$ structure, and 1.73 \AA\  bond length corresponding
to Fig. \ref{fig:optimized-geometries}(e). The resultant optimized
geometry has a distorted icosahedral structure with $D_{2h}$ symmetry,
with non-uniform nearest-neighbor distances in the range 1.71 Å\textemdash 1.80
Å, as shown in Fig. \ref{fig:optimized-geometries}(f). This distorted
structure has an energy 3.55 eV higher than the quasi-planar structure,
but 0.31 eV more stable as compared to the $I_{h}$ structure. To
explore the stability of this structure, we performed vibrational
frequency analysis on it at the CCSD level, using the same cc-pVDZ
basis set, and found that it is a transition state, with one imaginary
frequency. This, clearly reveals that perfect $I_{h}$ structure of
B$_{12}$ cluster with the singlet spin is totally unstable, and deforms
into a distorted icosahedral transition state corresponding to a saddle
point on the potential energy surface. This indicates that this $D_{2h}$
structure is itself unstable, and will undergo further distortions
to achieve stability. Thus, our calculations have confirmed the instability
of the $I_{h}$ structure of the B$_{12}$ cluster.

\subsection{Quasi-planar convex structure}

\begin{figure}
\begin{centering}
\includegraphics[scale=0.75]{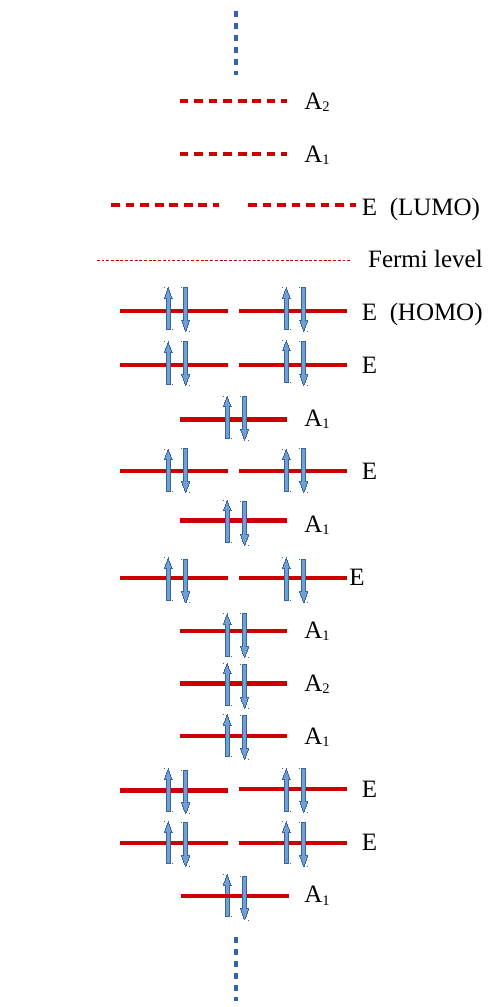} 
\par\end{centering}
\caption{A rough sketch of the energy levels of quasi-planar ($C_{3v}$) isomer,
along with a few low-lying virtual levels, obtained from the Hartree-Fock
calculations performed at the optimized geometry, using a cc-pVTZ
basis set. Irreducible representations of the orbitals corresponding
to these levels, are indicated next to them. It is obvious that the
doubly-degenerate HOMO is fully occupied for this isomer, indicating
that it is stable against a Jahn-Teller distortion. \textcolor{red}{\label{fig:orbitals-b12-quasi-planar}}}
\end{figure}

This lowest lying quasi-planar convex shaped isomer of B\textsubscript{12}
has $C_{3v}$ point group symmetry, with the irreducible representation
\textsuperscript{1}A\textsubscript{1} of the electronic ground state
wave function. Several authors have theoretically studied this isomer
earlier,\citep{Nature-mat-wang,Atis,Kiran} including one of us.\citep{BOUSTANI-cpl-1995,PRB97}
Our geometry optimization study reveals this to be the most stable
one amongst the eleven isomers studied (see Table \ref{tab:total-energies}),
a result in good agreement with several earlier studies performed
using lower levels of theory.\citep{kawai-JCP91,BOUSTANI-cpl-1995,PRB97}
The optimized geometry of this isomer is shown in Fig. \ref{fig:optimized-geometries}(a).
This structure is composed of one outer ring with nine boron atoms,
and an inner triangular ring with three boron atoms. The inner ring
is slightly out of plane compared with the outer ring, and has larger
bond length (1.71 Å) as compared to those in the outer ring (1.65
Å and 1.59 Å). Each inner boron atom is surrounded by six other boron
atoms, thus this quasi-planar isomer consists of three dovetailed
hexagonal pyramids. The reported bond lengths optimized using Hartree-Fock
level of theory by one of us earlier are in good agreement with our
present work.\citep{PRB97} The bond lengths obtained by us are in
good agreement with those reported by Atis \emph{et al}.\citep{Atis}
Existence of this isomer was verified experimentally by Wang \emph{et
al},\citep{Nature-mat-wang} and recently the unusual stability of
this isomer was explained theoretically by Kiran and co-workers.\citep{Kiran}
To understand the stability of this isomer from the Jahn-Teller perspective,
we present the orbital occupancy diagram in Fig. \ref{fig:orbitals-b12-quasi-planar}
based upon a single-point Hartree-Fock calculation performed at the
optimized geometry, using a cc-pVTZ basis set. From the figure it
is obvious that even though the HOMO is two-fold degenerate, it is
completely filled, and, therefore, as per Jahn-Teller theorem, it
will not undergo any further distortion. To confirm this, we performed
vibrational frequency analysis for this structure at the CCSD level
theory using cc-pVDZ basis, and found it to be completely stable,
with all the vibrational frequencies being real. The fact that this
structure has the lowest-energy of all the isomers considered, also
suggests that it is stable.

\begin{figure}
\begin{centering}
\includegraphics[scale=0.45]{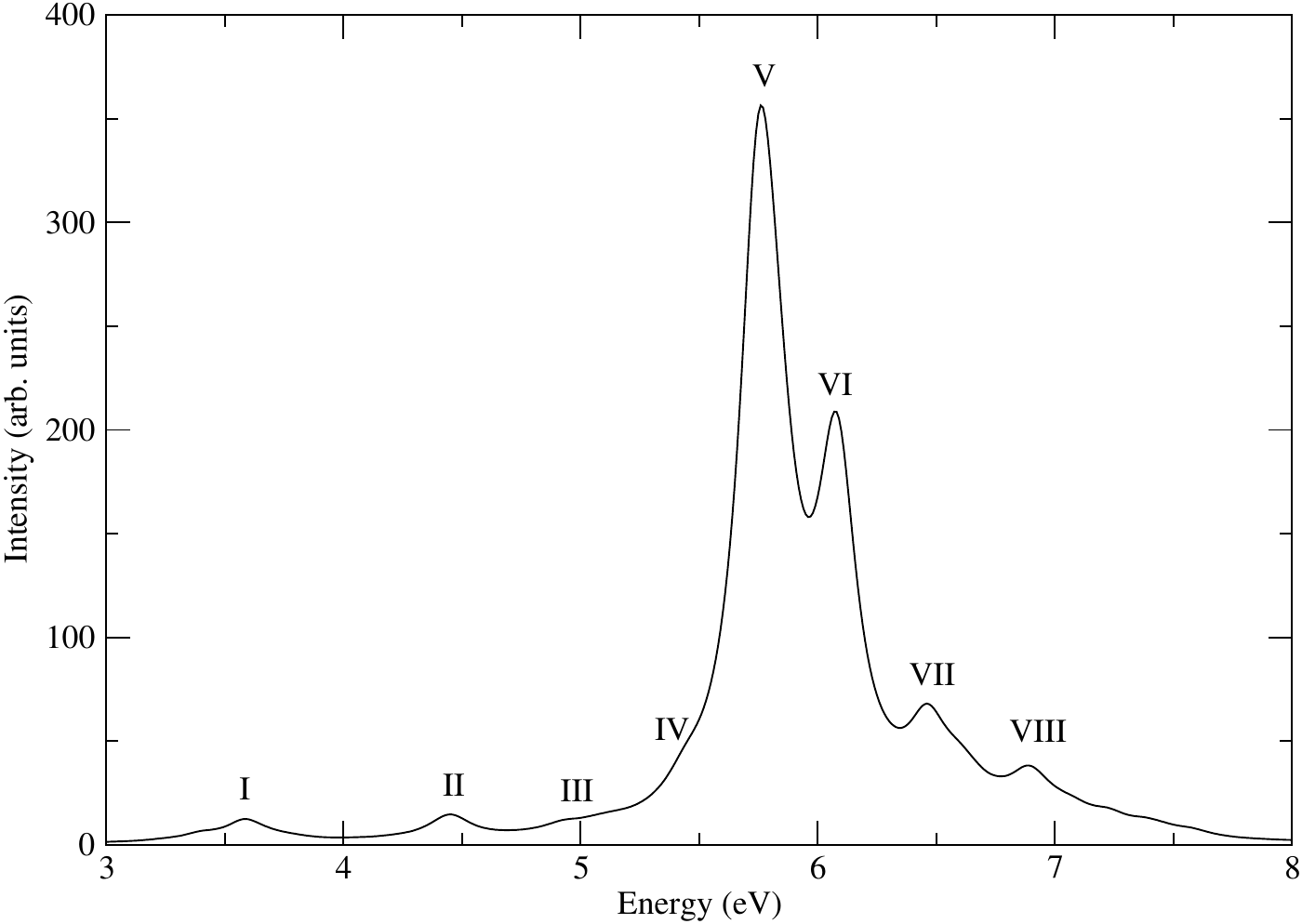} 
\par\end{centering}
\caption{Linear optical absorption spectrum of quasi-planar isomer computed
using the MRSDCI approach, and the cc-pVDZ basis set. During the calculations
forty six active orbitals were used, while all the twelve 1s core
orbitals were assumed frozen. To plot the spectrum, 0.1 eV uniform
line-width was used. \label{fig:opt-absorption-quasi-planar}}
\end{figure}

Because this is a stable isomer, we also calculated its ground state
linear optical absorption spectra presented in Fig. \ref{fig:opt-absorption-quasi-planar}.
During the electron-correlated MRSDCI calculations, C\textsubscript{s}
point group symmetry was utilized. HOMO orbitals of this isomer are
doubly degenerate, denoted as H\textsubscript{1} and H\textsubscript{2},
both of which are doubly occupied. From Fig. \ref{fig:opt-absorption-quasi-planar},
it is obvious that absorption in this isomer starts with a very small
peak (I) at 3.58 eV, followed by three weak peaks at 4.43 eV (II),
4.93 eV (III) and 5.44 eV (IV), which is a shoulder to the most intense
peak (V) located at 5.74 eV. The wave function of the excited state
corresponding to this peak is dominated by single excitation (H -
3)\textsubscript{2} $\rightarrow$ L\textsubscript{2} and (H - 3)\textsubscript{1}
$\rightarrow$ L\textsubscript{1,} with respect to the closed-shell
Hartree-Fock reference configuration. Detailed information about the
excited states contributing to the peaks, is presented in Table S1
of the supporting information.

\subsection{Double ring}

\begin{figure}
\begin{centering}
\includegraphics[width=3cm]{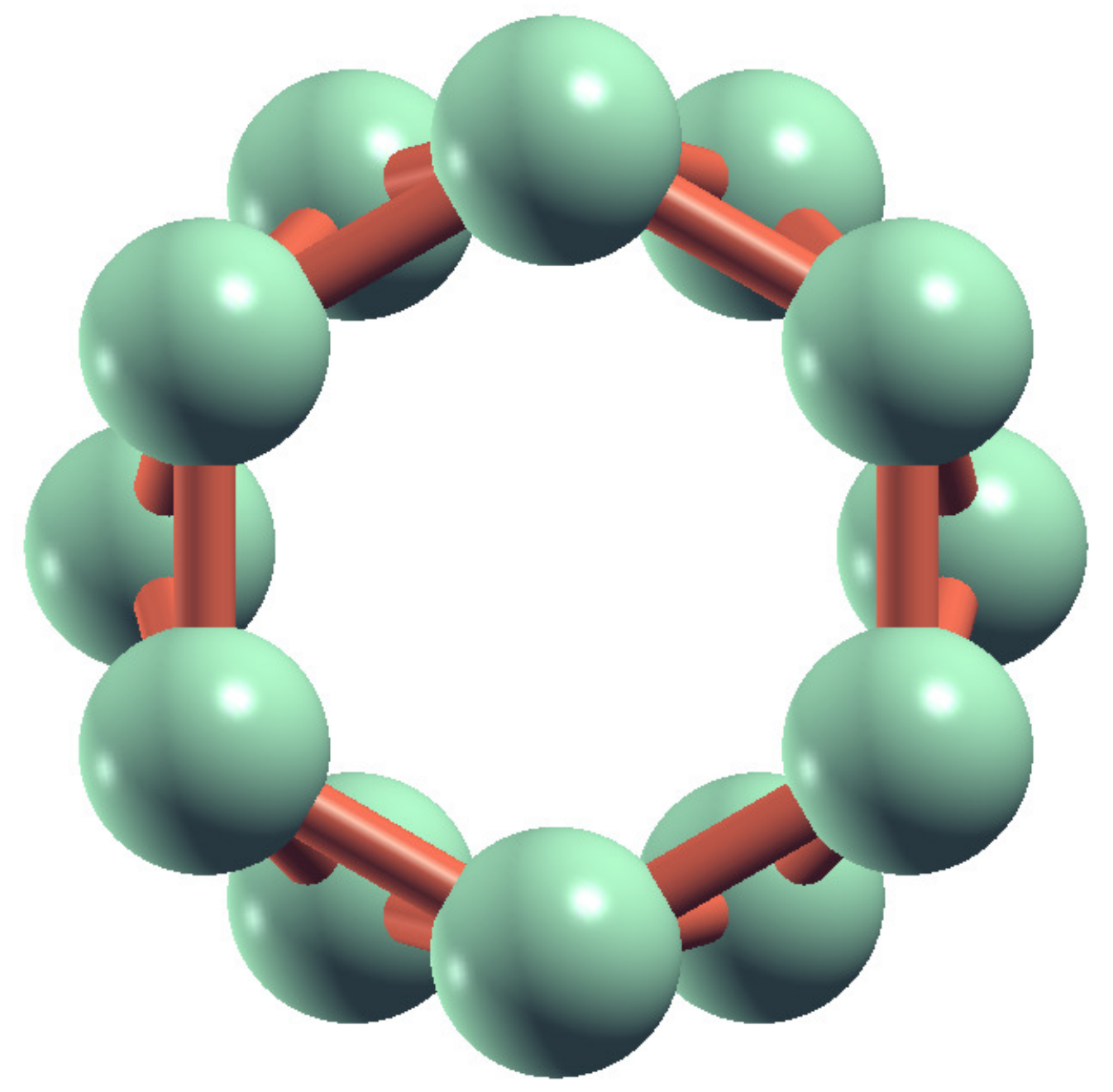} \vspace{1cm}
\par\end{centering}
\caption{Double ring structure of B$_{12}$ isomer (Top view). \label{fig:double-ring}}
\end{figure}

\begin{figure}
\begin{centering}
\includegraphics[scale=0.45]{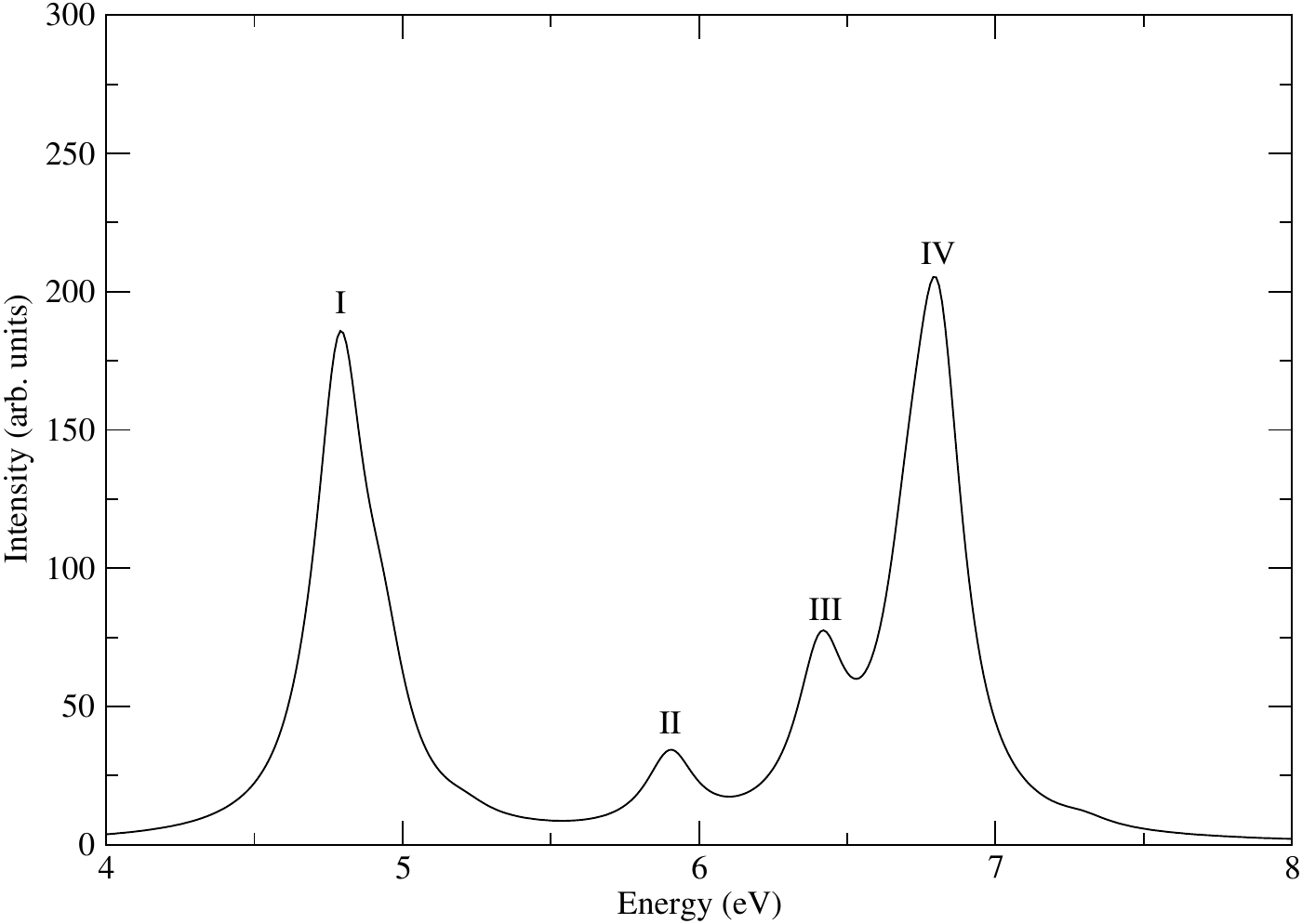} 
\par\end{centering}
\caption{Linear optical absorption spectrum of B$_{12}$ isomer with the double
ring structure, calculated using the MRSDCI approach, and the cc-pVDZ
basis set. During the calculations forty six active orbitals were
used, while all the twelve 1s core orbitals were assumed frozen. To
plot the spectrum, 0.1 eV uniform line-width is used.\label{fig:optics-double-ring}}
\end{figure}

This structure is composed of two regular hexagons of B atoms, displaced
from each other, without the boron atoms eclipsing each other, as
shown in Fig.\textcolor{red}{{} }\ref{fig:optimized-geometries}(b).
As a result, this isomer has the $D_{6d}$ point-group symmetry, along
with a closed-shell $^{1}A_{1}$ electronic ground state, with its
total energy 1.86 eV higher as compared to the lowest energy quasi-planar
isomer. The bond length of two closest in-plain and out of plain boron
atoms are 1.6458 Å and 1.7349 Å, in good agreement with recent work
of Atis\emph{ et al.} \citep{Atis} The top view of this isomer, with
its two hexagonal rings visible, is presented in Fig. \ref{fig:double-ring}.
Based upon the vibrational frequency analysis of this structure, we
conclude that this isomer is stable, and as a result we have computed
its optical absorption spectrm using the MRSDCI approach, presented
in Fig. \ref{fig:optics-double-ring}.

For the MRSDCI calculations $C_{s}$ (Abelian) point group symmetry
was employed, and in the Hartree-Fock ground state, doubly degenerate
HOMO orbitals denoted as $H_{1}$ and $H_{2}$ were fully occupied.
Other degenerate occupied and virtual molecular orbitals were also
denoted by suffixes 1 and 2 etc. The detailed information about the
wave functions of the excited states contributing to various peaks,
along with the frontier orbitals involved in those excitations, are
presented in Table S2, and Fig. 2, respectively, of the supporting
information. The computed spectrum starts with a very intense peak
(I), which is due to two nearly degenerate excited states around 4.8
eV, with wave functions dominated by singly-excited configurations
|(H - 1)\textsubscript{2} $\rightarrow$ (L+1)\textsubscript{2}$\rangle$,
|(H - 1)\textsubscript{1} $\rightarrow$ (L+1)\textsubscript{1}$\rangle$,
|(H -1)\textsubscript{1} $\rightarrow$ (L+1)\textsubscript{2}$\rangle$,
and |(H - 1)\textsubscript{2} $\rightarrow$ (L+1)\textsubscript{1}$\rangle$.
It is followed by two comparatively less intense peaks around 5.9
eV (II) and 6.4 eV (III), whose wave functions also consist mainly
of the single excitations.. The peak IV is the most intense peak near
6.8 eV, which is due to two nearly degenerate excited states, largely
due to singly-excited electrons. We note that the absorption spectrum
of this structure is sufficiently different, both qualitatively, and
quantitatively, as compared to that of the quasi-planar isomer.

\subsection{Chain}

\begin{figure}
\begin{centering}
\includegraphics[scale=0.45]{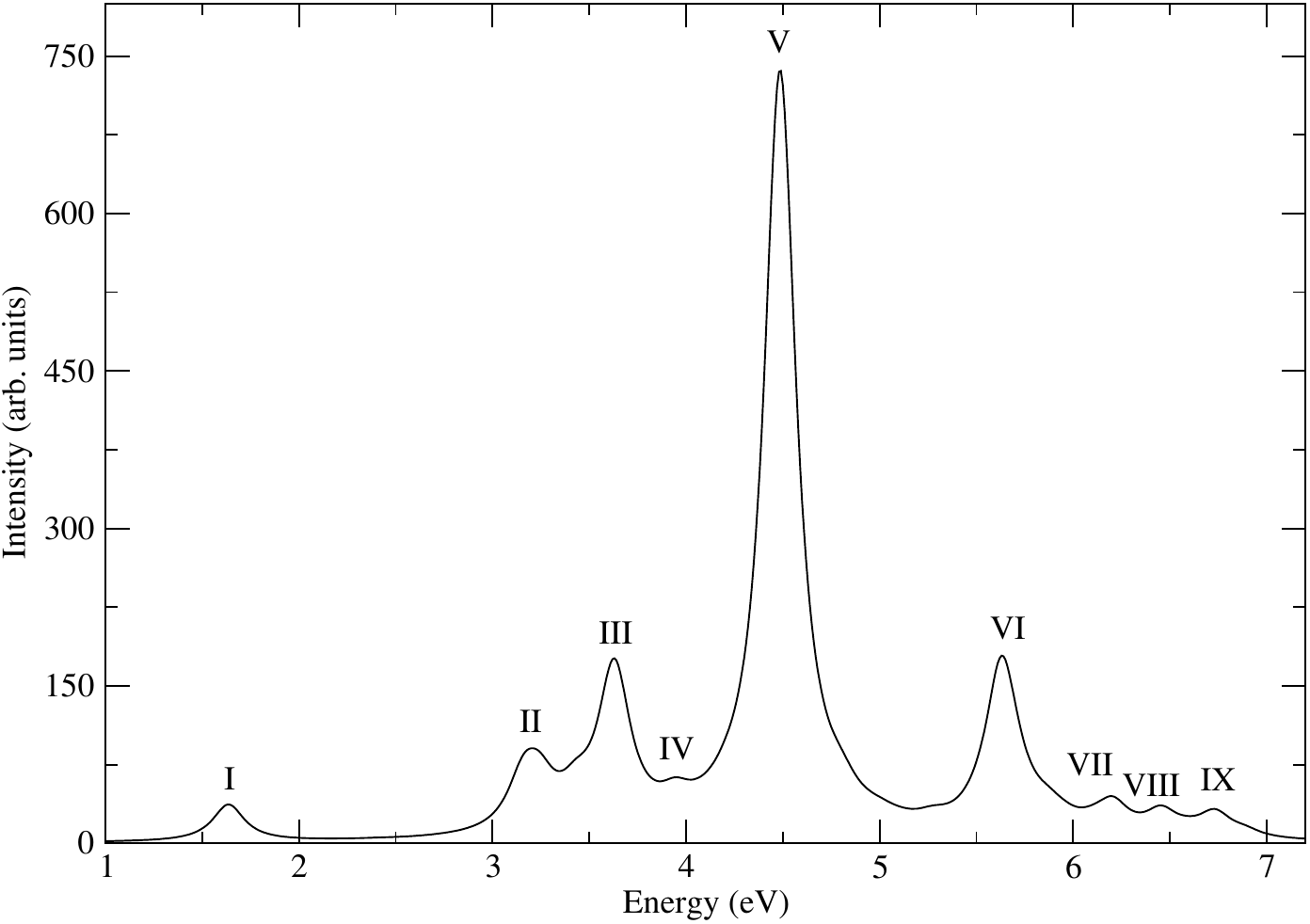} 
\par\end{centering}
\caption{Linear optical absorption spectrum of chain-like B$_{12}$ isomer
using the MRSDCI approach, and the cc-pVDZ basis set, for the photons
polarized in the plane of the chain, \emph{i.e.}, for excited states
of $^{1}B_{u}$ symmetry. During the calculations forty six active
orbitals were used, while all the twelve 1s core orbitals were assumed
frozen. To plot the spectrum, 0.1 eV uniform line-width is considered.\label{fig:optics-chain-bu}}
\end{figure}

\begin{figure}
\begin{centering}
\includegraphics[scale=0.45]{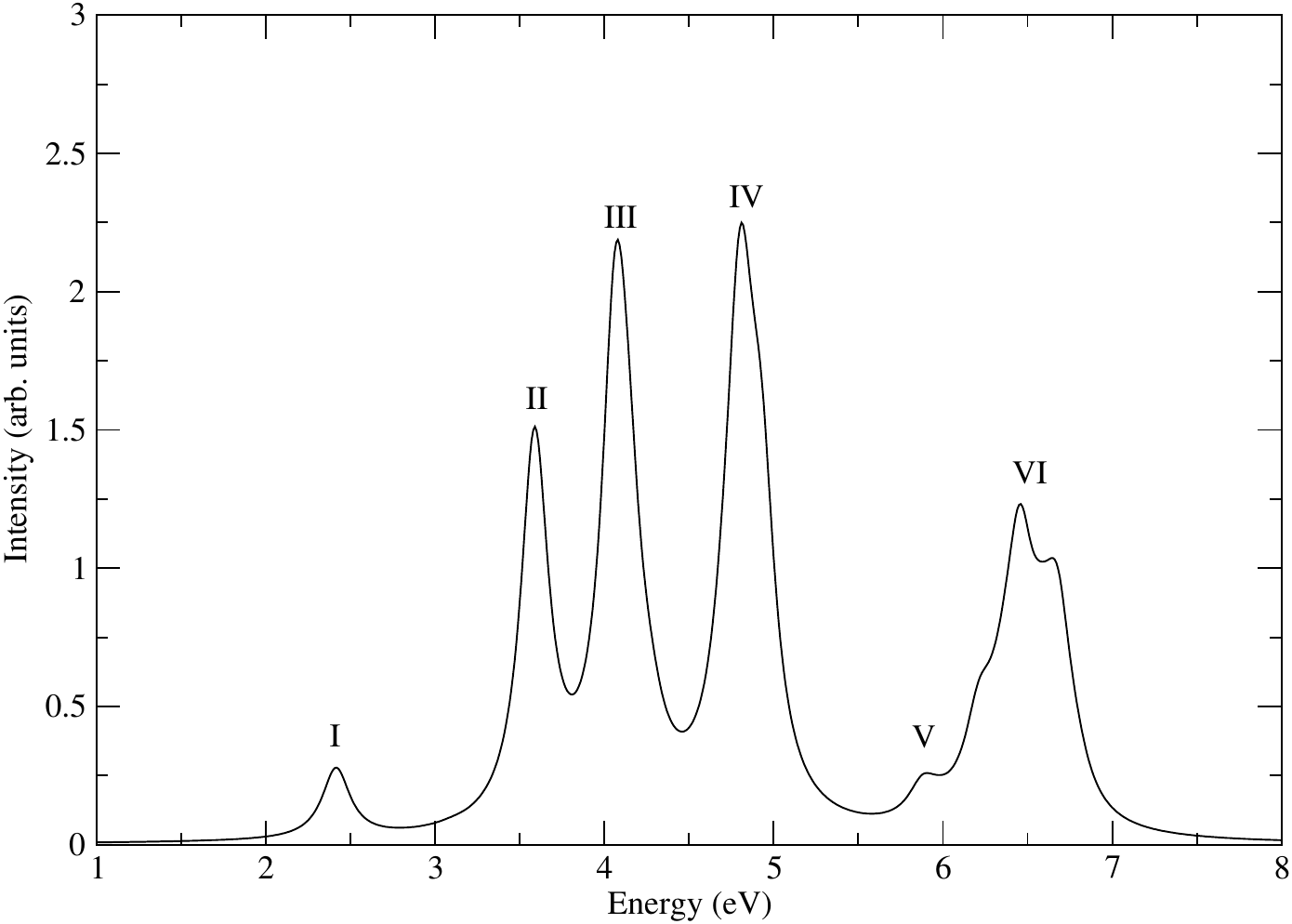} 
\par\end{centering}
\caption{Linear optical absorption spectrum of chain-like B$_{12}$ isomer,
corresponding to the excited states of \textbf{$A_{u}$ }symmetry\textbf{,}
with the photons polarized perpendicular to plane of the chain. Spectrum
was computed using the MRSDCI approach, and the cc-pVDZ basis set.
During the calculations forty six active orbitals were used, while
all the twelve 1s core orbitals were assumed frozen. To plot the spectrum,
0.1 eV uniform line-width is considered. \label{fig:chain-optics-au}}
\end{figure}

Chain-like isomer of B$_{12}$ is 2.86 eV higher in energy as compared
to the lowest-energy quasi-planar structure, and has the structure
of a ladder with rungs of the ladder not being parallel to each other
(see Fig. \ref{fig:optimized-geometries}(c)), with bond lengths ranging
from 1.55 Å\ to 1.87 Å. The point group symmetry of the isomer is
$C_{2h}$, with the electronic ground state symmetry \textsuperscript{1}A\textsubscript{g}.
We also performed the vibrational frequency analysis on this isomer,
revealing it to be stable. As a result, we have calculated the optical
absorption spectrum of this structure using the MRSDCI approach, and
the results are presented in Figs. \ref{fig:optics-chain-bu} and
\ref{fig:chain-optics-au}. In Fig. \ref{fig:optics-chain-bu}, we
have plotted the component corresponding to the photons polarized
in the plane of the chain, leading to the absorption to the $^{1}B_{u}$
type excited states, and this component carries the bulk of the absorption
intensity. In Fig. \ref{fig:chain-optics-au}, we plot the absorption
spectrum of the photons polarized perpendicular to the plane of the
chain, corresponding to $^{1}A_{u}$ excited states. We have plotted
two absorption components separately because of the huge difference
in the intensities involved. It is quite understandable that the bulk
of the oscillator strength is carried by absorption polarized in the
plane of the chain, because that is where the atoms, and hence, electrons
are, leading to large transition dipole moments with respect to the
ground state.

The in-plane absorption begins with a small peak (I) at 1.64 eV dominated
by |H $\rightarrow$ L$\rangle$ and |H-1 $\rightarrow$ L+1$\rangle$.
The most intense peak (V) appears near about 4.5 eV dominated by |H
$\rightarrow$ L$\rangle$ and |H-1 $\rightarrow$ L+1$\rangle$ single
excitations. From Fig. 3 of supporting information it is obvious that
the orbitals involved $H-1$/$L+1$ have $\pi/\pi^{*}$ character,
while $H/L$ have $\sigma/\sigma^{*}$ character, thus these transition
have a mixed $\pi-\pi^{*}+\sigma-\sigma*$ character. The detailed
many-particle wave functions of the excited states associated with
the peaks of the optical absorption spectra are presented in Table
S3 of the supporting information.

As far as the out-of-plane component of the absorption spectrum is
concerned, it is obvious from Fig. \ref{fig:chain-optics-au} and
from Table S4 of the supporting information, that oscillator strengths
corresponding to peaks are at least two orders of magnitude smaller
than those of ``in-plane'' component. This component of the absorption
starts at 2.42 eV, with its most intense peak IV located at 4.85 eV.
From Table S4, it is clear that the wave functions of the excited
states contributing to the peaks are dominated by singly excited configurations
involving orbitals away from the Fermi level, and doubly excited configurations
involving orbitals close to the Fermi level. However, detection of
these resonances will be a difficult task unless it is possible to
orient the isomer perpendicular to the polarization direction of the
incident photons.

Comparison of the optical absorption spectrum of the chain isomer
with those of the quasi-planar and the double ring structures, clearly
reveals qualitative and quantitative differences which can be utilized
as the fingerprints of these structures, and thus can be used in their
optical detection.

\subsection{Hexagons}

This planar isomer belongs to D\textsubscript{2h} point group with
the electronic ground state symmetry \textsuperscript{1}A\textsubscript{g}
and has the appearance of a boron dimer surrounded by ten other boron
atoms. One of us discovered this structure earlier using Hartree-Fock
level geometry optimization, and a 3-21G basis set.\citep{PRB97}
In this work the optimized geometry obtained using the CCSD level
of theory, and cc-pVDZ basis set, is shown in Fig. \ref{fig:optimized-geometries}(d).
For this structure, the average bond length in between two consecutive
peripheral boron atoms is 1.56 Å, while the bond length between any
of the central atom, and the closest peripheral atom is near about
1.8 Å.

\subsection{Quad}

This isomer has a three-dimensional structure shown in Fig. \ref{fig:optimized-geometries}(g),
with a D\textsubscript{2h} point group, along with the electronic
ground state of symmetry \textsuperscript{1}A\textsubscript{g}.
The structure consists of two six-membered rings of boron atoms, lying
in mutually perpendicular planes. In one plane the B-B bond lengths
are almost equal, i.e., near about 1.65 Å, but in the other plane
it has two distinct bond lengths of 1.72 Åand 1.83 Å. This isomer
is predicted to be 5.17 eV higher in energy as compared to the lowest
energy one.

\subsection{Pris-3-square-1}

This isomer also has a three-dimensional structure of a cuboidal shape,
whose two faces are squares, while the remaining four are rectangles.
The length of the sides of square shaped faces is 1.73 Å, and the
two squares are 2.53 Å\ apart from each other. Furthermore, four
opposite faces as shown in Fig. \ref{fig:optimized-geometries}(h)
are capped by a boron atom each in a symmetric manner, with the nearest
distance between the capping atom and the cube atom being 1.69 Å.
This structure has D\textsubscript{4h} point group symmetry, and
it is 8.5 eV higher in energy than the lowest energy one, with the
electronic ground state symmetry \textsuperscript{1}A\textsubscript{1g}. 

\subsection{Pris-3-square-2}

This isomer has a very similar structure as Pris-3-square isomer discussed
in the previous section, except that the length of the sides of the
square-shaped faces is slightly larger at 1.79 Å, and the distance
between the opposite squares is slightly shorter at 1.98 Å. As shown
in Fig. \ref{fig:optimized-geometries}(i), distance of the capping
atoms from the corresponding faces is also slightly shorter at 1.66
Å. With a structure similar to that Pris-3-square-1, this isomer also
has D\textsubscript{4h} point group symmetry, with the electronic
ground state symmetry of \textsuperscript{1}A\textsubscript{1g}.
In spite of a structure similar to that of Pris-3-square-1 isomer,
it is energetically about 2 eV higher than that, and 10.58 eV higher
than the lowest energy quasi-planar structure.

\subsection{Parallel triangular}

B\textsubscript{12} - parallel triangular isomer has D\textsubscript{3h}
point group symmetry, with the electronic ground state symmetry \textsuperscript{1}A$^{\prime}$\textsubscript{1}.
As shown in Fig. \ref{fig:optimized-geometries}(j), it consists of
four parallel equilateral triangles arranged symmetrically about a
central plane. The first/last of those triangles has edge length 1.71
Å, while the two middle triangles have equal bond lengths of 1.85
Å. The connecting bond length in between two middle triangles have
edge length 1.85 Å. The interplanar distances between the first two
and the last two triangles is close to 1.6 Å, while that between the
middle triangles is 1.85 Å. This structure is roughly 12 eV higher
in energy as compared to the lowest energy structure.

\subsection{Planar square}

This isomer which is about 13.5 eV higher than the lowest energy structure,
is strictly a planar one, with the D\textsubscript{4h} point group,
and the electronic ground state symmetry \textsuperscript{1}A\textsubscript{1g}.
As shown in Fig. \ref{fig:optimized-geometries}(k), it is composed
of four isosceles trapezoids connecting with each other at a common
side, resulting in a square-like symmetry. The equal sides of isosceles
trapezoid have the bond length of 1.75 Å, while the other two sides
have the bond lengths of 1.56 Å, and 1.66 Å.

\section{Conclusions}

\label{sec:conclusions}

In conclusion, we presented a detailed \emph{ab initio} electron-correlated
study of the electronic structure and the ground state geometries
of several isomers of B$_{12}$ cluster. In agreement with the earlier
works, quasi-planar convex structure was found to be the lowest energy
structures, however, several other structures corresponding to the
local minima on the potential energy surface were also identified.
In particular, our calculations confirm that B$_{12}$ perfect icosahedral
structure with $I_{h}$ symmetry is unstable against structural distortions.
Furthermore, Jahn-Teller analysis of the icosahedral isomer revealed
that it may be unstable because at the single-electron level of theory,
it has several unfilled degenerate HOMO orbitals. On the other hand,
a similar analysis of the lowest-energy quasi-planar structure reveals
that the isomer has completely filled doubly-degenerate HOMO, signaling
its stability against Jahn-Teller distortion. These results are also
in agreement with the vibrational frequency analysis which also predicts
the singlet $I_{h}$ structure to be unstable, and the quasi-planar
structure to be stable. However, whether Jahn-Teller distortion causes
B$_{12}$ perfect icosahedron into the quasi-planar structure, can
be resolved by other calculations, such as the ones based upon linear-transit
theory connecting the potential energy surface of the $I_{h}$ structure,
to that of the quasi-planar one, which at present is beyond the scope
of this work. Vibrational frequency analysis also predicted two other
isomers, namely double ring, and the chain structures, to be stable.
For these three stable structures, namely, quasi-planar, double ring,
and the chain isomer, we also calculated the optical absorption spectra
employing the state-of-the-art MRSDCI methodology, utilizing large-scale
CI expansions. The calculated spectra revealed strong structure-property
relationship which can be utilized for detection of different isomers
using optical techniques.

\section*{Acknowledgements}

Work of P.B. was supported by a Senior Research Fellowship offered
by University Grants Commission, India.

\bibliographystyle{unsrt}
\addcontentsline{toc}{section}{\refname}\bibliography{b12}

\end{document}


\begin{frontmatter} 

\title{Supporting Information: First Principles Study of Structural and
Optical Properties of B$_{12}$ Isomers}

\author{Pritam Bhattacharyya}

\address{Department of Physics, Indian Institute of Technology Bombay, Powai,
Mumbai 400076, India}

\ead{pritambhattacharyya01@gmail.com}

\author{Ihsan Boustani }

\address{Theoretical and Physical Chemistry, Faculty of Mathematics and Natural
Sciences, Bergische Universit\"at, Wuppertal, Gauss Strasse 20, D-42097
Wuppertal, Germany }

\ead{boustani@uni-wuppertal.de}

\author{Alok Shukla}

\address{Department of Physics, Indian Institute of Technology Bombay, Powai,
Mumbai 400076, India}

\ead{shukla@phy.iitb.ac.in}

\end{frontmatter}

Here we present the plots of the frontier orbitals, and the wave functions
of important excited states contributing to the calculated optical
absorption spectra of three B$_{12}$ isomers, namely: (a) quasi-planar,
(b) double ring, and (c) chain. 

\section{Quasi-Planar Isomer}

\begin{figure}[H]
\includegraphics[scale=0.04]{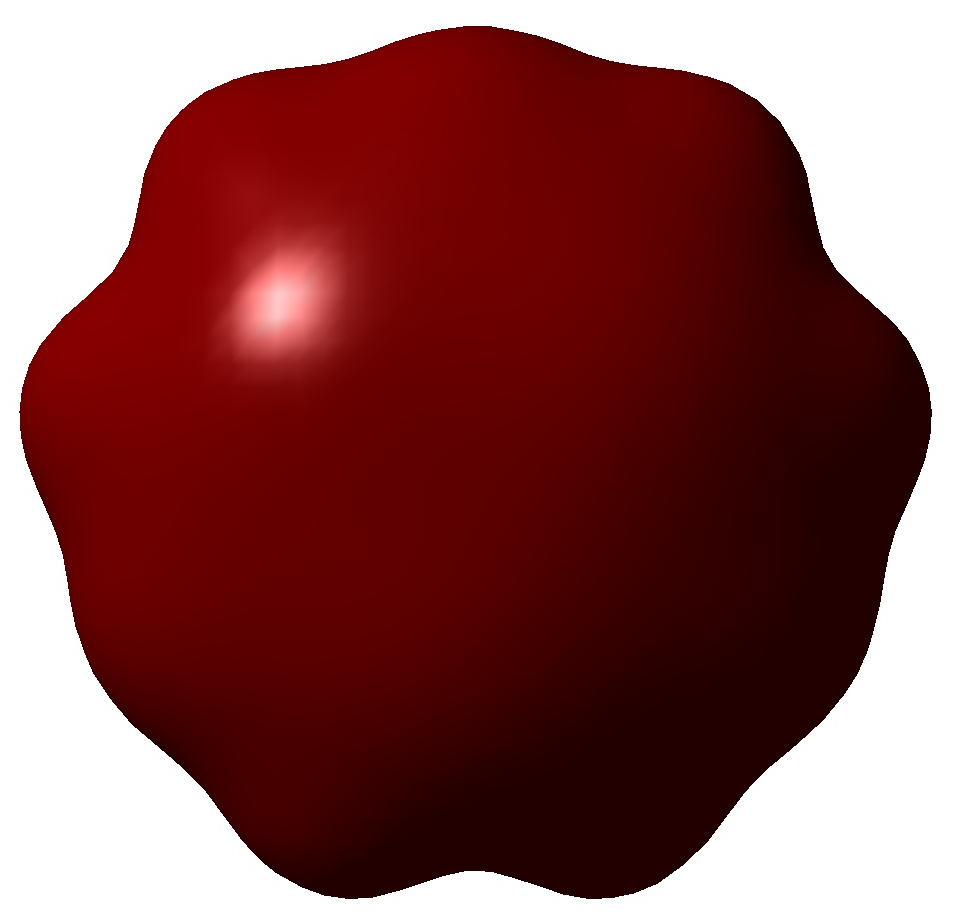}~~~\includegraphics[scale=0.04]{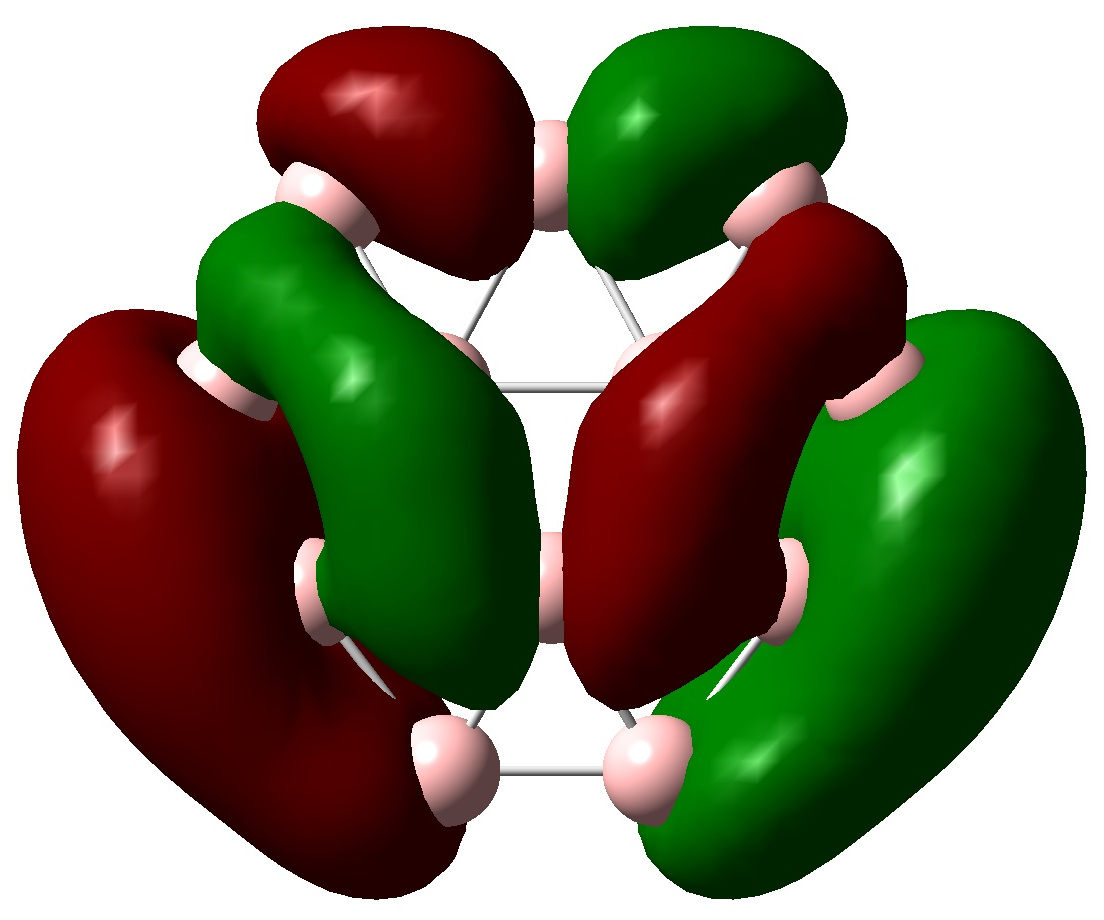}~~~\includegraphics[scale=0.04]{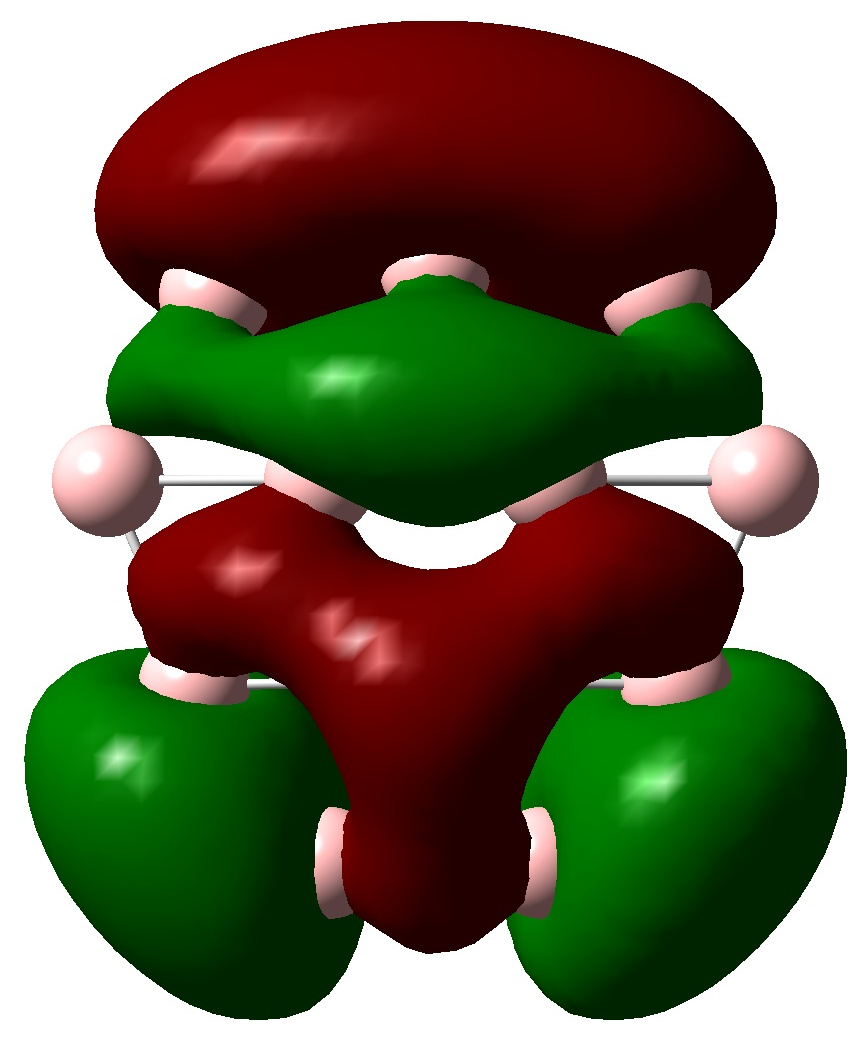}~~~\includegraphics[scale=0.04]{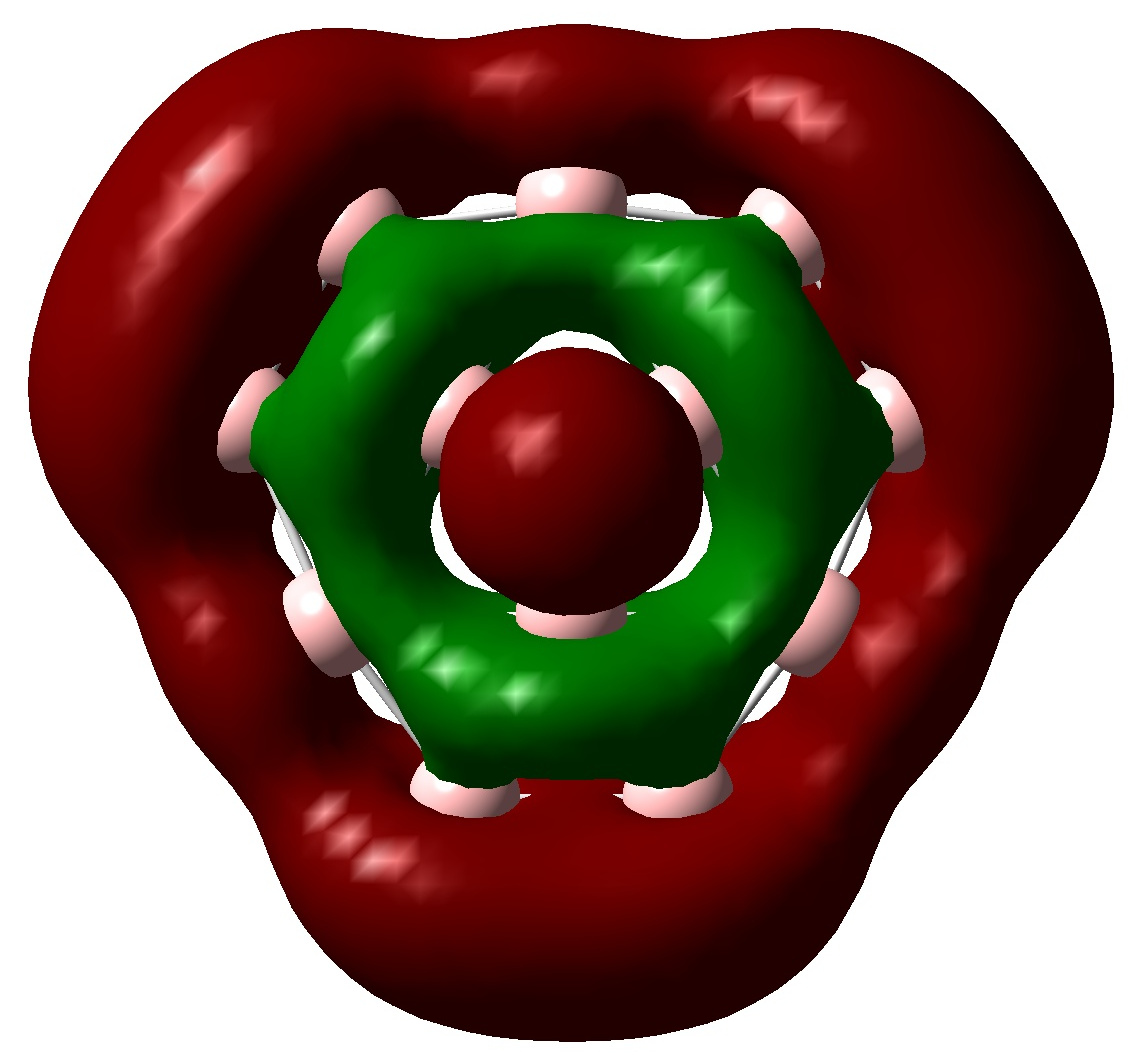}~~~\includegraphics[scale=0.04]{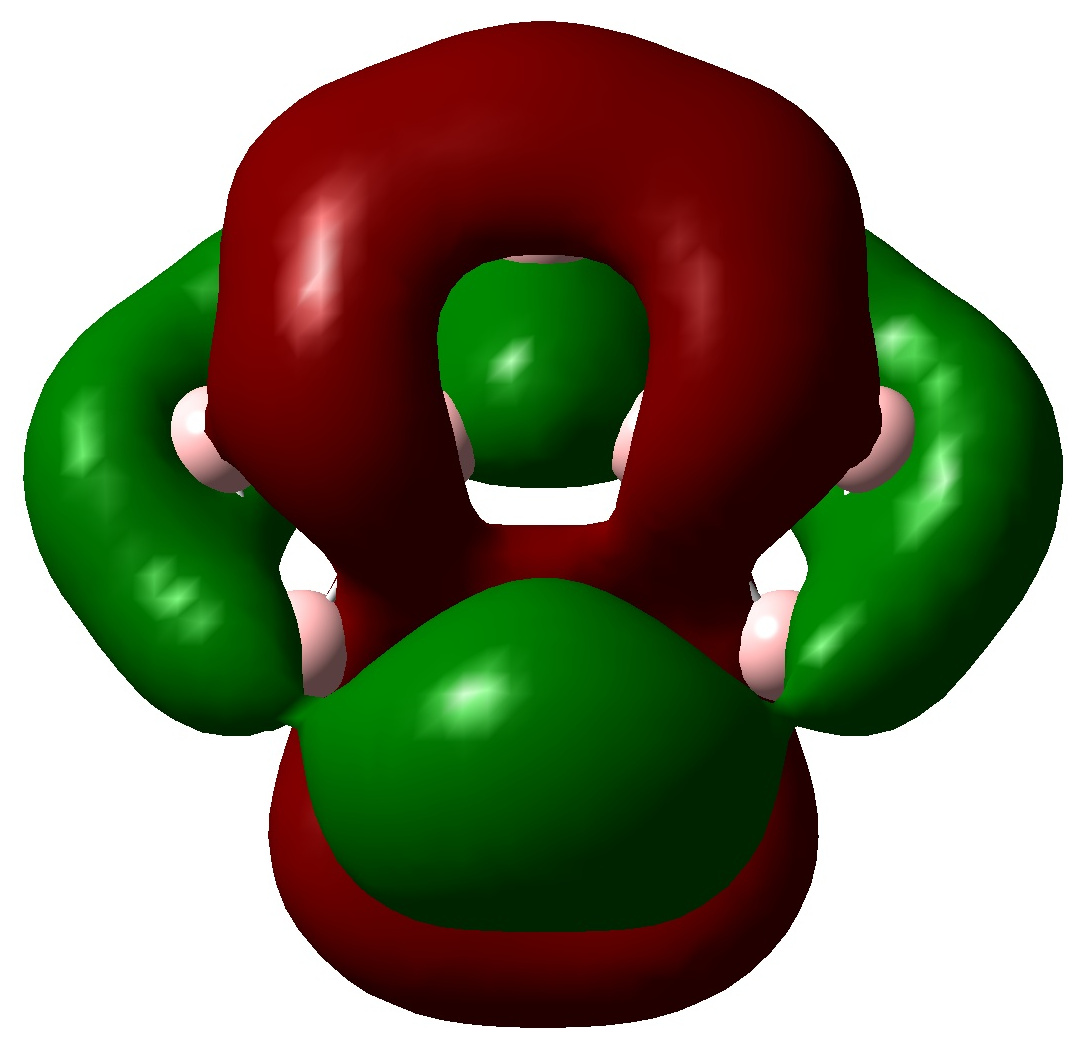}~~~\includegraphics[scale=0.04]{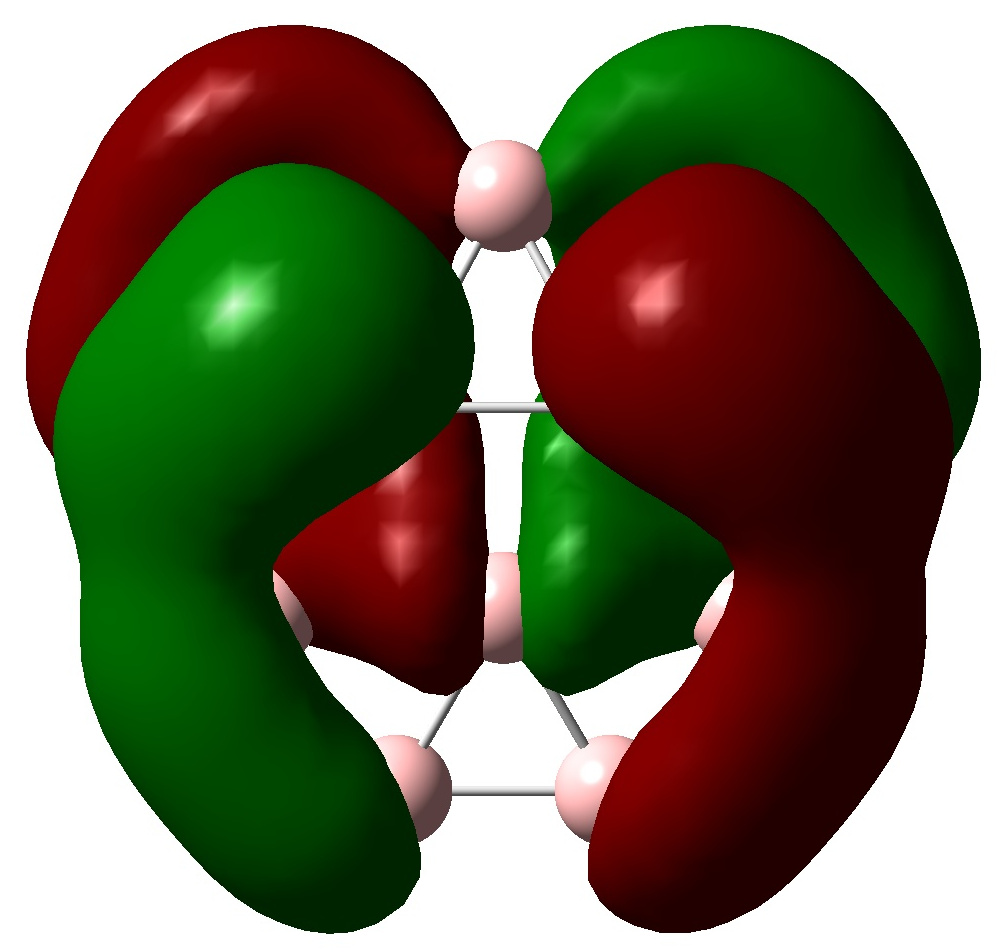}~~~\includegraphics[scale=0.04]{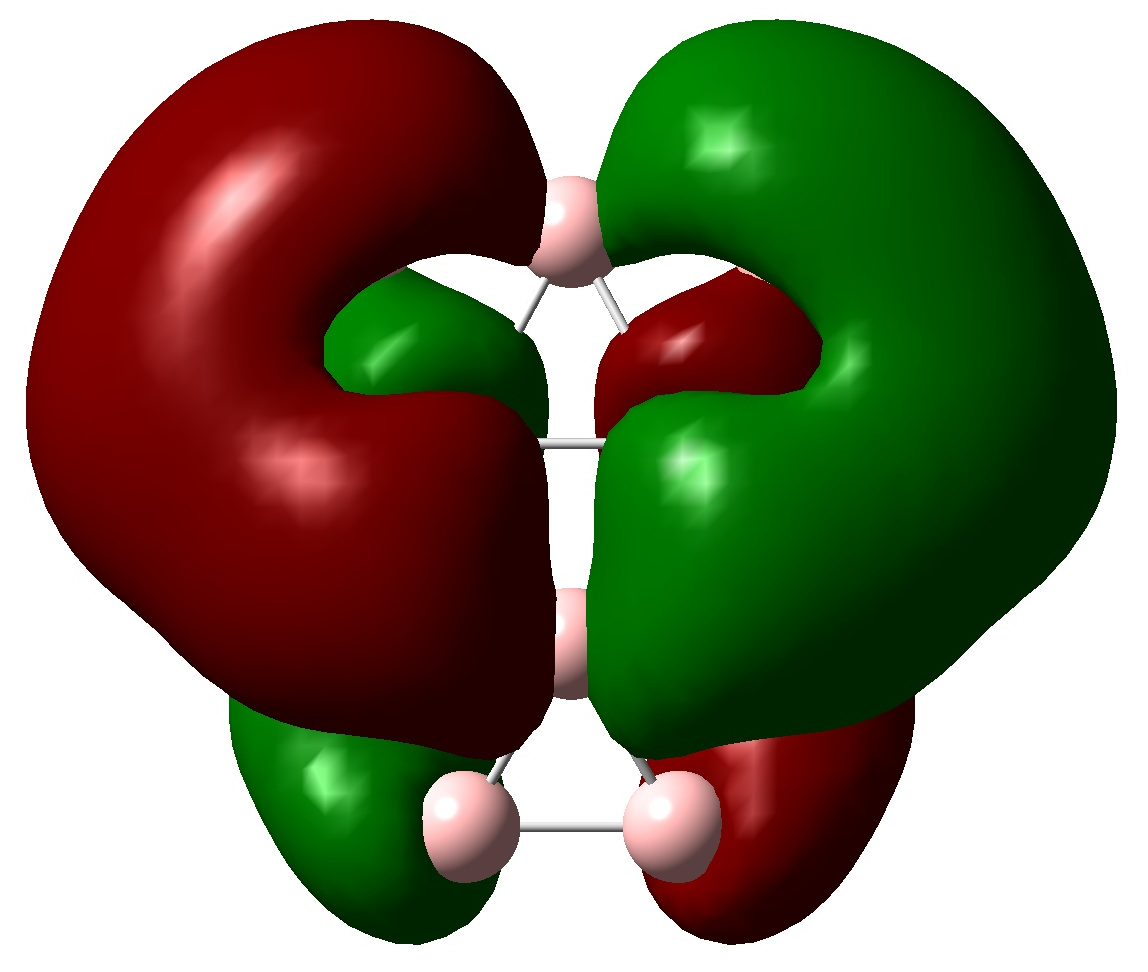}~~~\includegraphics[scale=0.04]{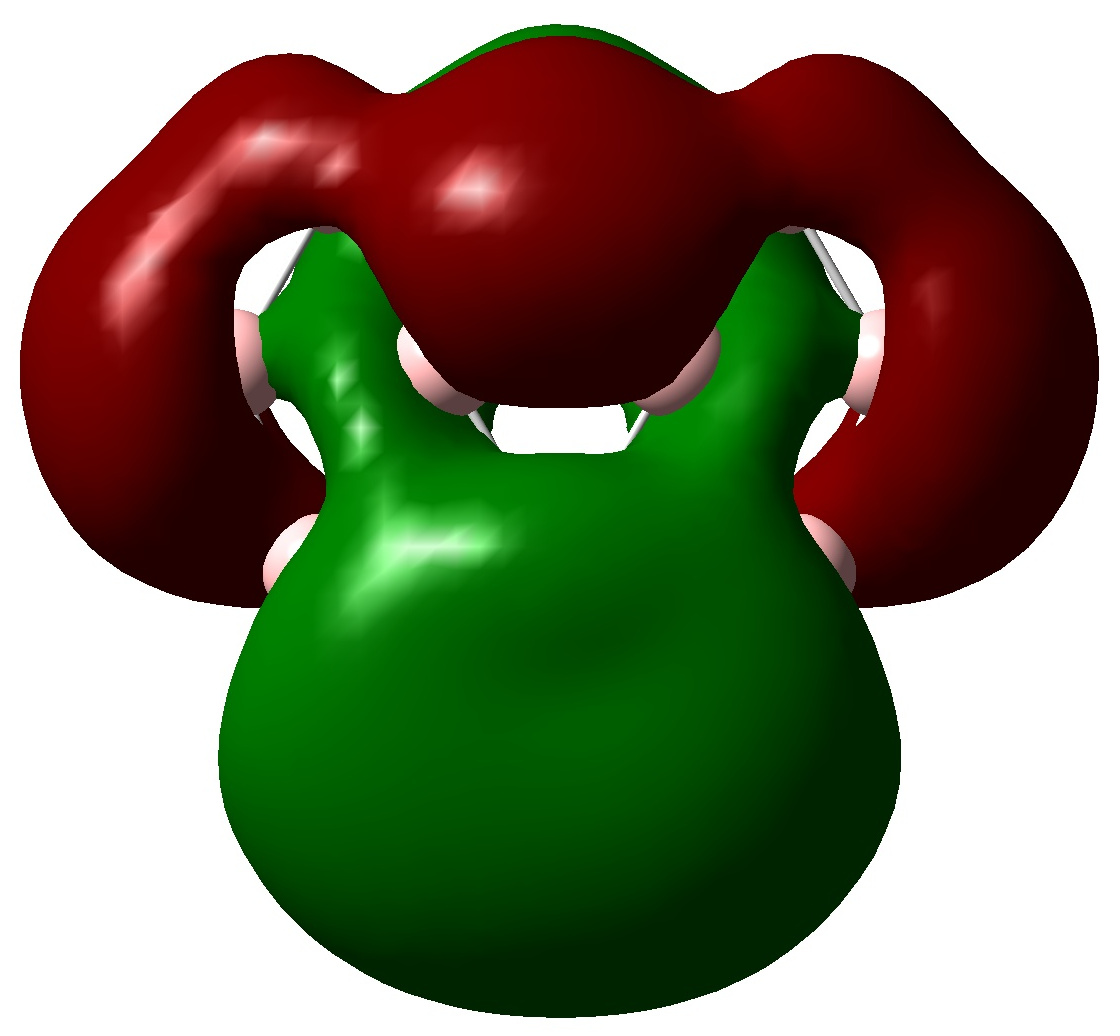}

~~\ \textbf{H-4}~\textbf{~~~~~~~(H-3)}\textsubscript{\textbf{2}}~~~~\textbf{~(H-3)}\textsubscript{\textbf{1}}~~~~~\textbf{~H-2}~~~~~~~\textbf{~~(H-1)}\textsubscript{\textbf{2}}~~~~~~\textbf{(H-1)}\textsubscript{\textbf{1}}\textbf{~~~~~~~~H}\textsubscript{\textbf{2}}\textbf{~~~~~~~~~~~H}\textsubscript{\textbf{1}}

\vspace{1cm}

\includegraphics[scale=0.04]{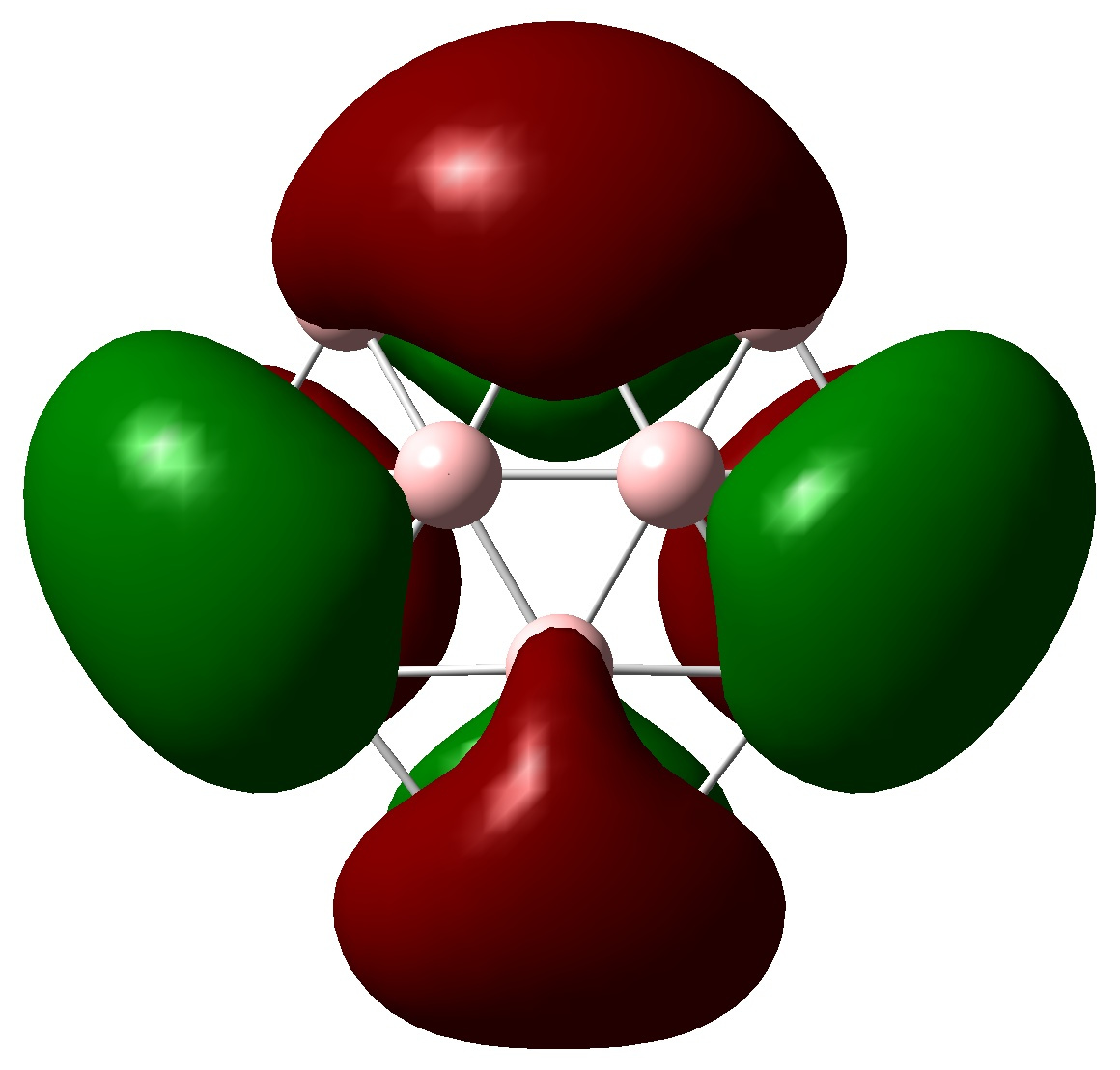}~~~\includegraphics[scale=0.04]{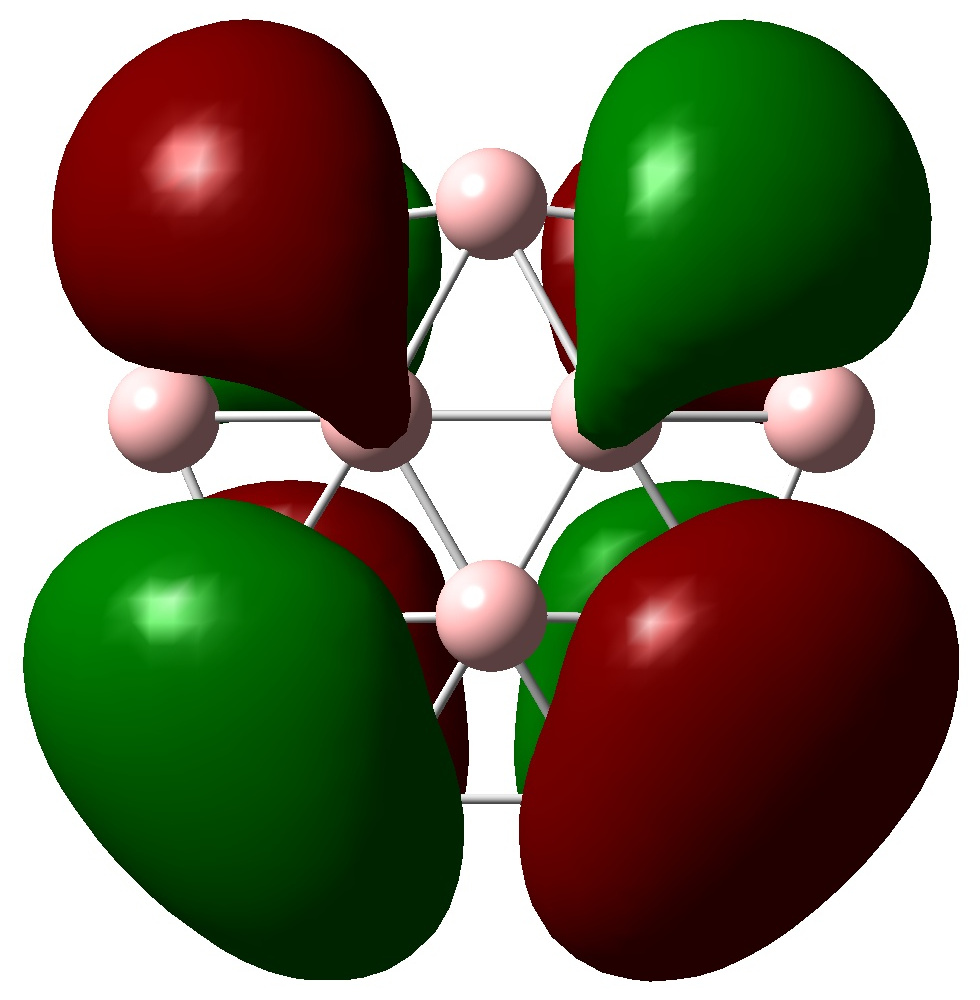}~~~\includegraphics[scale=0.04]{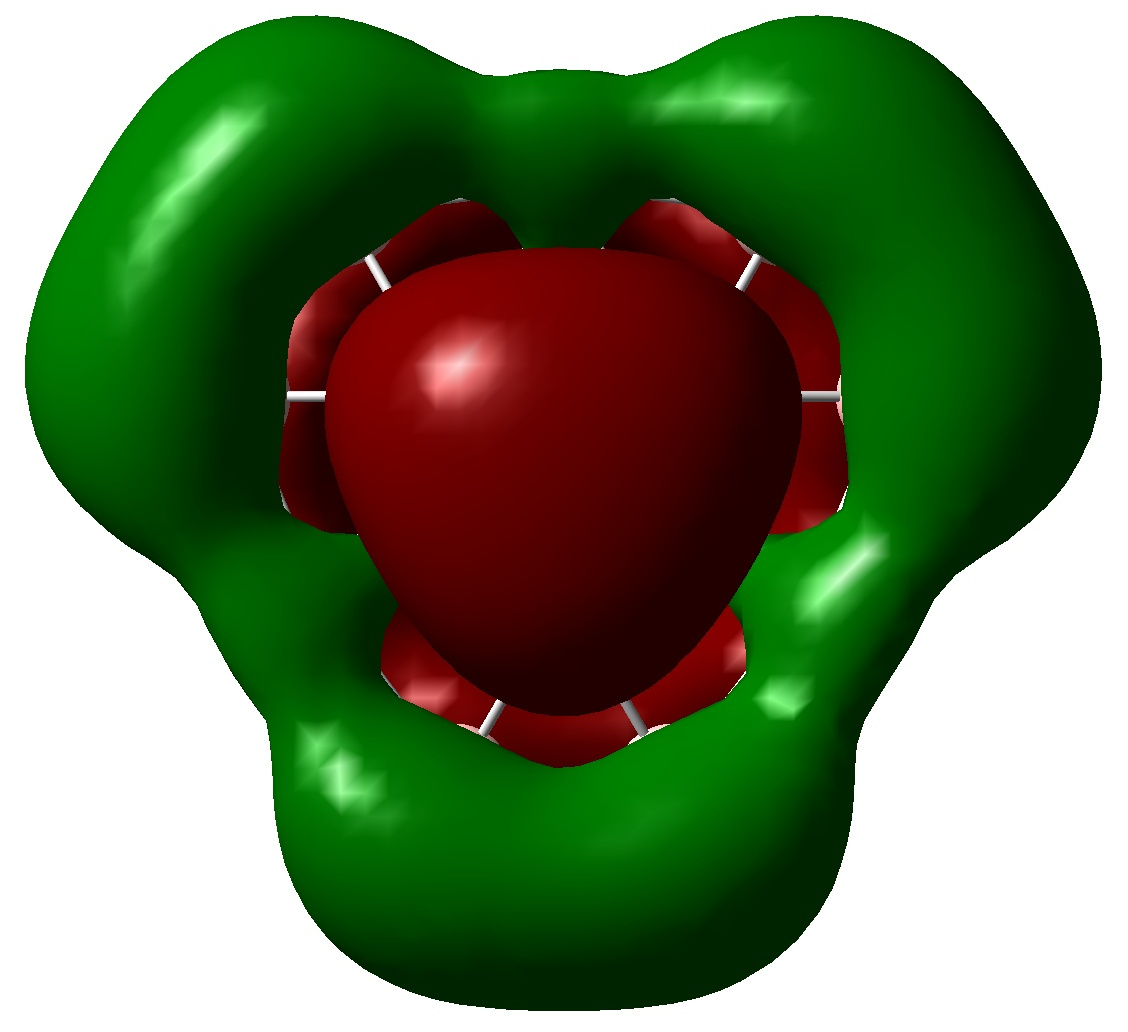}~~~\includegraphics[scale=0.04]{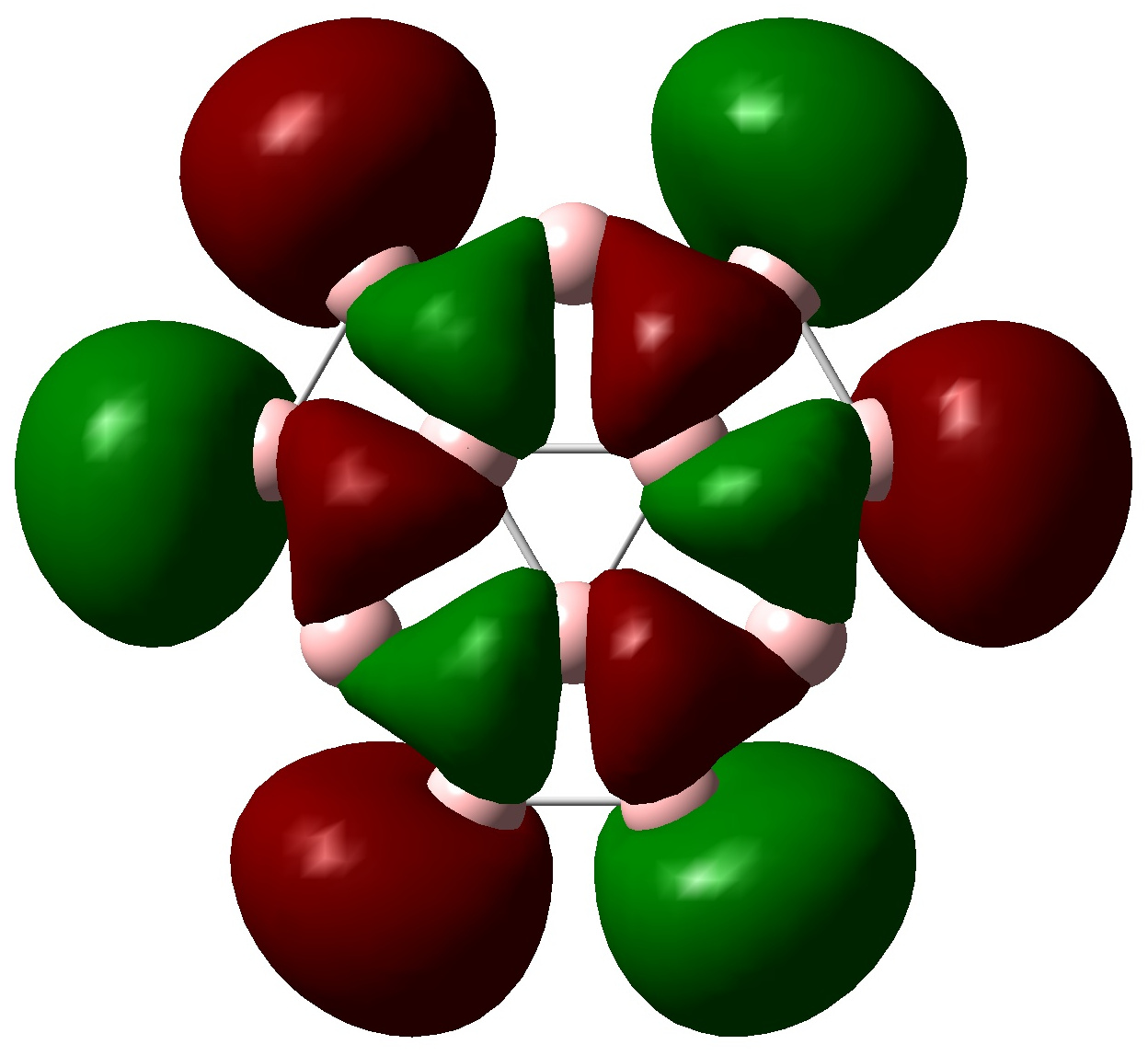}~~~\includegraphics[scale=0.04]{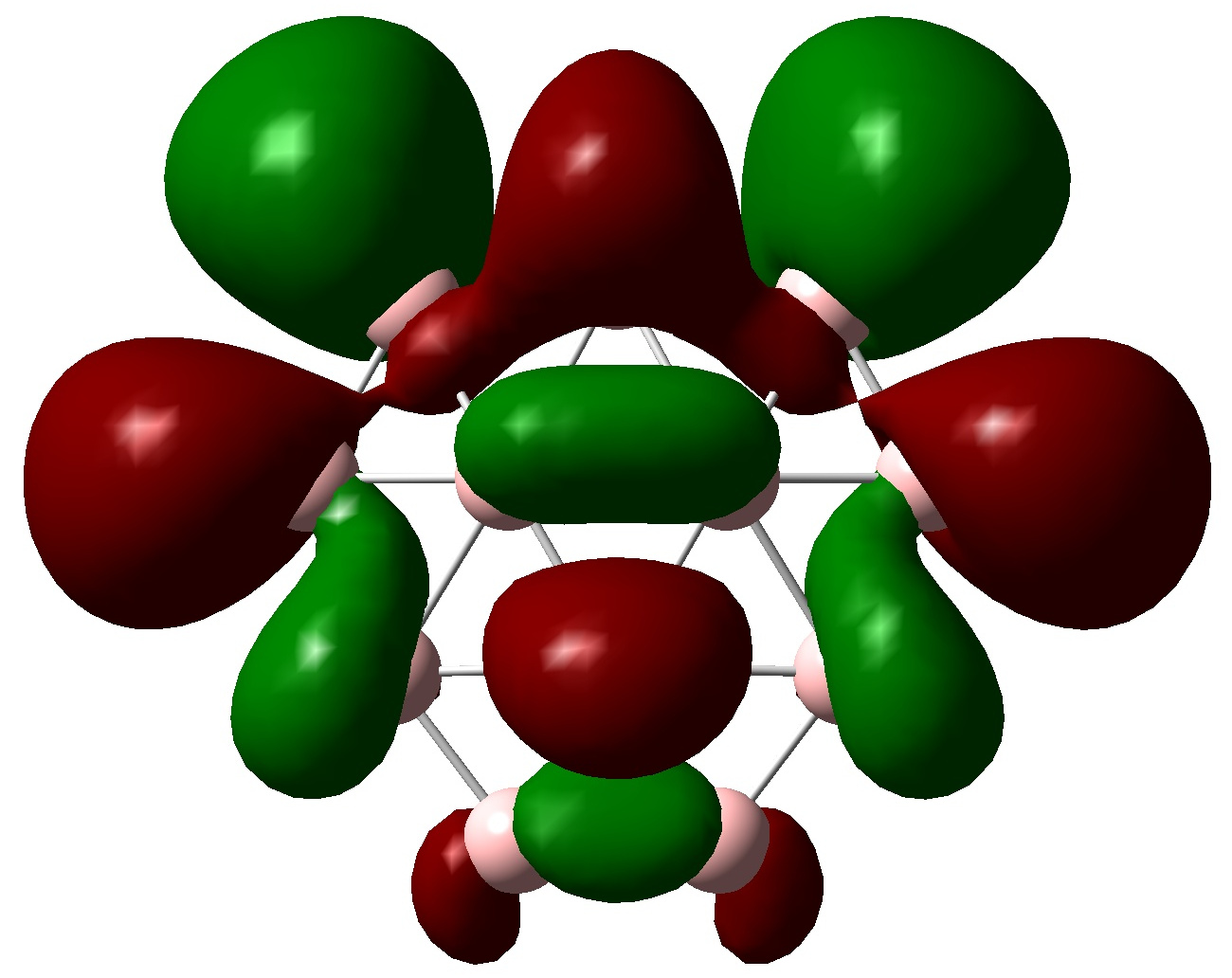}~~~\includegraphics[scale=0.04]{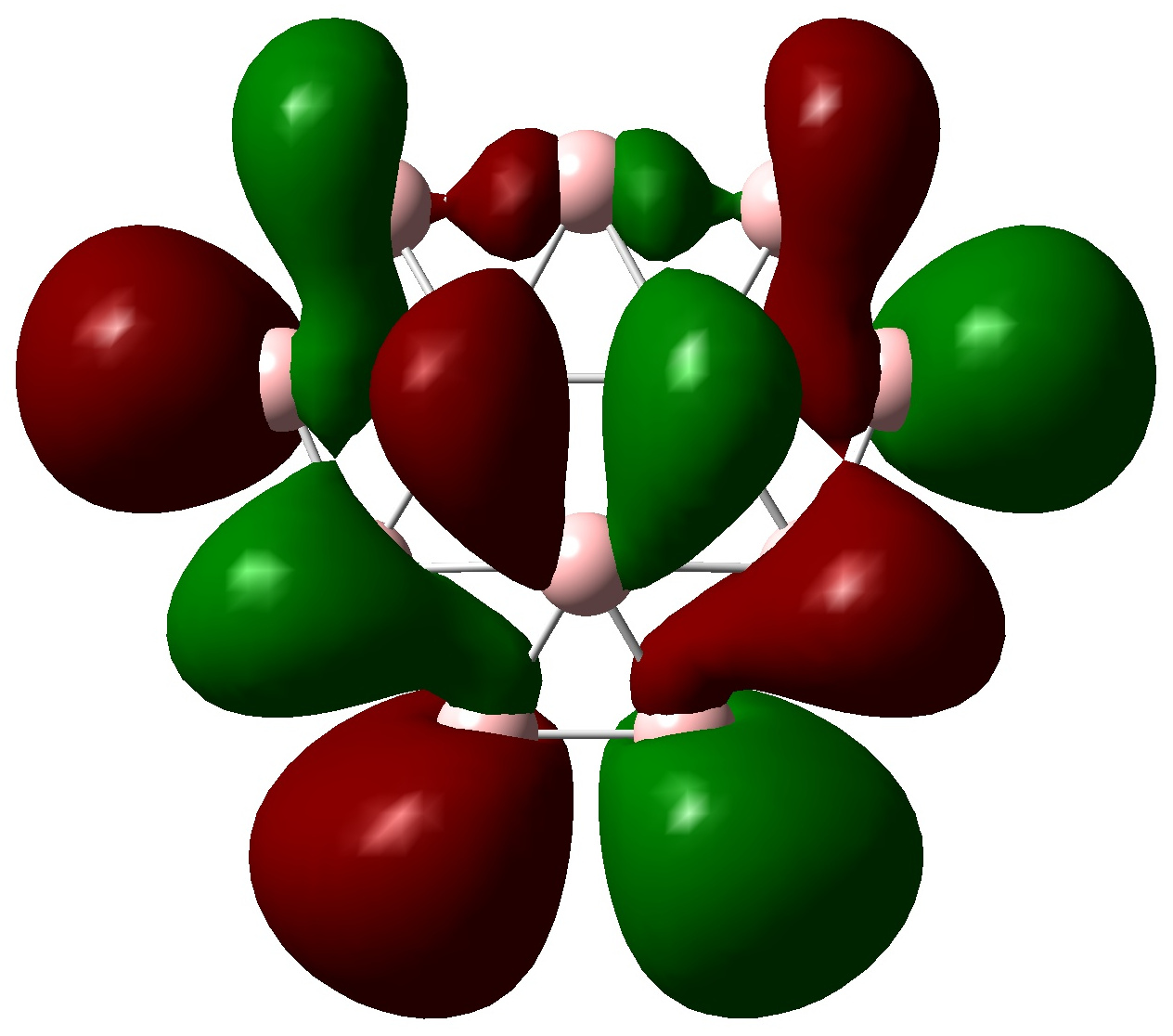}~~~\includegraphics[scale=0.04]{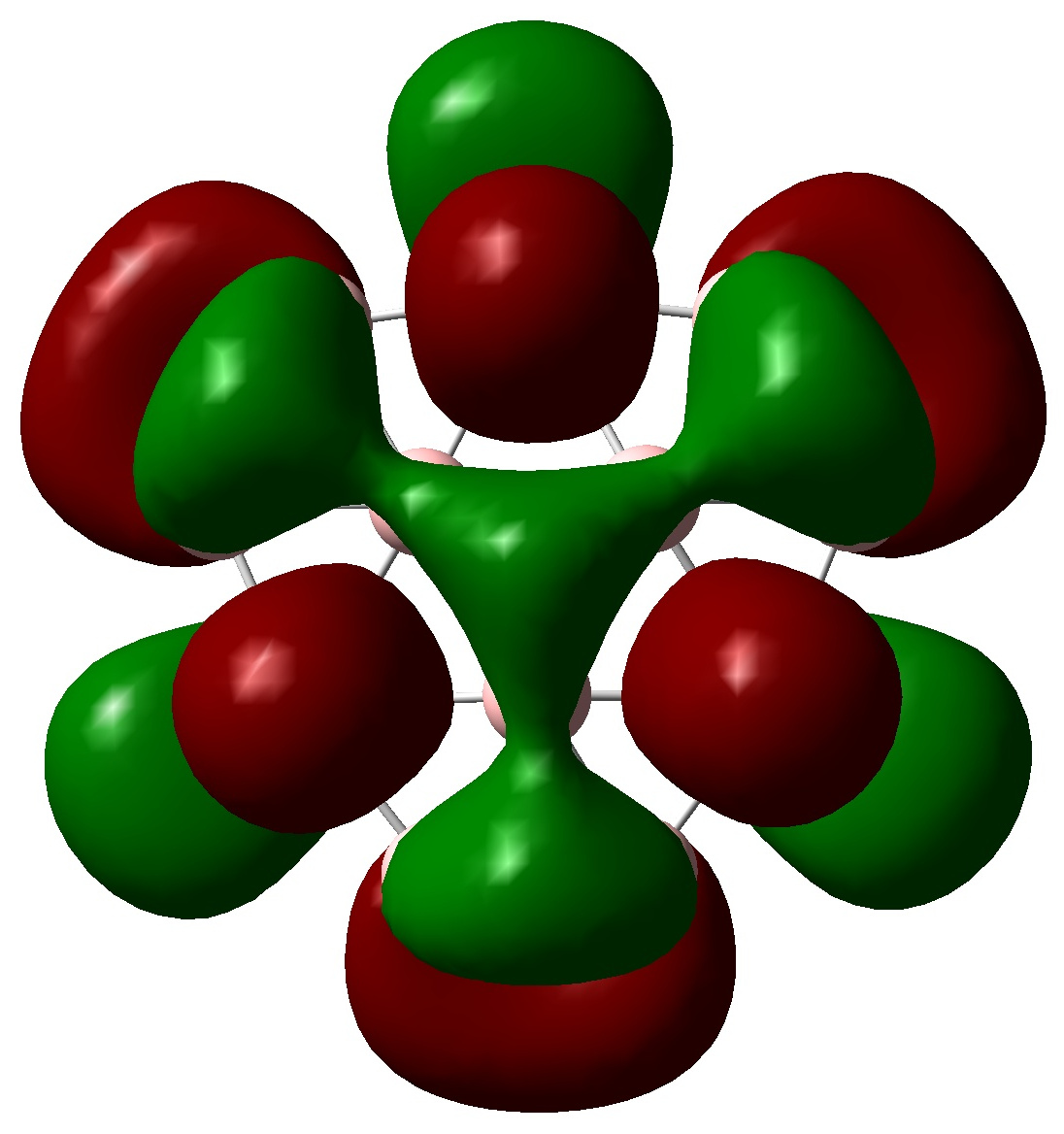}~~~\includegraphics[scale=0.04]{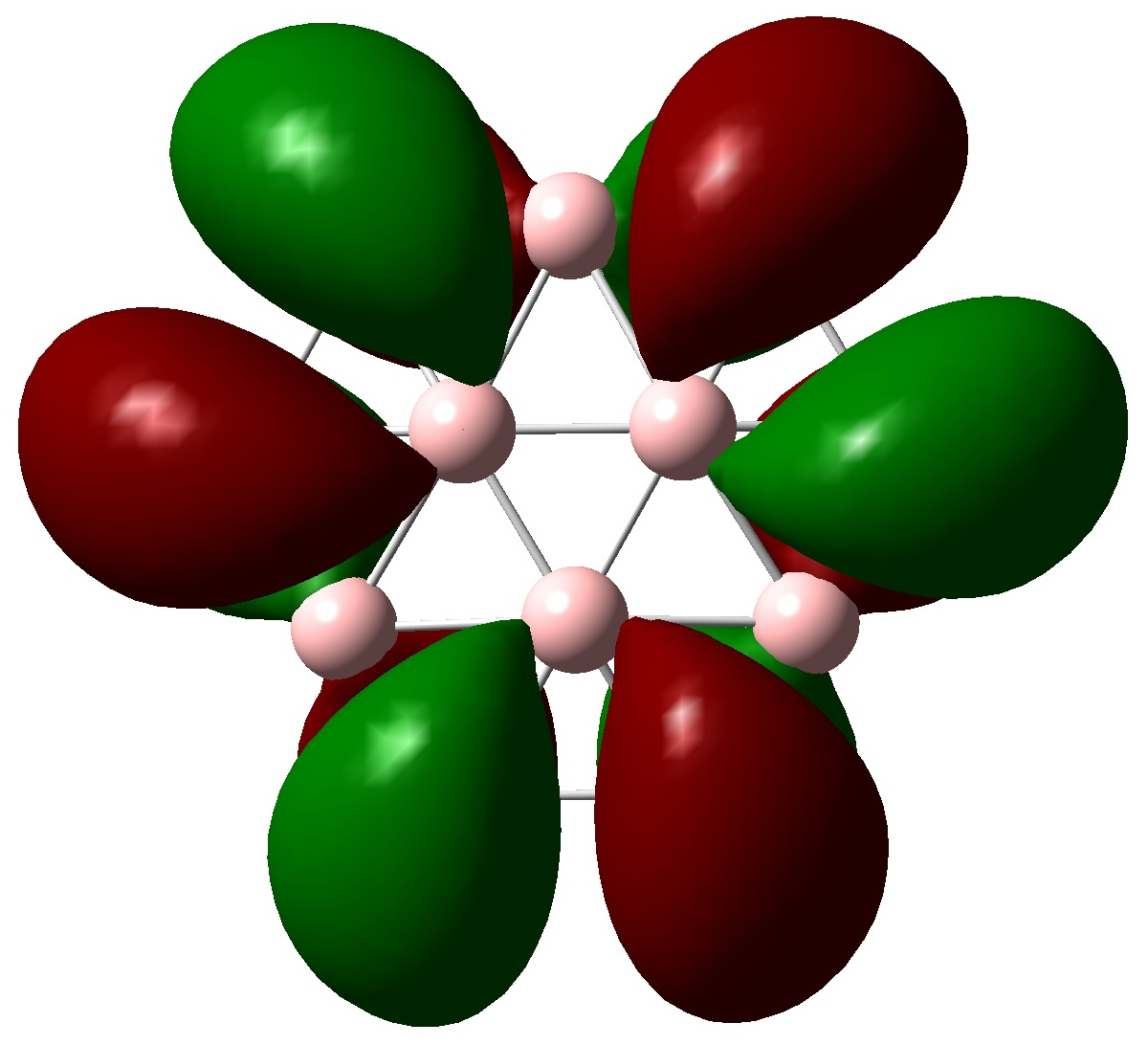}~~~\includegraphics[scale=0.04]{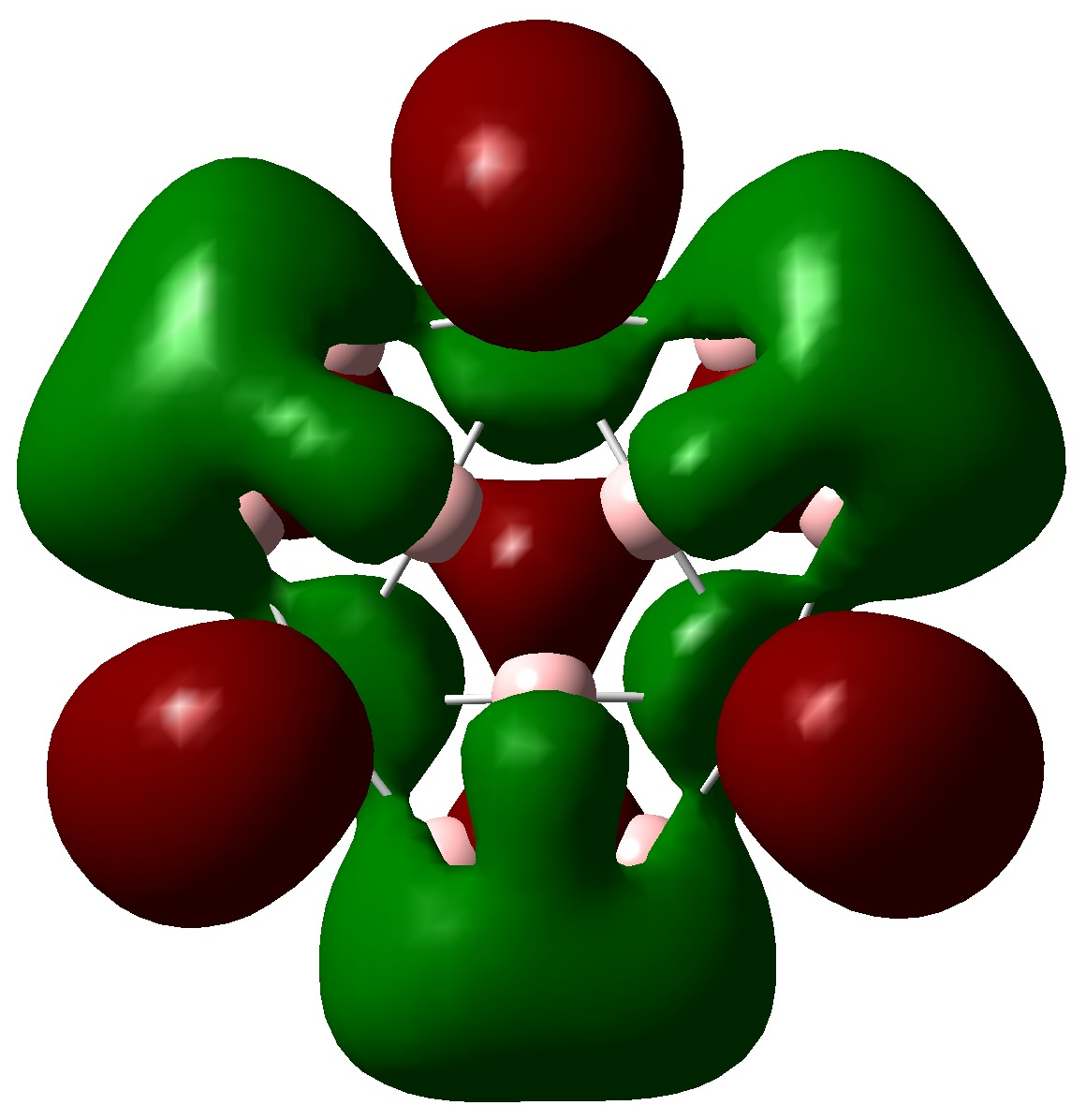}~~~

~~\textbf{~~~~L}\textsubscript{\textbf{1}}~~~\textbf{~~~~~~~L}\textsubscript{\textbf{2}}~~~~~~~~~~\textbf{L+1}~~~~~~~~~~\textbf{L+2}~~~~~~~~~~~\textbf{(L+3)}\textsubscript{\textbf{1}}~~~~~~~\textbf{(L+3)}\textsubscript{\textbf{2}}\textbf{~~~~~~L+4~~~~~~~~(L+6)}\textsubscript{\textbf{1}}\textbf{~~~~~(L+6)}\textsubscript{\textbf{2}}~

\caption{Frontier molecular orbitals of quasi-planar ($C_{3v}$) isomer of
B\protect\textsubscript{12} (see Fig. 1a, of the main article), contributing
to important peaks in its optical absorption spectra. The symbols
H and L corresponds to HOMO and LUMO orbitals. Subscripts 1, 2 etc.
are used to label the degenerate orbitals.}
\end{figure}

\begin{table}[H]
\caption{Many-particle wave functions of the excited states associated with
the peaks in the optical absorption spectra of quasi-planar ($C_{3v}$)
isomer of B$_{12}$ (Fig. 1a of the main article). $E$ denotes the
excitation energy (in eV) of the state involved, and $f$ stands for
oscillator strength corresponding to the electric-dipole transition
to that state, from the ground state. In the ``Polarization'' column,
x, y, and z denote the the direction of polarization of the absorbed
light, with respect to the Cartesian coordinate axes defined in the
figure below. In the ``Wave function'' column, \textbf{$|HF\rangle$}
indicates the Hartree-Fock (HF) configuration, while other configurations
are defined as virtual excitations with respect to it. $H$ and $L$
denote HOMO and LUMO orbitals, while the degenerate orbitals are denoted
by additional subscripts 1, 2 etc. In parentheses next to each configuration,
its coefficient in the CI expansion is indicated. Below, GS does not
indicate a peak, rather it denotes the ground states wave function
of the isomer. ~~~~~~~~~~~~~~~~~~~~~~~~~~~~~~~~~~~~~~~~~~~~~~~~~~~~~~~~~~~~~~~~~~~~~~~~~~~~~~~\protect\includegraphics[scale=0.12]{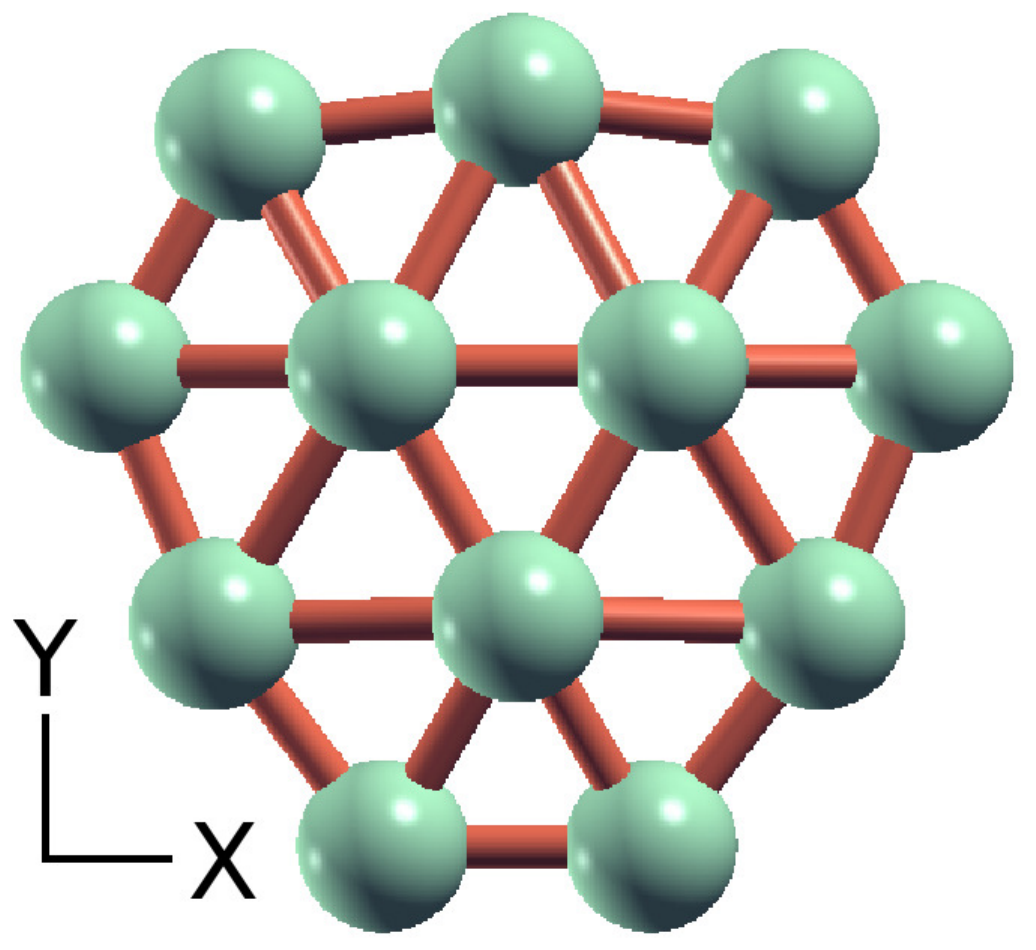}}

\begin{tabular}{ccccccccc}
\hline 
Peak & \hspace{0.4cm} & $E$ (eV) & \hspace{0.4cm} & $f$ & \hspace{0.4cm} & Polarization & \hspace{0.4cm} & Wave function\tabularnewline
\hline 
\hline 
GS\textsuperscript{} &  &  &  &  &  &  &  & $\arrowvert HF\rangle(0.8882)$\tabularnewline
 &  &  &  &  &  &  &  & \tabularnewline
I &  & 3.58 &  & 0.6317 &  & y/z &  & $\arrowvert H_{2}\rightarrow L_{2}\rangle(0.4915)$\tabularnewline
 &  &  &  &  &  &  &  & $\arrowvert H_{1}\rightarrow L{}_{1}\rangle(0.4898)$\tabularnewline
 &  &  &  &  &  &  &  & \tabularnewline
II &  & 4.43 &  & 0.3851 &  & y/z &  & $\arrowvert(H-1)_{1}\rightarrow L+1\rangle(0.4715)$\tabularnewline
 &  &  &  &  &  &  &  & $\arrowvert H_{2}\rightarrow L+2\rangle(0.4434)$\tabularnewline
 &  & 4.46 &  & 0.3518 &  & x &  & $\arrowvert(H-1)_{2}\rightarrow L+1\rangle(0.5284)$\tabularnewline
 &  &  &  &  &  &  &  & $\arrowvert H_{1}\rightarrow L+2\rangle(0.4022)$\tabularnewline
 &  &  &  &  &  &  &  & \tabularnewline
III &  & 4.93 &  & 0.1464 &  & y/z &  & $\arrowvert(H-1)_{2}\rightarrow L+2\rangle(0.6251)$\tabularnewline
 &  &  &  &  &  &  &  & $\arrowvert H_{2}\rightarrow L+2\rangle(0.5838)$\tabularnewline
 &  & 5.07 &  & 0.1238 &  & x &  & $\arrowvert H-2\rightarrow L_{2}\rangle(0.7529)$\tabularnewline
 &  &  &  &  &  &  &  & $\arrowvert H_{1}\rightarrow L+2\rangle(0.3081)$\tabularnewline
 &  &  &  &  &  &  &  & \tabularnewline
IV &  & 5.44 &  & 0.7118 &  & y/z &  & $\arrowvert H-2\rightarrow L+1\rangle(0.8518)$\tabularnewline
 &  &  &  &  &  &  &  & $\arrowvert H_{1}\rightarrow L+1;H_{2}\rightarrow L_{2}\rangle(0.0586)$\tabularnewline
 &  &  &  &  &  &  &  & \tabularnewline
V &  & 5.74 &  & 8.5755 &  & x &  & $\arrowvert(H-3)_{1}\rightarrow L_{2}\rangle(0.6017)$\tabularnewline
 &  &  &  &  &  &  &  & $\arrowvert(H-1)_{1}\rightarrow L+2\rangle(0.2998)$\tabularnewline
 &  & 5.76 &  & 11.1191 &  & y/z &  & $\arrowvert(H-3)_{2}\rightarrow L_{2}\rangle(0.3757)$\tabularnewline
 &  &  &  &  &  &  &  & $\arrowvert(H-3)_{1}\rightarrow L_{1}\rangle(0.3679)$\tabularnewline
 &  & 5.83 &  & 4.4994 &  & x &  & $\arrowvert(H-3)_{2}\rightarrow L_{1}\rangle(0.5156)$\tabularnewline
 &  &  &  &  &  &  &  & $\arrowvert(H-3)_{1}\rightarrow L_{2}\rangle(0.4430)$\tabularnewline
\hline 
\end{tabular}

\medskip{}

$\qquad\qquad\qquad\qquad\qquad\qquad\qquad\qquad\qquad\qquad\qquad\qquad\qquad\qquad\qquad\qquad\qquad\qquad$(CONTINUED)
\end{table}

\begin{table}[H]
$\qquad\qquad\qquad\qquad\qquad\qquad\qquad\qquad$ Table S1 -- CONTINUES
FROM THE PREVIOUS PAGE

\vspace{1cm}

\begin{tabular}{ccccccccc}
\hline 
Peak & \hspace{0.4cm} & E (eV) & \hspace{0.4cm} & $f$ & \hspace{0.4cm} & Polarization & \hspace{0.4cm} & Wave function\tabularnewline
\hline 
\hline 
VI &  & 6.07 &  & 6.5321 &  & y/z &  & $\arrowvert(H-3)_{2}\rightarrow L_{2}\rangle(0.4341)$\tabularnewline
 &  &  &  &  &  &  &  & $\arrowvert(H-3)_{1}\rightarrow L_{1}\rangle(0.4305)$\tabularnewline
 &  & 6.09 &  & 4.6717 &  & x &  & $\arrowvert(H-3)_{2}\rightarrow L_{1}\rangle(0.5747)$\tabularnewline
 &  &  &  &  &  &  &  & $\arrowvert(H-3)_{1}\rightarrow L_{2}\rangle(0.2940)$\tabularnewline
 &  &  &  &  &  &  &  & \tabularnewline
VII &  & 6.46 &  & 1.3019 &  & y/z &  & $\arrowvert H_{1}\rightarrow(L+3)_{1}\rangle(0.5743)$\tabularnewline
 &  &  &  &  &  &  &  & $\arrowvert H-4\rightarrow L_{1}\rangle(0.3075)$\tabularnewline
 &  & 6.46 &  & 1.1134 &  & x &  & $\arrowvert H-4\rightarrow L_{2}\rangle(0.5028)$\tabularnewline
 &  &  &  &  &  &  &  & $\arrowvert(H-3)_{2}\rightarrow L+1\rangle(0.2557)$\tabularnewline
 &  & 6.59 &  & 0.8096 &  & x &  & $\arrowvert H_{1}\rightarrow(L+3)_{2}\rangle(0.5504)$\tabularnewline
 &  &  &  &  &  &  &  & $\arrowvert H_{2}\rightarrow(L+3)_{1}\rangle(0.3998)$\tabularnewline
 &  &  &  &  &  &  &  & \tabularnewline
VIII &  & 6.88 &  & 0.9803 &  & y/z &  & $\arrowvert H_{1}\rightarrow(L+6)_{1}\rangle(0.4043)$\tabularnewline
 &  &  &  &  &  &  &  & $\arrowvert(H-1)_{1}\rightarrow(L+3)_{1}\rangle(0.3240)$\tabularnewline
 &  & 6.92 &  & 0.4733 &  & x &  & $\arrowvert H_{2}\rightarrow(L+6)_{1}\rangle(0.3334)$\tabularnewline
 &  &  &  &  &  &  &  & $\arrowvert H_{2}\rightarrow L_{1};H_{2}\rightarrow L_{2}\rangle(0.3249)$\tabularnewline
\hline 
\end{tabular}
\end{table}

\newpage{}

\section{Double Ring Isomer}

\begin{figure}[H]
\includegraphics[scale=0.04]{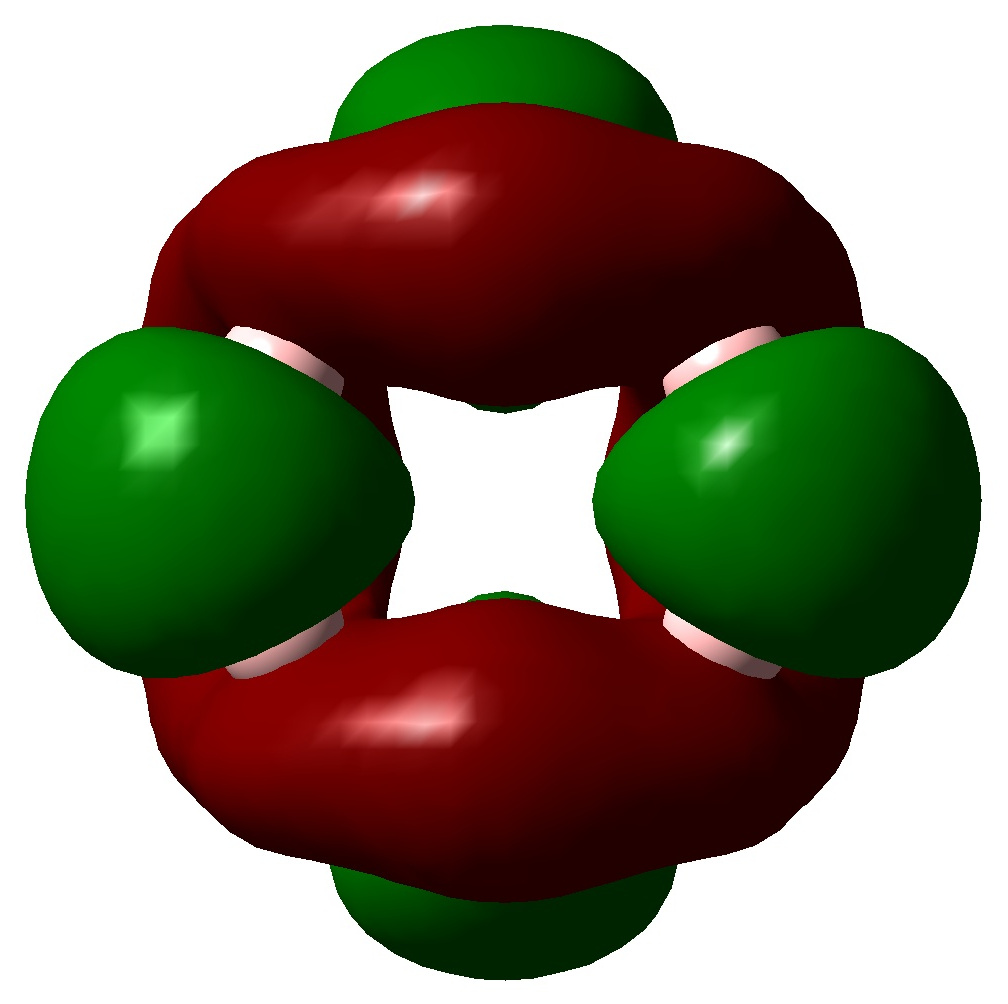}~~~\includegraphics[scale=0.04]{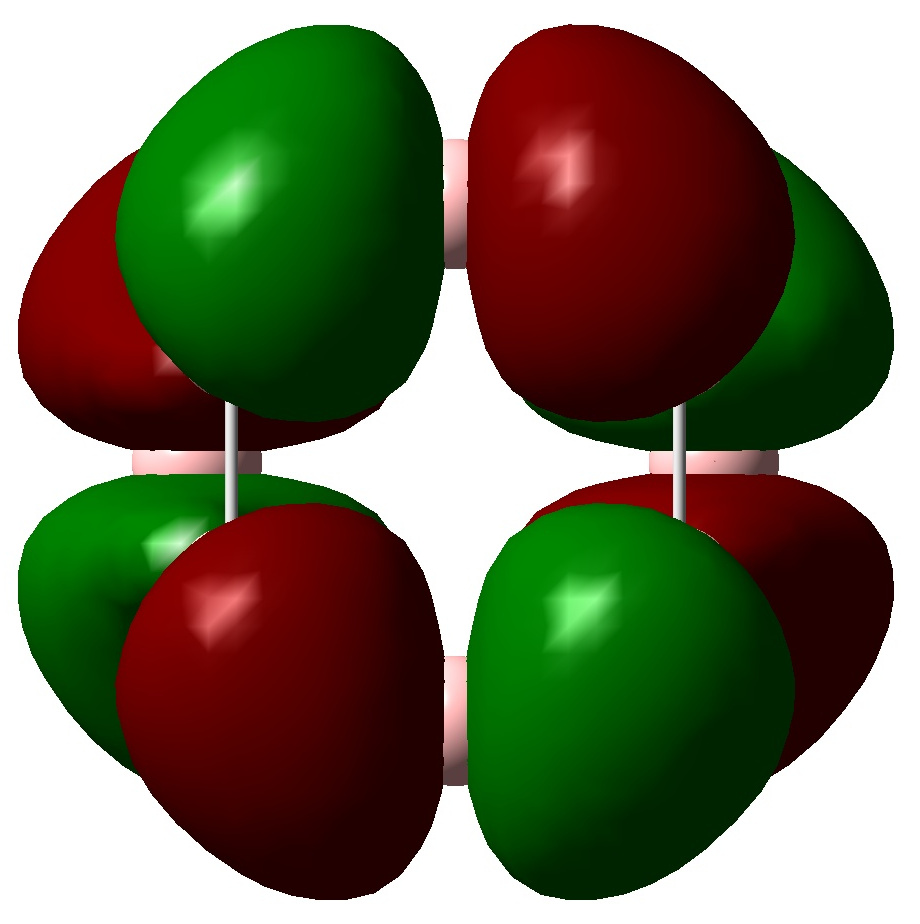}~~~\includegraphics[scale=0.04]{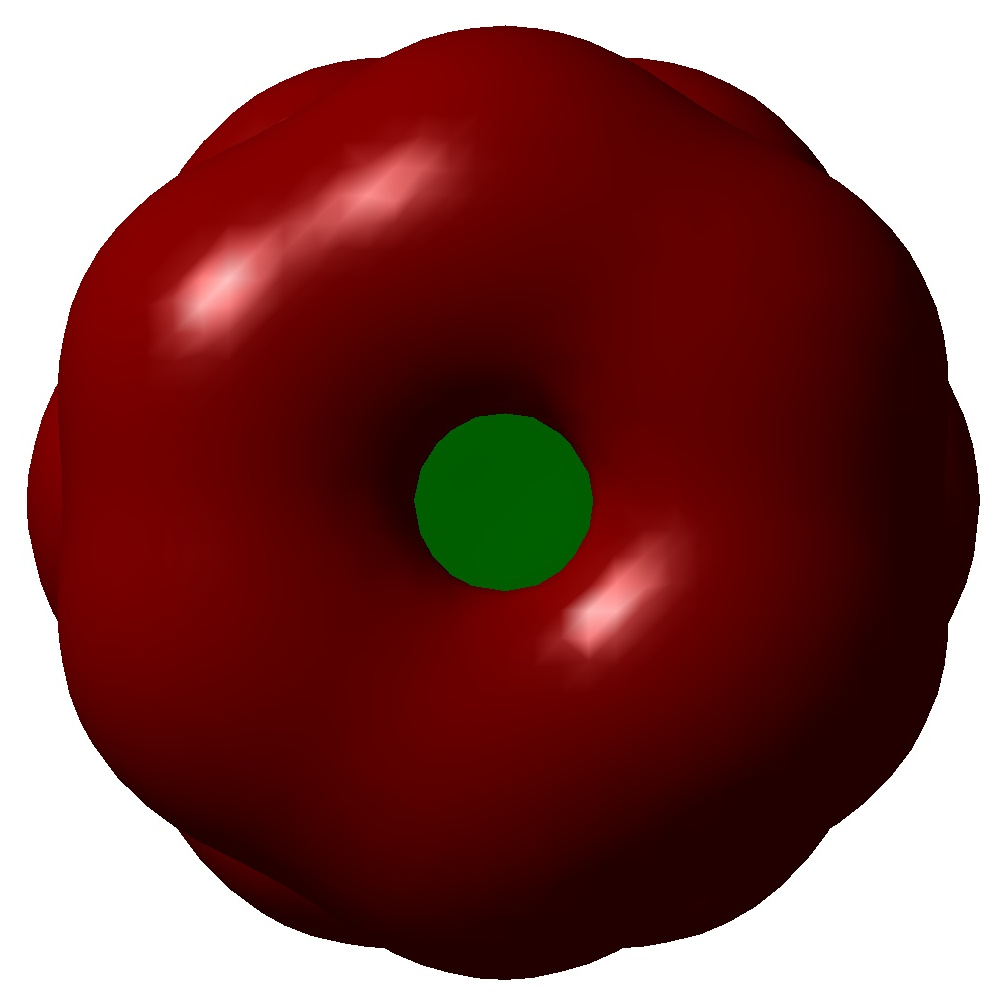}~~~\includegraphics[scale=0.04]{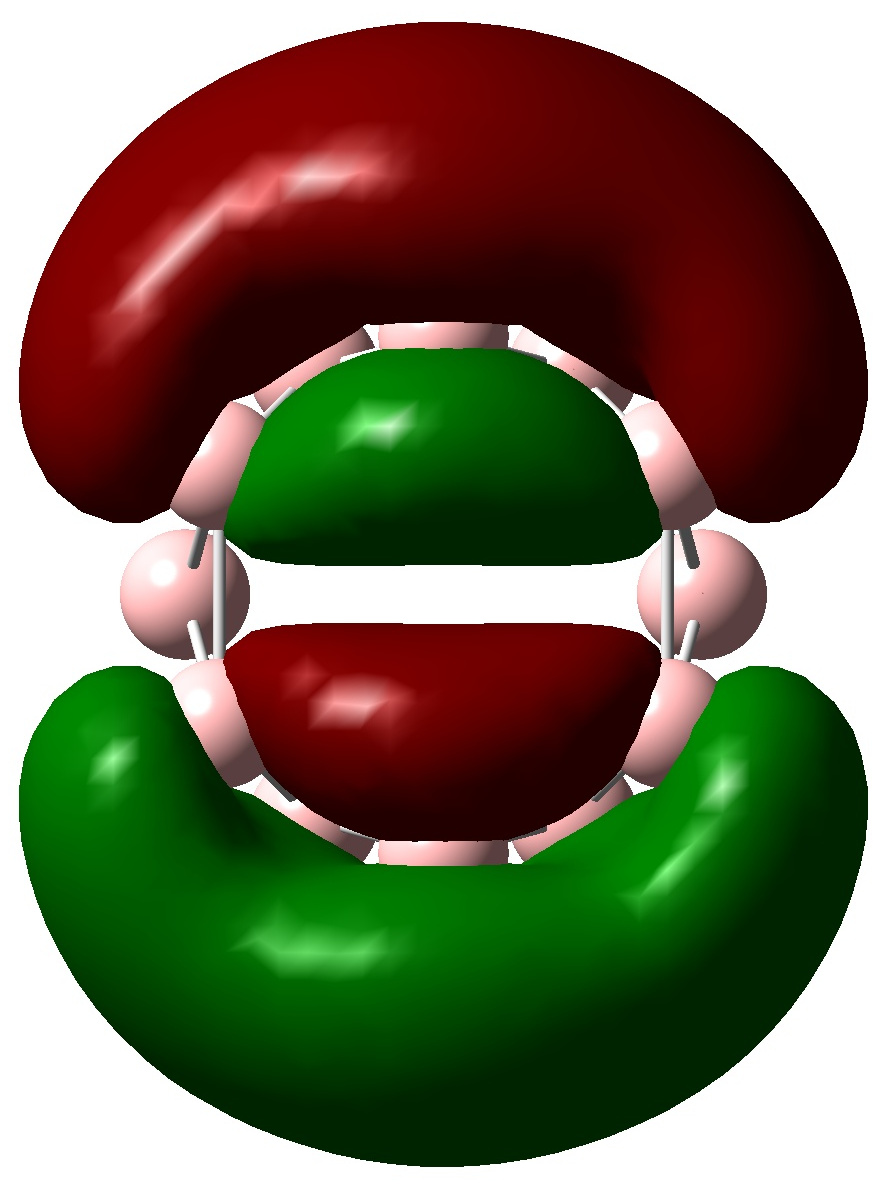}~~~\includegraphics[scale=0.04]{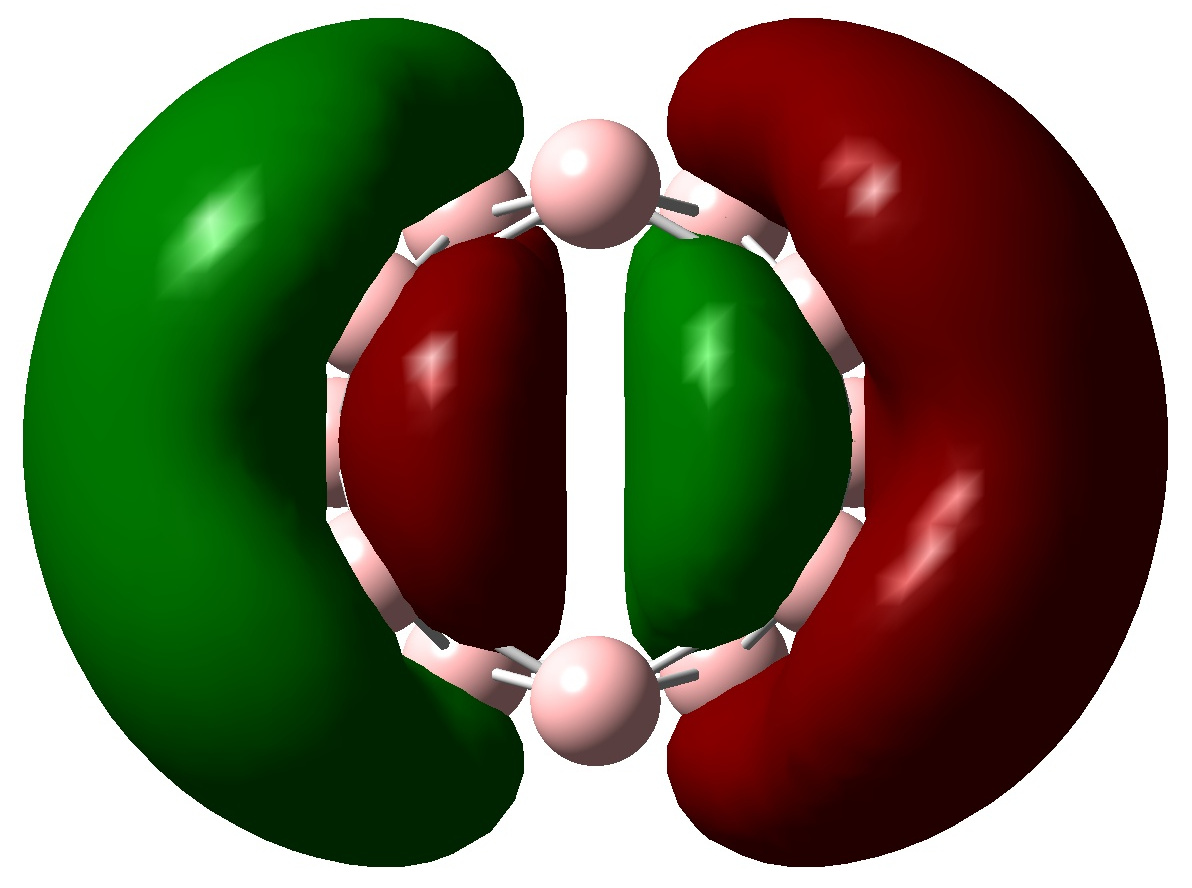}~~~\includegraphics[scale=0.04]{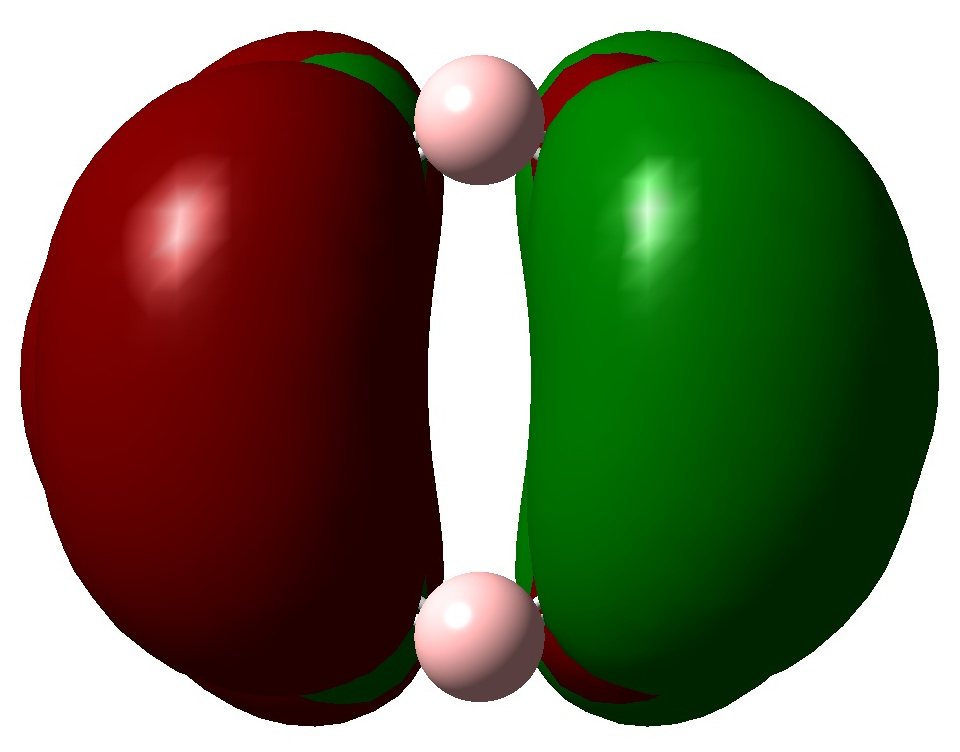}~~~\includegraphics[scale=0.04]{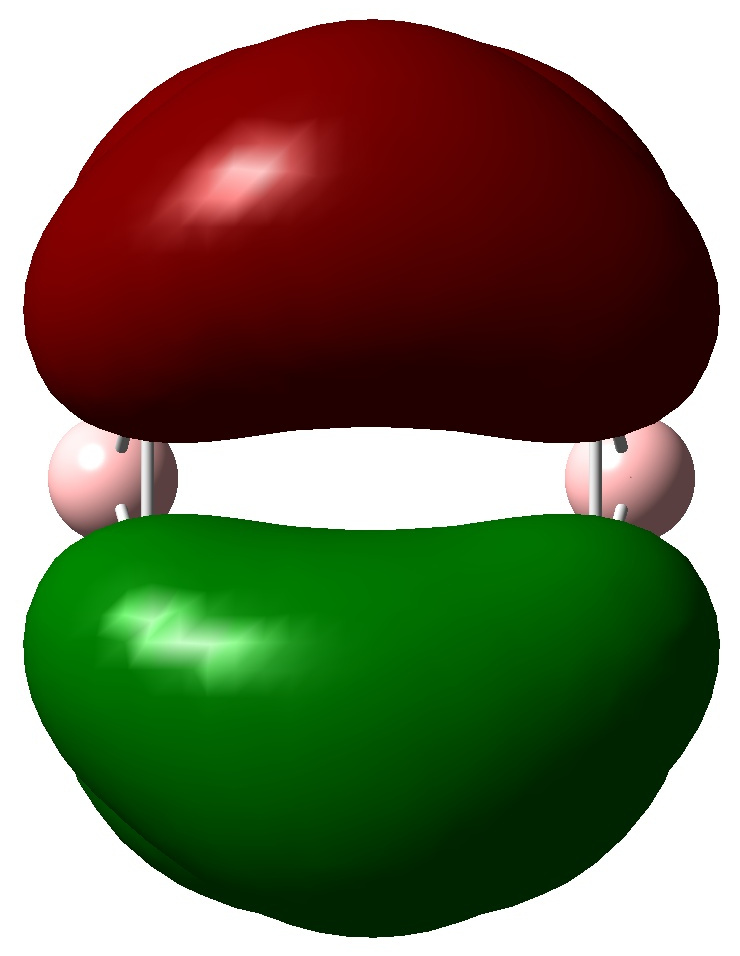}~~~\includegraphics[scale=0.04]{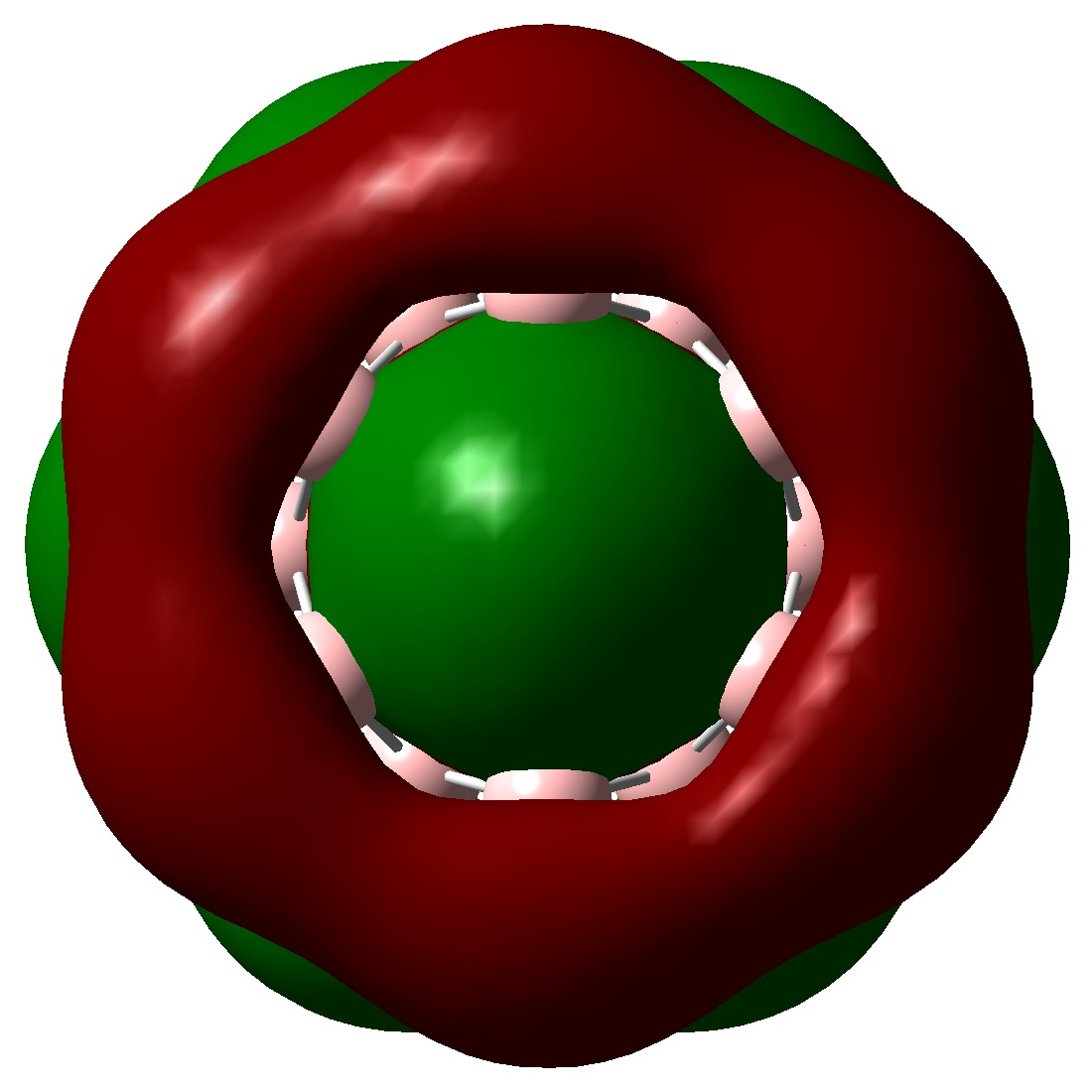}
~~~\includegraphics[scale=0.04]{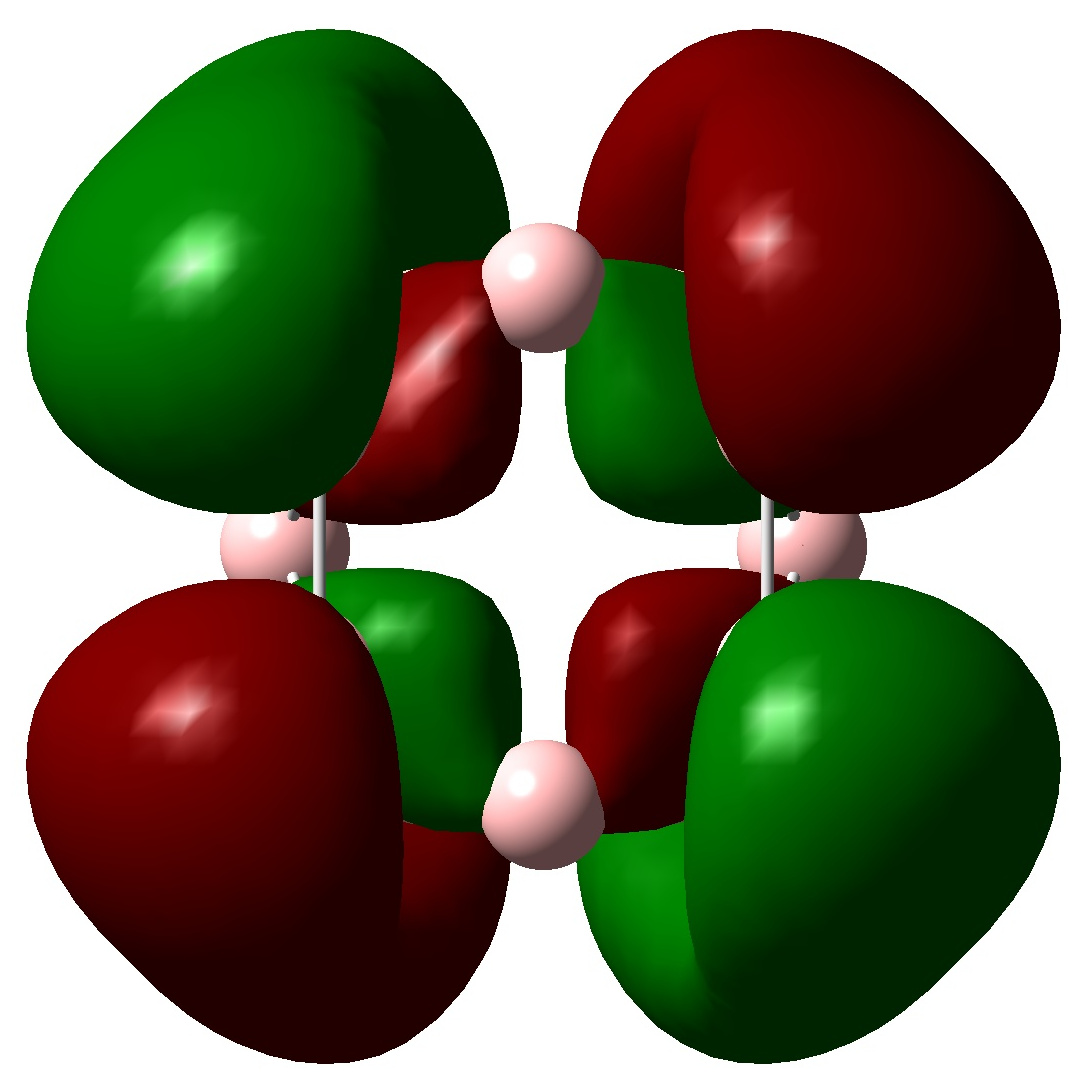}

\textbf{~(H-3)}\textsubscript{\textbf{2}}~~~~\textbf{~(H-3)}\textsubscript{\textbf{1}}~~~~~~\textbf{~H-2}~~~~~~\textbf{~(H-1)}\textsubscript{\textbf{2}}~~~~\textbf{~~(H-1)}\textsubscript{\textbf{1}}\textbf{~~~~~~~H}\textsubscript{\textbf{2}}\textbf{~~~~~~~~H}\textsubscript{\textbf{1}}\textbf{~~~~~~~~~~L~~~~~~~~~(L+1)}\textsubscript{\textbf{1}}~~~

\vspace{1cm}

\includegraphics[scale=0.04]{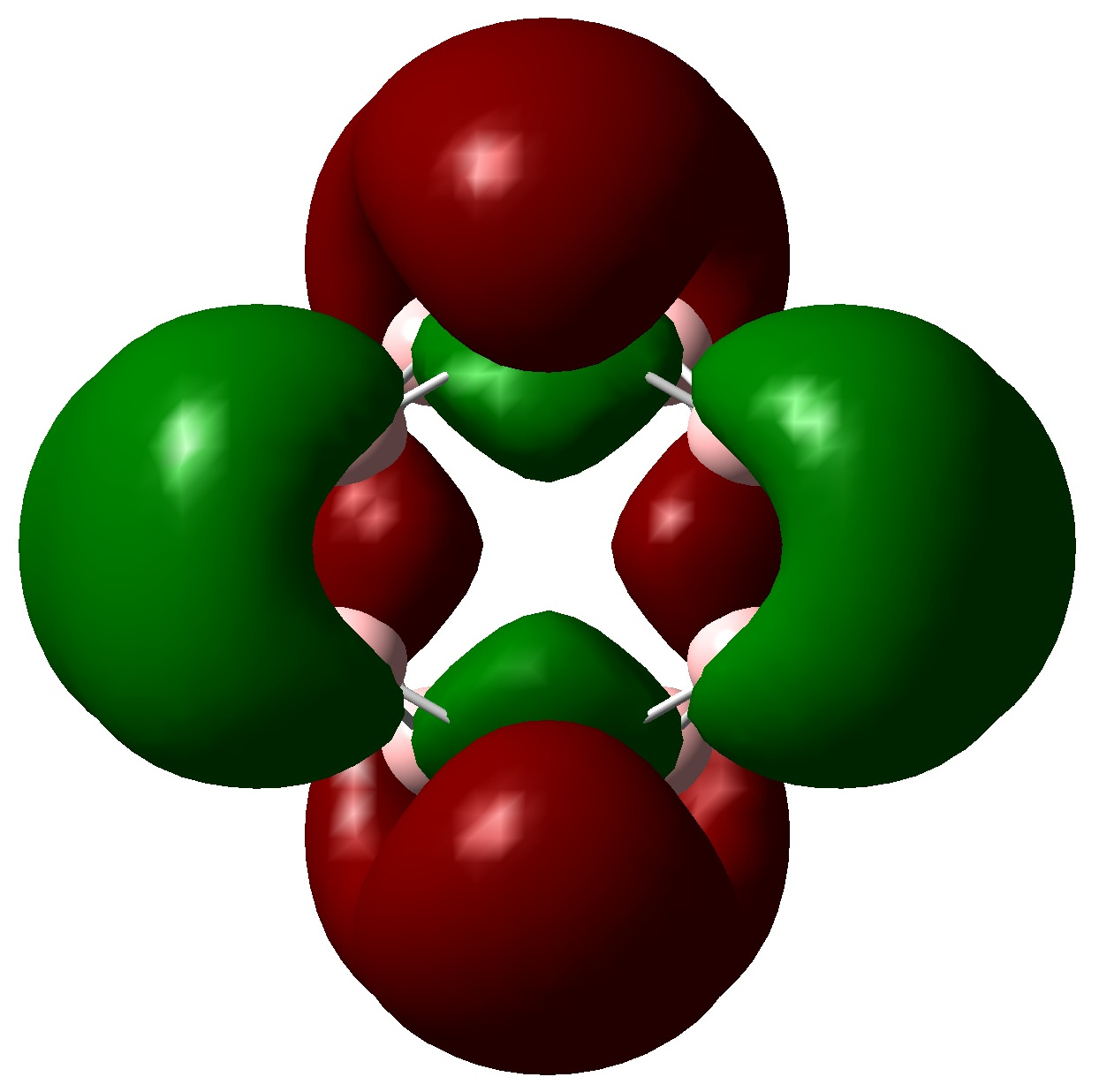}~~~\includegraphics[scale=0.04]{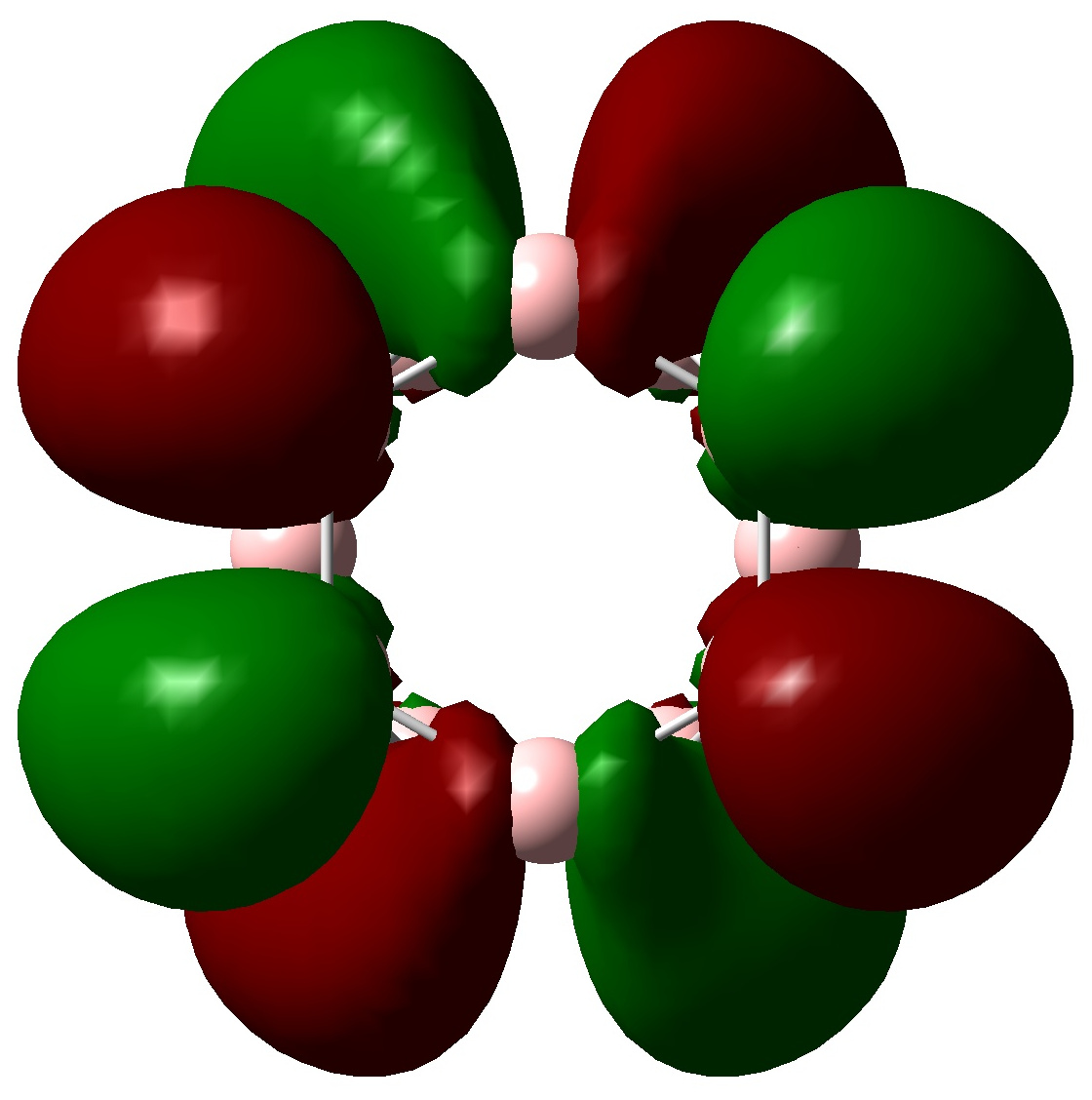}~~~\includegraphics[scale=0.04]{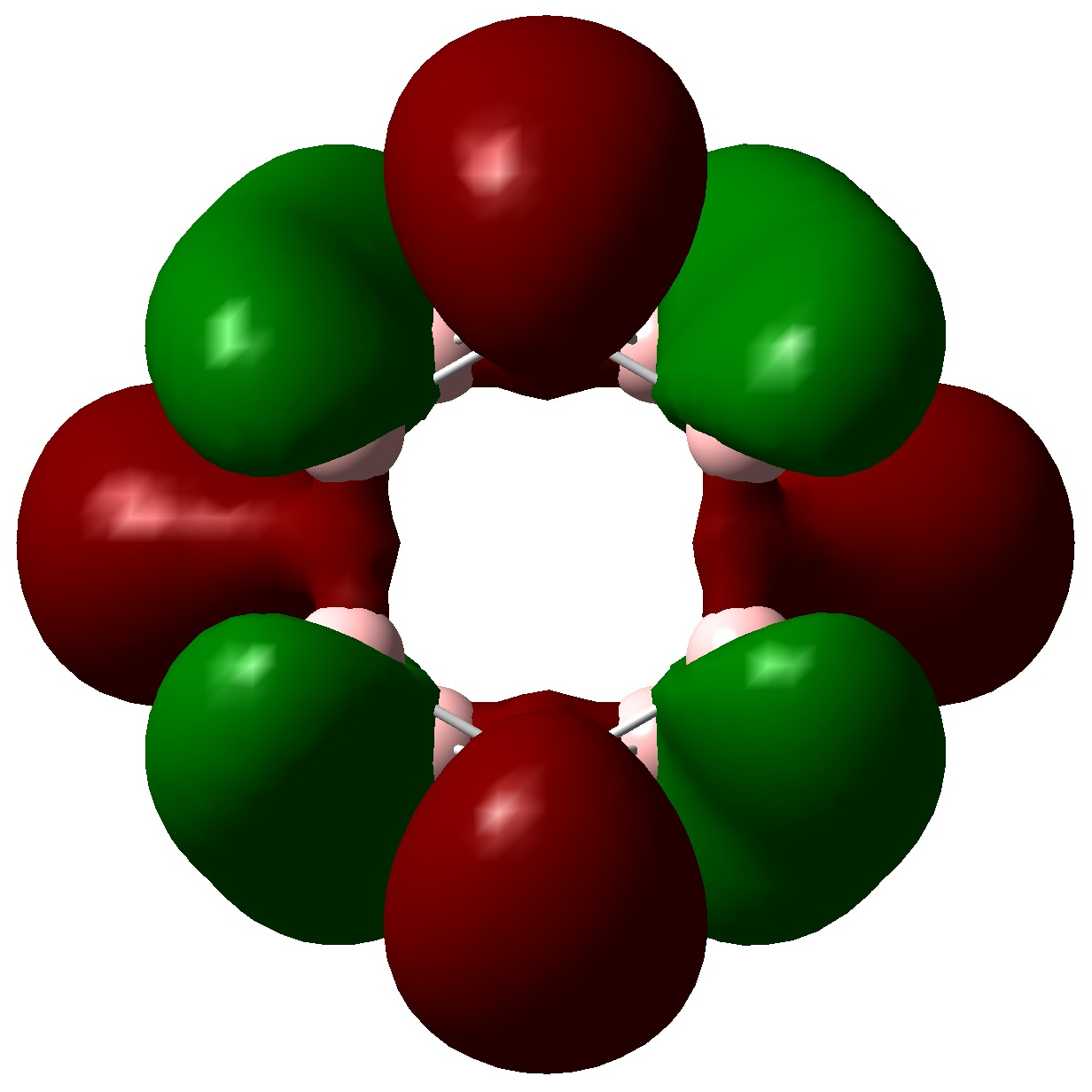}~~~\includegraphics[scale=0.04]{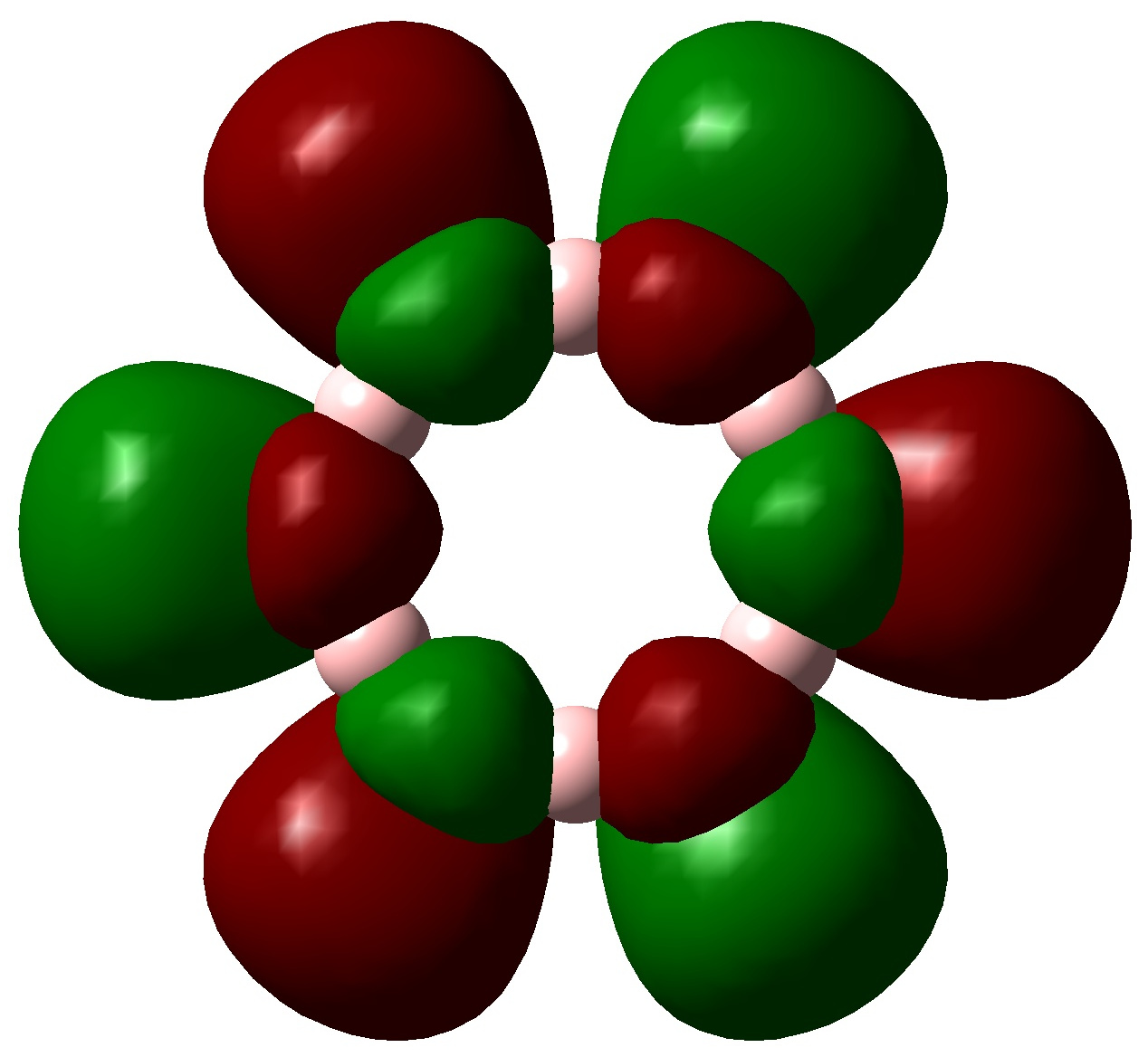}~~~\includegraphics[scale=0.04]{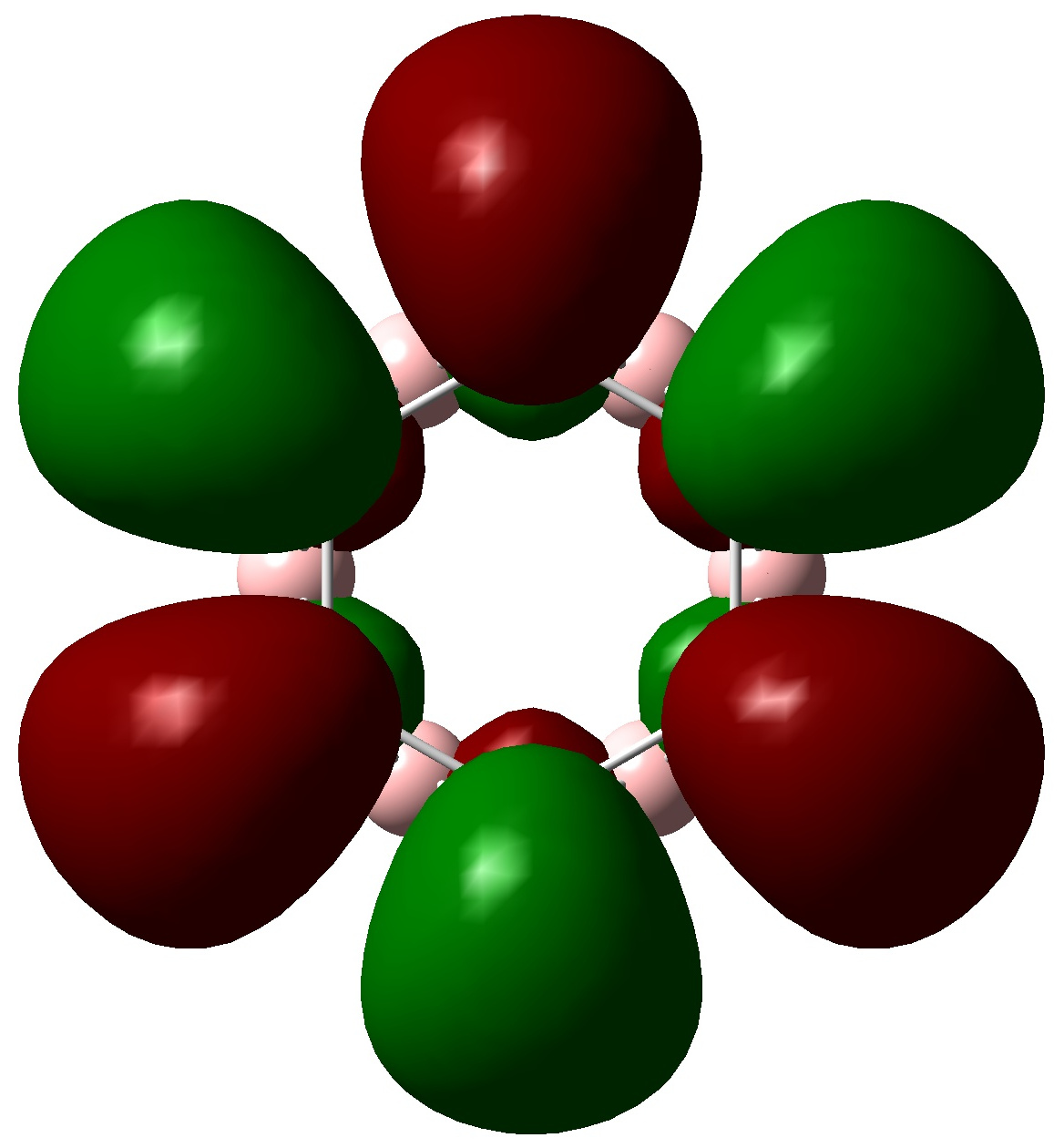}~~~\includegraphics[scale=0.04]{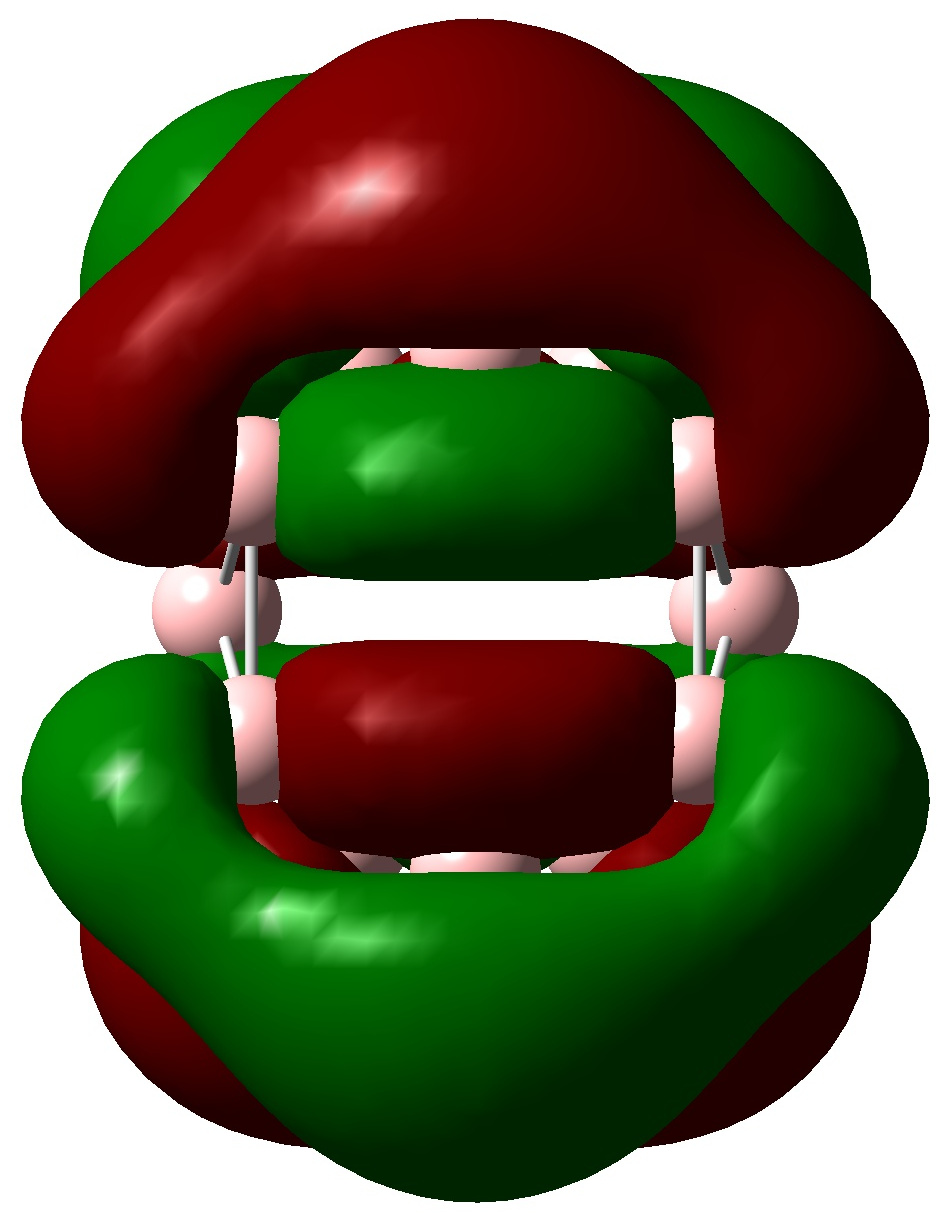}~~~\includegraphics[scale=0.04]{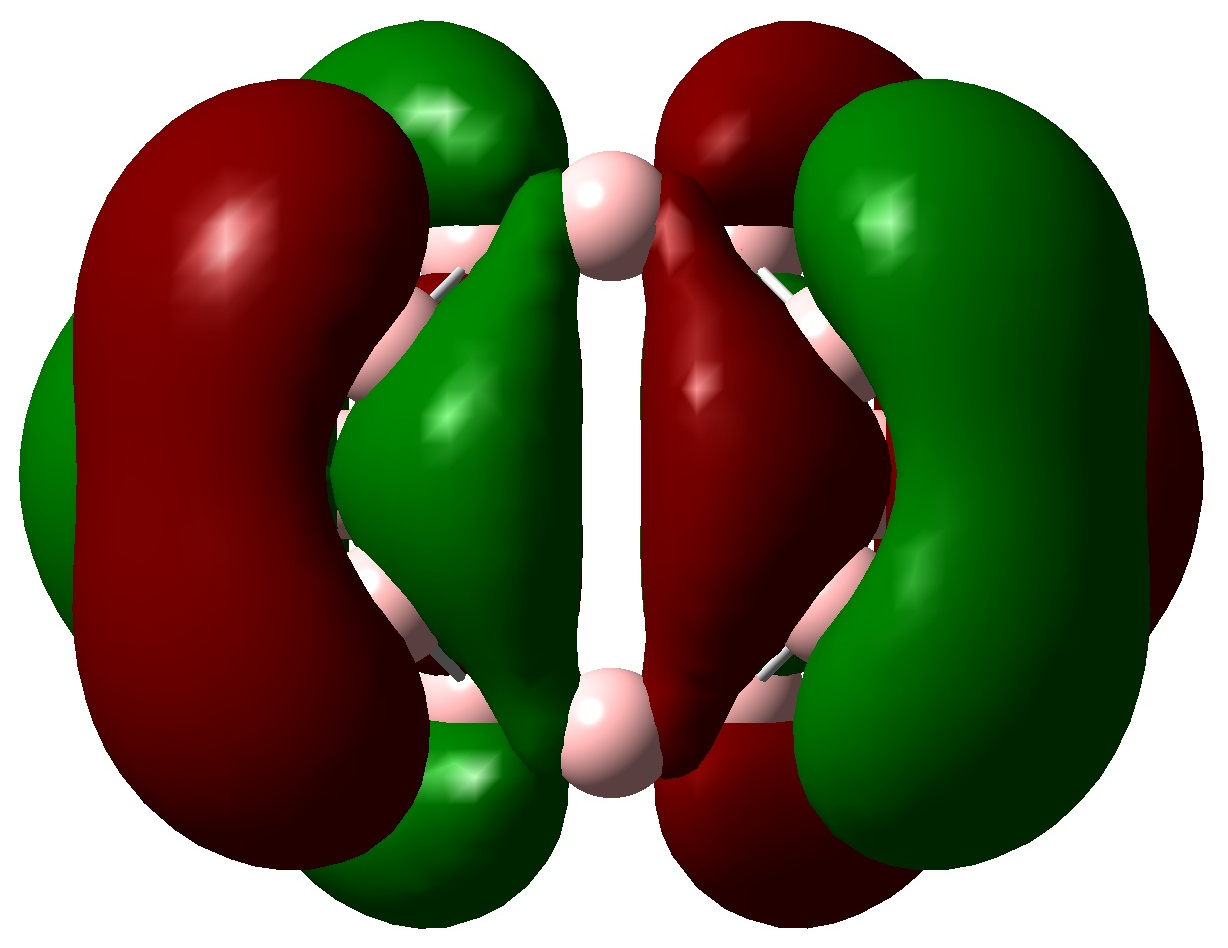}~~~\includegraphics[scale=0.04]{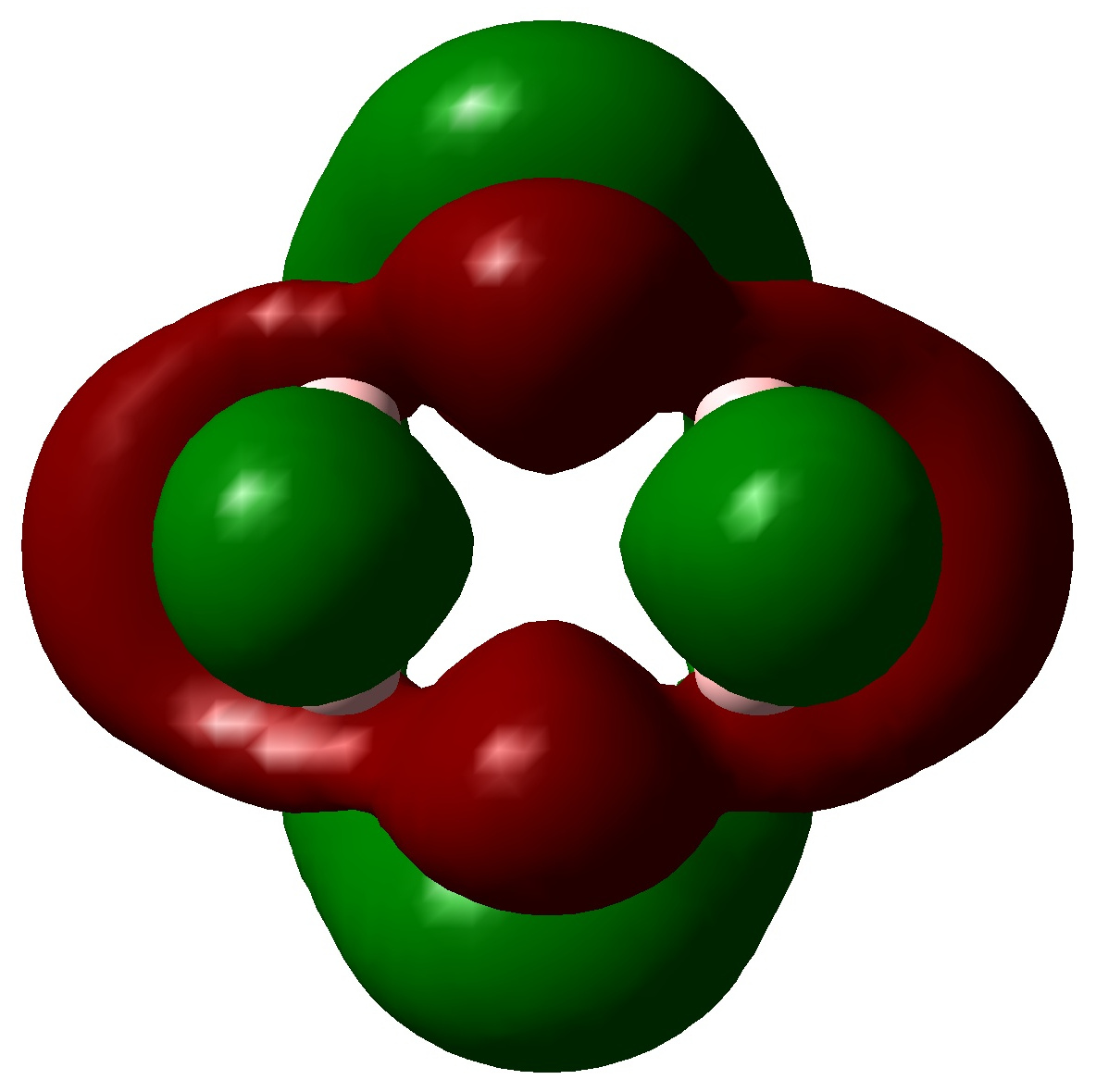}~~~\includegraphics[scale=0.04]{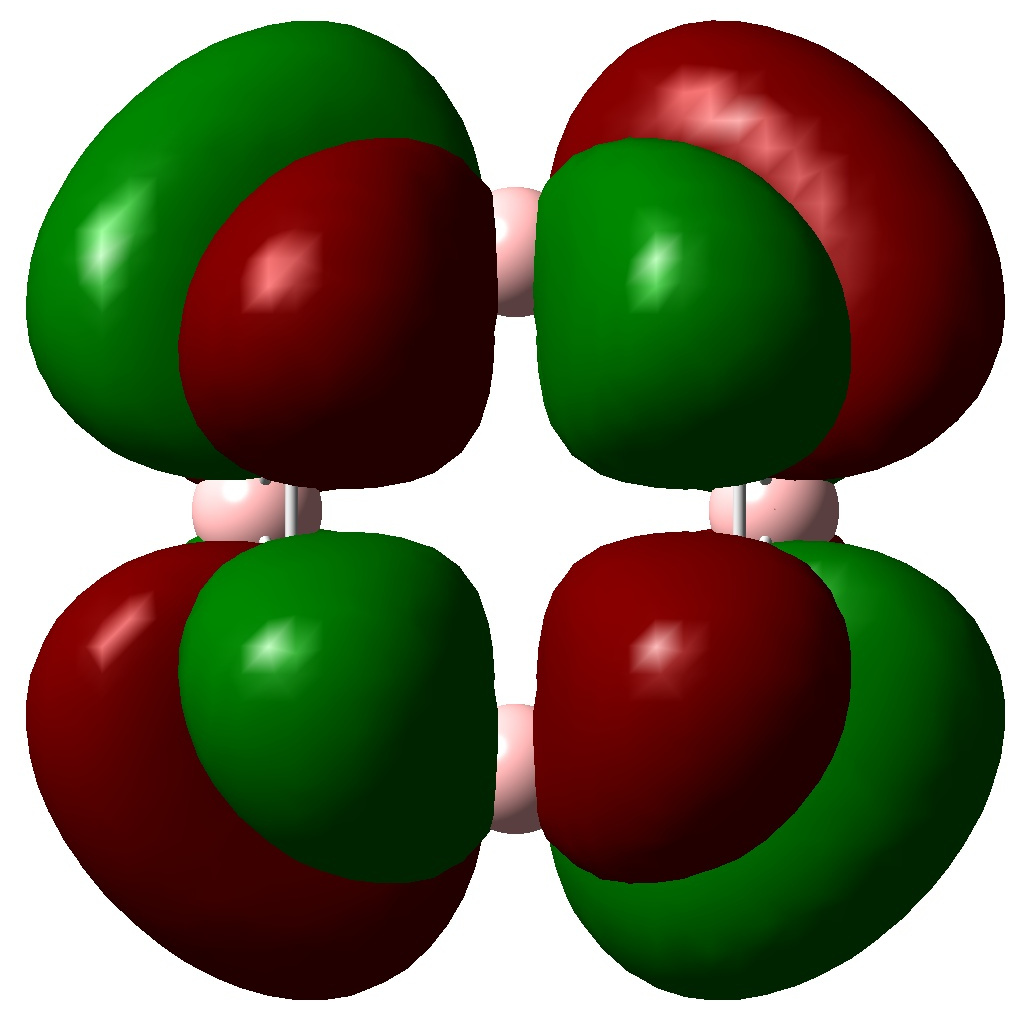}~~~

~~\textbf{~(L+1)}\textsubscript{\textbf{2}}~~~\textbf{~~~(L+2)}\textsubscript{\textbf{1}}~~~~~~\textbf{(L+2)}\textsubscript{\textbf{2}}~~~~~~~\textbf{(L+3)}\textsubscript{\textbf{1}}~~~~~~~\textbf{(L+3)}\textsubscript{\textbf{2}}~~~~~\textbf{(L+4)}\textsubscript{\textbf{1}}\textbf{~~~~(L+4)}\textsubscript{\textbf{2}}\textbf{~~~~~~(L+5)}\textsubscript{\textbf{1}}\textbf{~~~~~(L+5)}\textsubscript{\textbf{2}}

\caption{Frontier molecular orbitals of double ring ($D_{6d}$) isomer of B\protect\textsubscript{12}
(see Fig. 1b, of the main article), contributing to important peaks
in its optical absorption spectra. The symbols H and L corresponds
to HOMO and LUMO orbitals. Subscripts 1,2 etc. are used to label the
degenerate orbitals.}
\end{figure}

$\newline$

\begin{table}[H]
\caption{Many particle wave functions of the excited states associated with
the peaks in the optical absorption spectra of the double ring ($D_{6d}$)
isomer of B$_{12}$ (Fig. 1b, main article). $E$ denotes the excitation
energy (in eV) of the state involved, and $f$ stands for oscillator
strength corresponding to the electric-dipole transition to that state,
from the ground state. In the ``Polarization'' column, x, y, and
z denote the the direction of polarization of the absorbed light,
with respect to the Cartesian coordinate axes defined in the figure
below. In the ``Wave function'' column, \textbf{$|HF\rangle$} indicates
the Hartree-Fock (HF) configuration, while other configurations are
defined as virtual excitations with respect to it. $H$ and $L$ denote
HOMO and LUMO orbitals, while the degenerate orbitals are denoted
by additional subscripts 1, 2 etc. In parentheses next to each configuration,
its coefficient in the CI expansion is indicated. Below, GS does not
indicate a peak, rather it denotes the ground states wave function
of the isomer. ~~~~~~~~~~~~~~~~~~~~~~~~~~~~~~~~~~~~~~~~~~~~~~~~~~~~~~~~~~~~~~~~~~~~~~~~~~~~~~~\protect\includegraphics[scale=0.15]{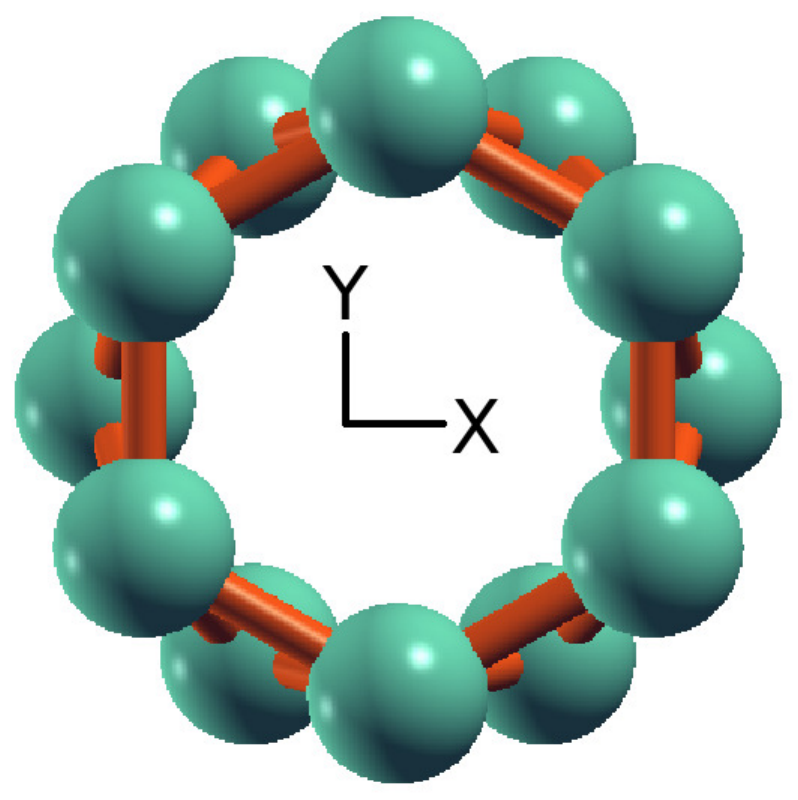}}

\begin{tabular}{ccccccccc}
\hline 
Peak & \hspace{0.4cm} & E (eV) & \hspace{0.4cm} & $f$ & \hspace{0.4cm} & Polarization & \hspace{0.4cm} & Wave function\tabularnewline
\hline 
\hline 
GS\textsuperscript{} &  &  &  &  &  &  &  & $\arrowvert HF\rangle(0.8971)$\tabularnewline
 &  &  &  &  &  &  &  & \tabularnewline
I &  & 4.78 &  & 5.7331 &  & $y$ &  & $\arrowvert(H-1)_{2}\rightarrow(L+1)_{2}\rangle(0.5539)$\tabularnewline
 &  &  &  &  &  &  &  & $\arrowvert(H-1)_{1}\rightarrow(L+1)_{1}\rangle(0.5442)$\tabularnewline
 &  & 4.79 &  & 5.4136 &  & $x$ &  & $\arrowvert(H-1)_{1}\rightarrow(L+1)_{2}\rangle(0.5360)$\tabularnewline
 &  &  &  &  &  &  &  & $\arrowvert(H-1)_{2}\rightarrow(L+1)_{1}\rangle(0.5224)$\tabularnewline
 &  & 4.93 &  & 2.7899 &  & $z$ &  & $\arrowvert H{}_{2}\rightarrow(L+4)_{2}\rangle(0.6124)$\tabularnewline
 &  &  &  &  &  &  &  & $\arrowvert H-2\rightarrow L\rangle(0.4341)$\tabularnewline
 &  &  &  &  &  &  &  & \tabularnewline
II &  & 5.90 &  & 1.8173 &  & $z$ &  & $\arrowvert(H-1){}_{1}\rightarrow(L+4)_{1}\rangle(0.4577)$\tabularnewline
 &  &  &  &  &  &  &  & $\arrowvert(H-1)_{2}\rightarrow(L+4)_{2}\rangle(0.4178)$\tabularnewline
 &  &  &  &  &  &  &  & \tabularnewline
III &  & 6.41 &  & 3.9347 &  & $z$ &  & $\arrowvert(H-3){}_{1}\rightarrow(L+1)_{1}\rangle(0.5615)$\tabularnewline
 &  &  &  &  &  &  &  & $\arrowvert(H-3){}_{2}\rightarrow(L+1)_{2}\rangle(0.5587)$\tabularnewline
 &  &  &  &  &  &  &  & \tabularnewline
VI &  & 6.80 &  & 6.1390 &  & $y$ &  & $\arrowvert(H{}_{2}\rightarrow(L+5)_{2}\rangle(0.4131)$\tabularnewline
 &  &  &  &  &  &  &  & $\arrowvert(H-3){}_{2}\rightarrow(L+3)_{2}\rangle(0.3927)$\tabularnewline
 &  & 6.81 &  & 5.3021 &  & $x$ &  & $\arrowvert(H-3){}_{2}\rightarrow(L+3)_{1}\rangle(0.4368)$\tabularnewline
 &  &  &  &  &  &  &  & $\arrowvert H{}_{1}\rightarrow(L+5)_{2}\rangle(0.3998)$\tabularnewline
\hline 
\end{tabular}
\end{table}

\section{Chain Isomer}

\begin{figure}[H]
\includegraphics[scale=0.038]{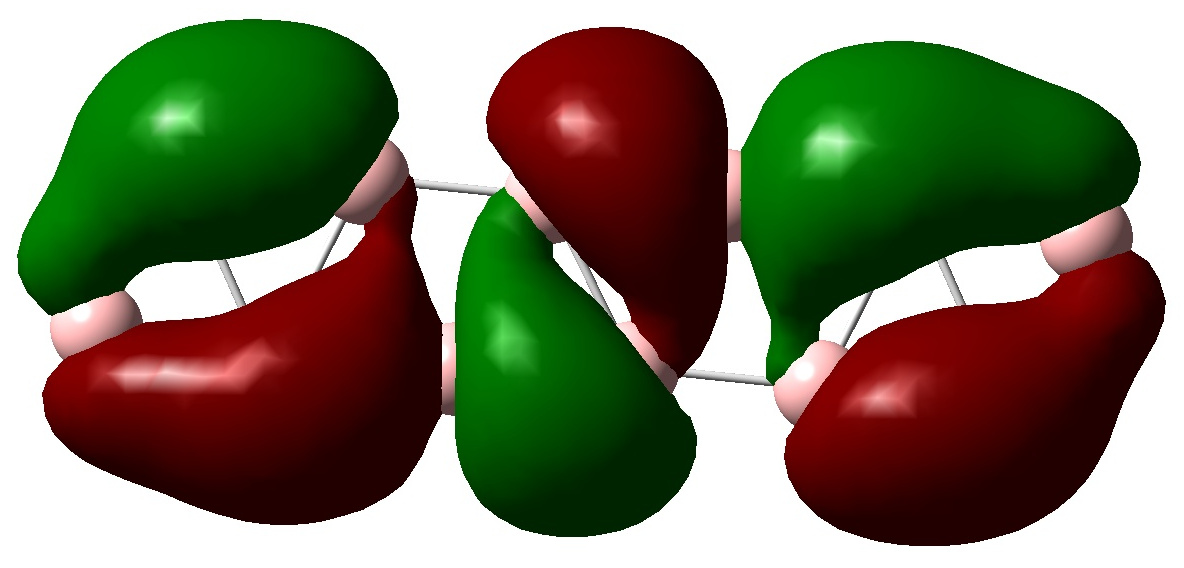}~~~\includegraphics[scale=0.038]{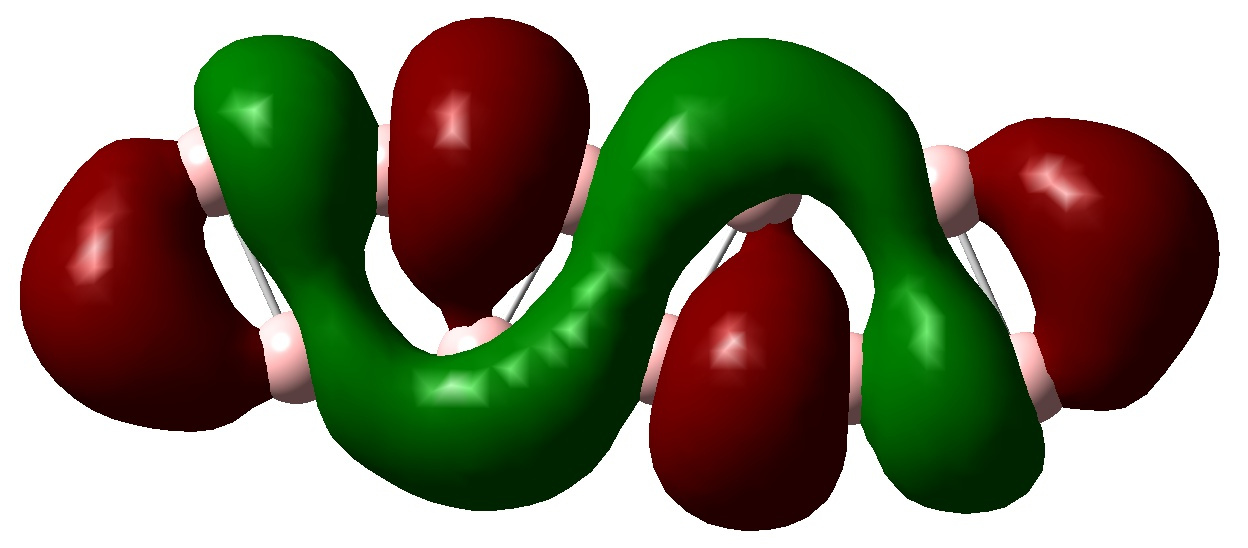}~~~\includegraphics[scale=0.038]{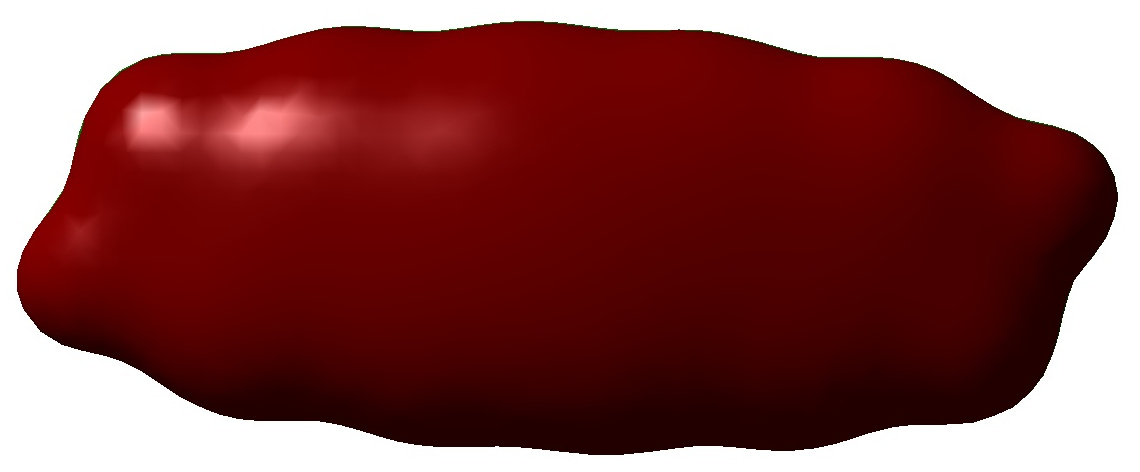}~~~\includegraphics[scale=0.038]{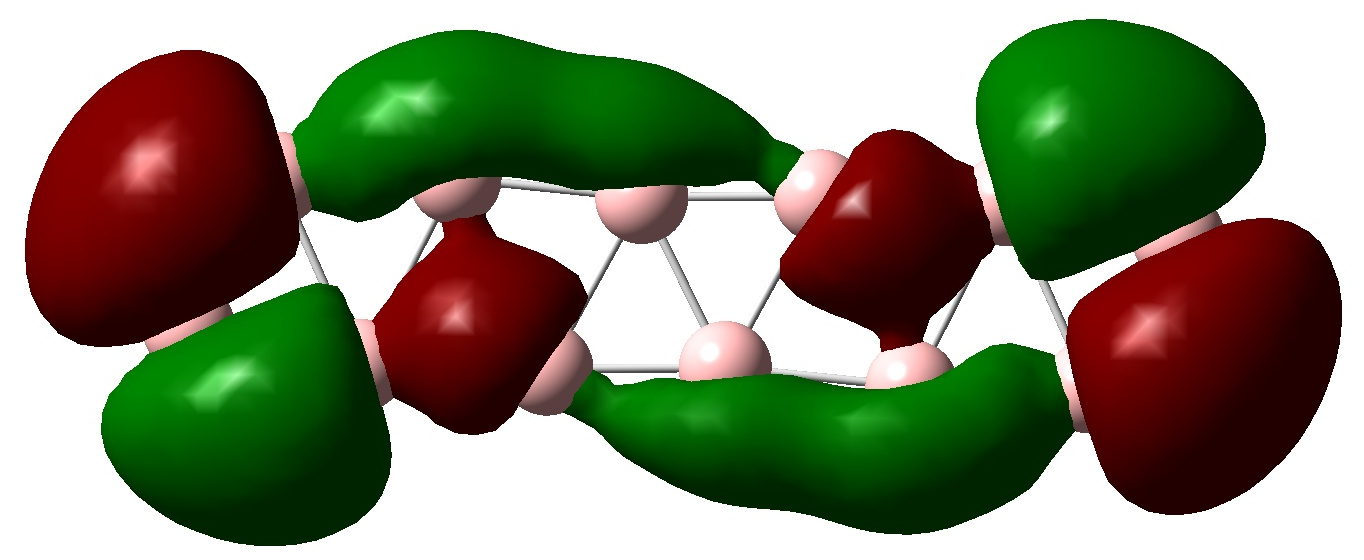}~~~\includegraphics[scale=0.038]{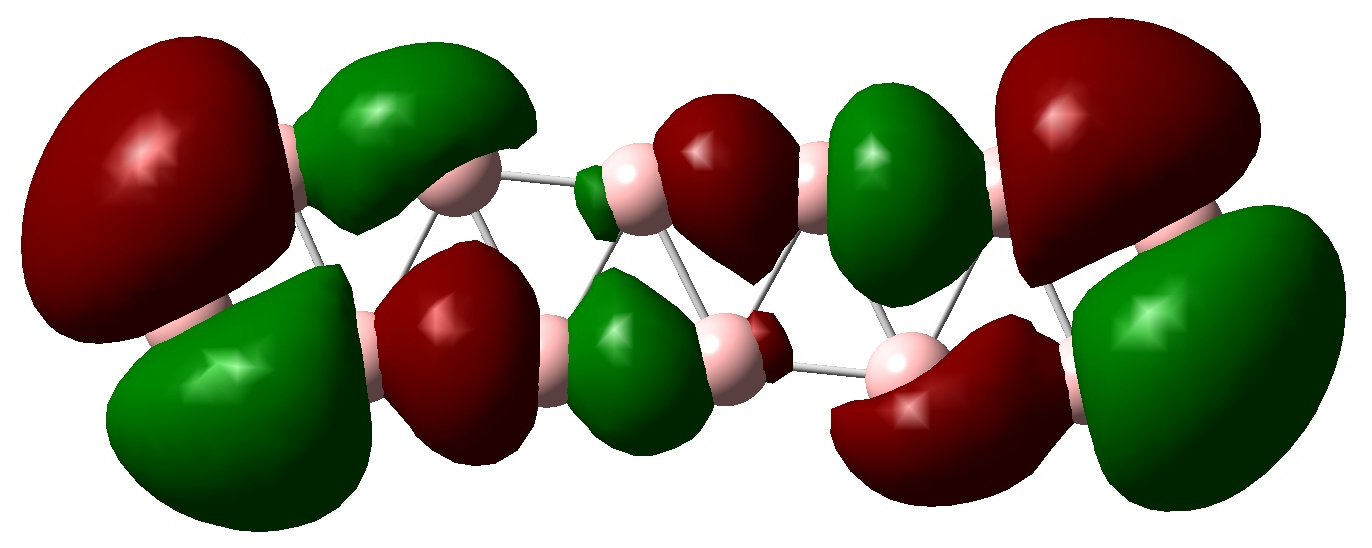}~~~\includegraphics[scale=0.037]{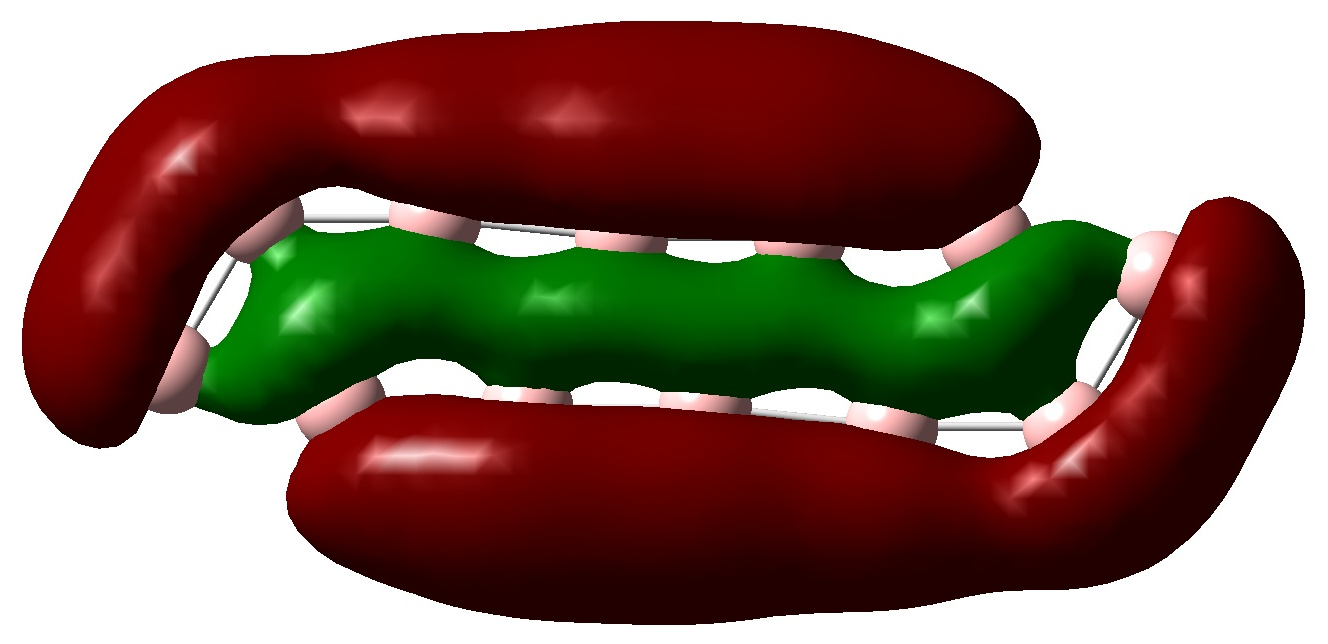}~~~\includegraphics[scale=0.038]{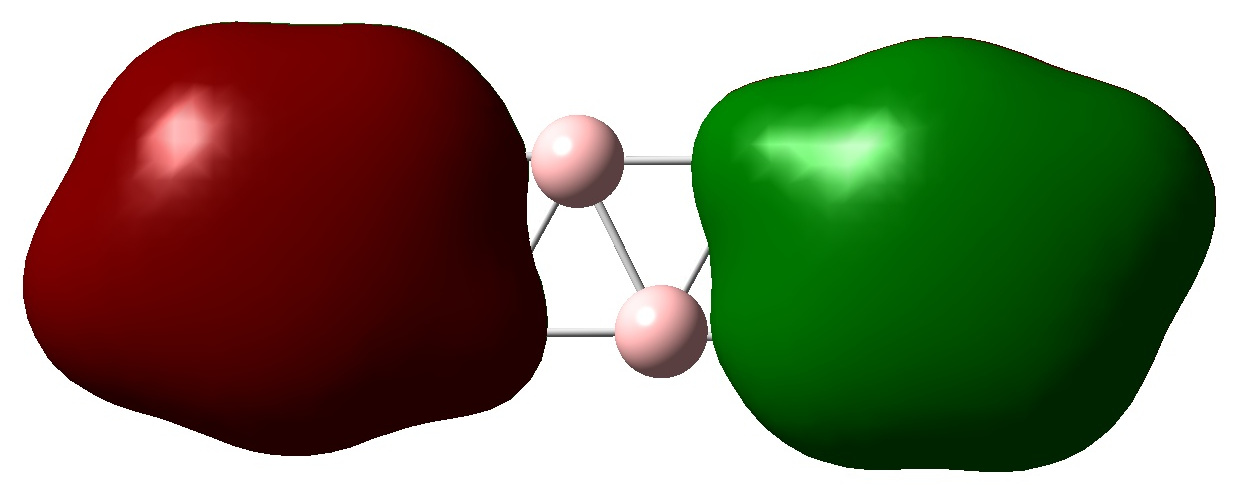}~~~\includegraphics[scale=0.037]{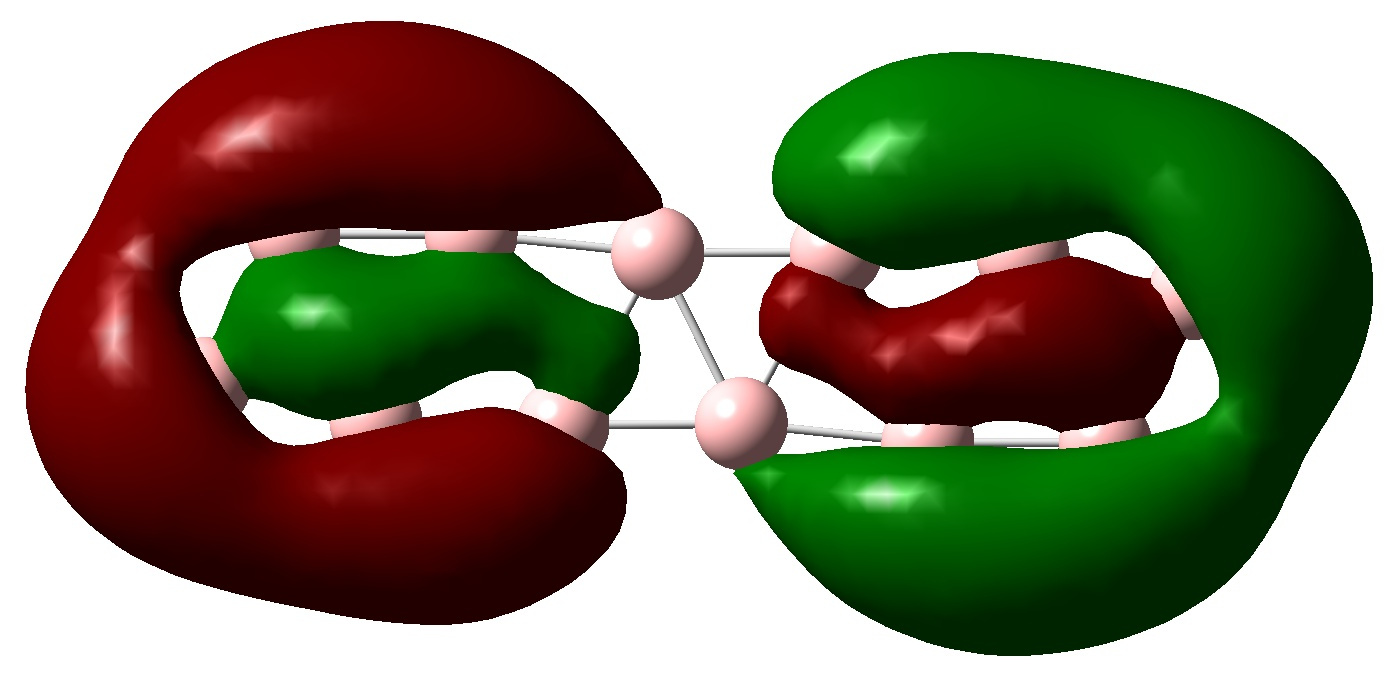}

\textbf{~~~~H-9}~~~~~~~\textbf{~~~H-8}~~~~~~~~\textbf{~~~H-7}~~~~~~~~~~~\textbf{~H-6}~~~~~~~\textbf{~~~~~H-5~~~~~~~~~~~H-4~~~~~~~~~~H-3~~~~~~~~~~~H-2~~~}

\vspace{1cm}

~~~\includegraphics[scale=0.038]{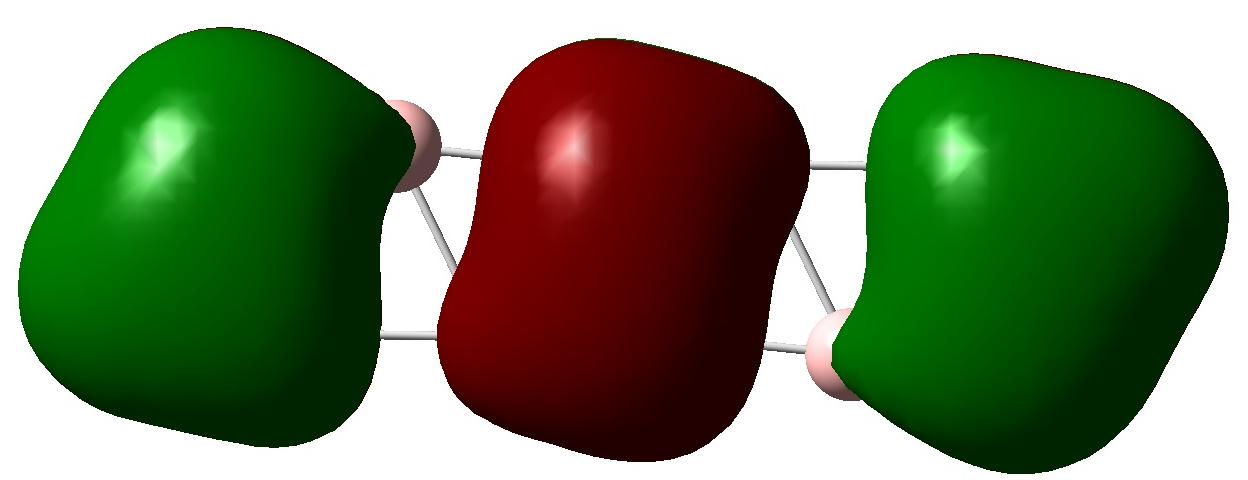}~~~\includegraphics[scale=0.036]{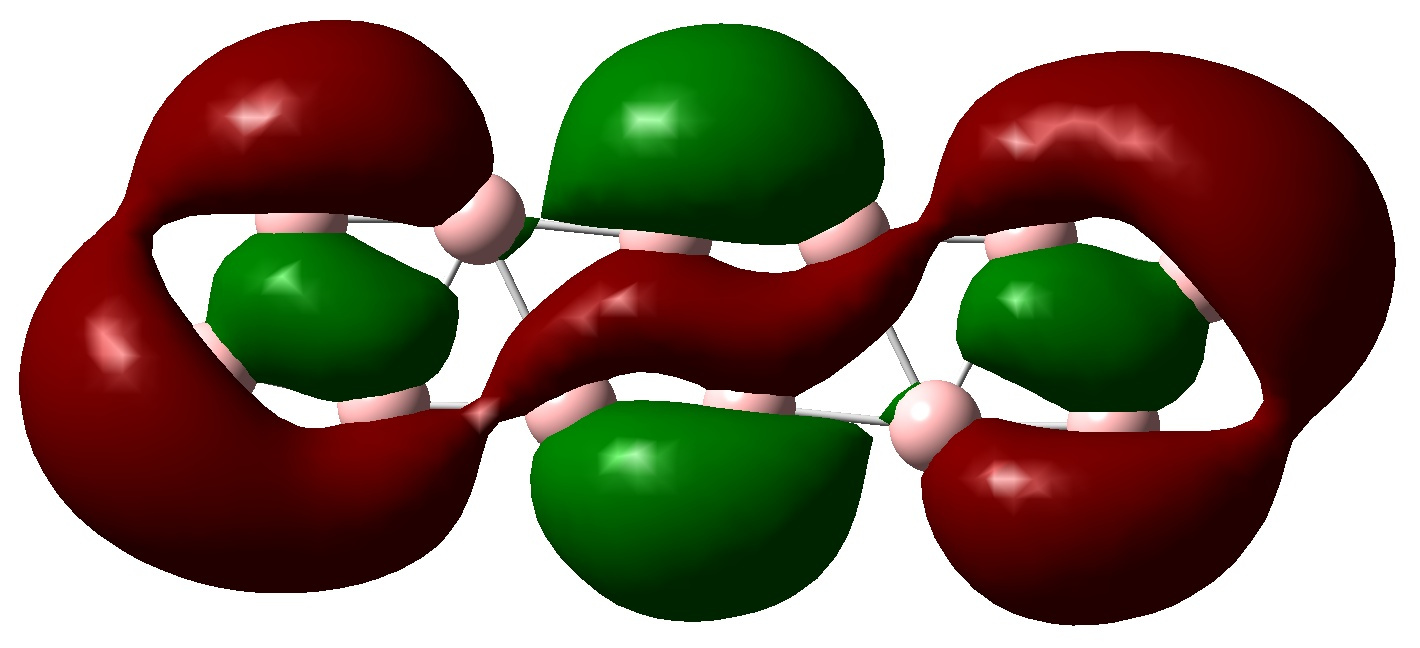}~~~\includegraphics[scale=0.036]{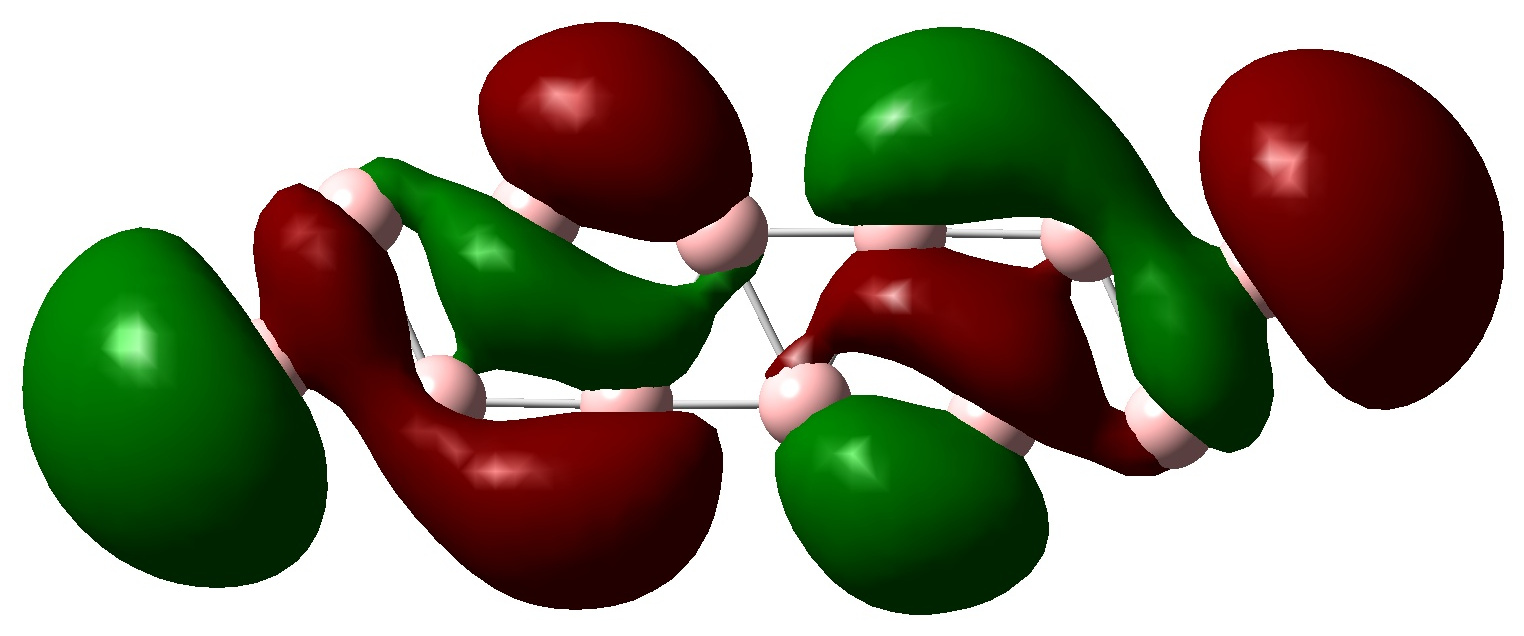}~~~\includegraphics[scale=0.037]{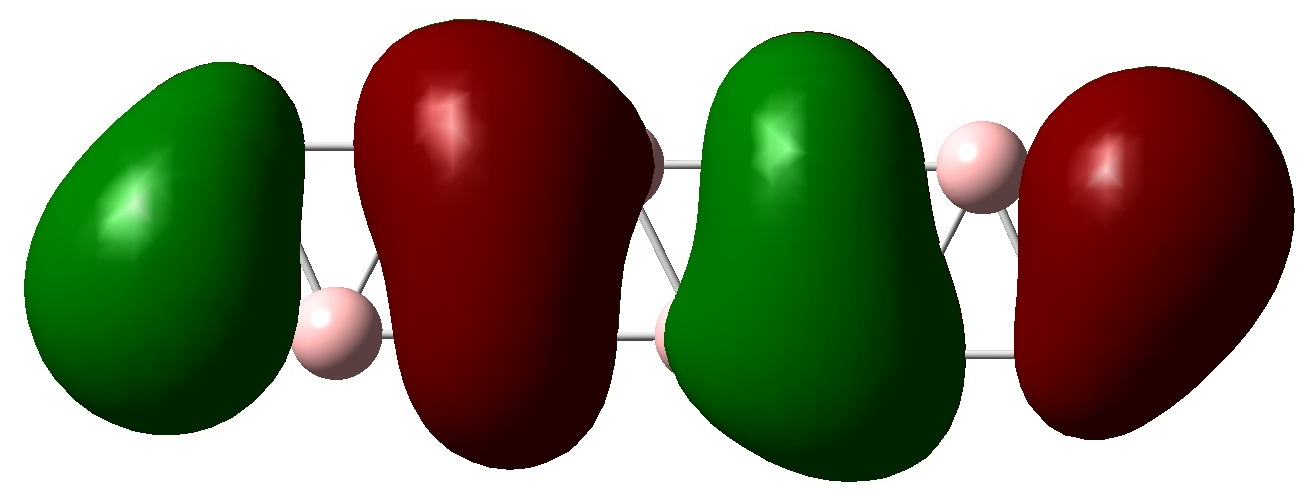}~~~\includegraphics[scale=0.036]{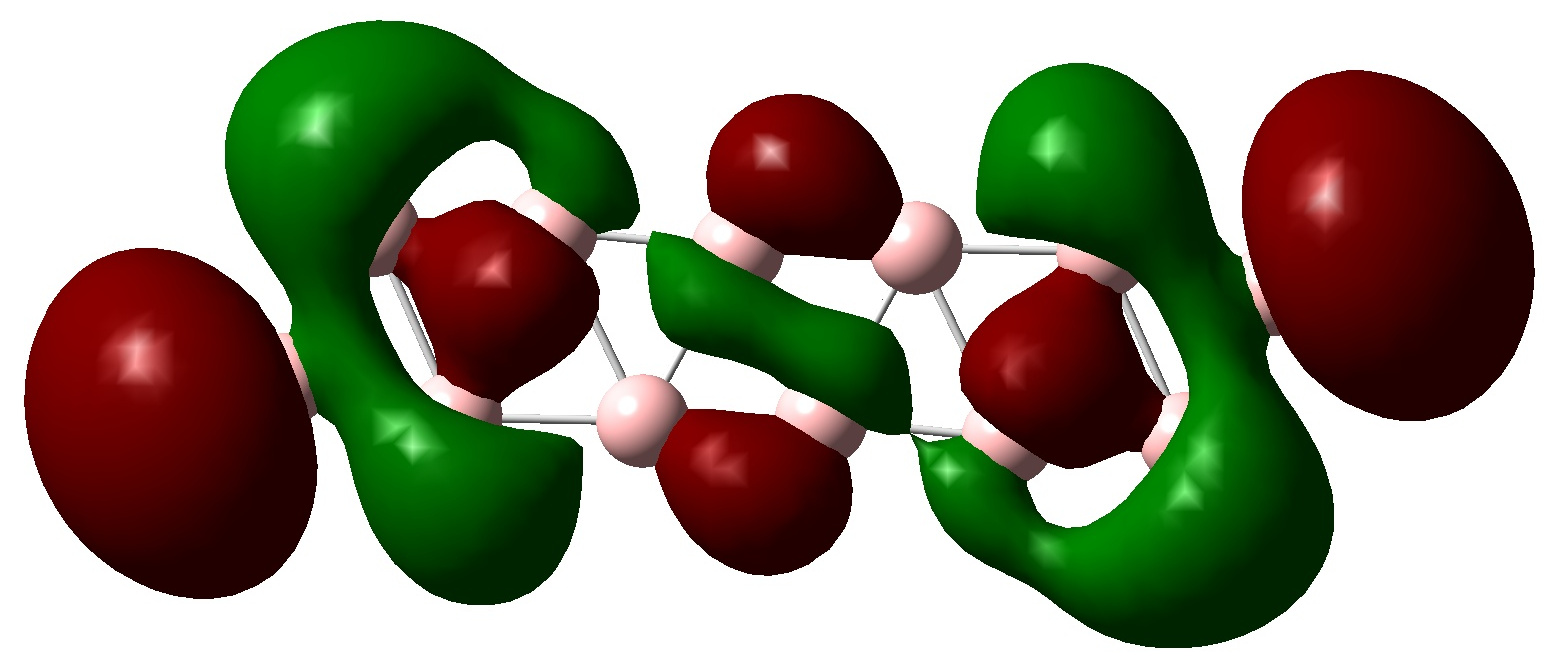}~~~\includegraphics[scale=0.037]{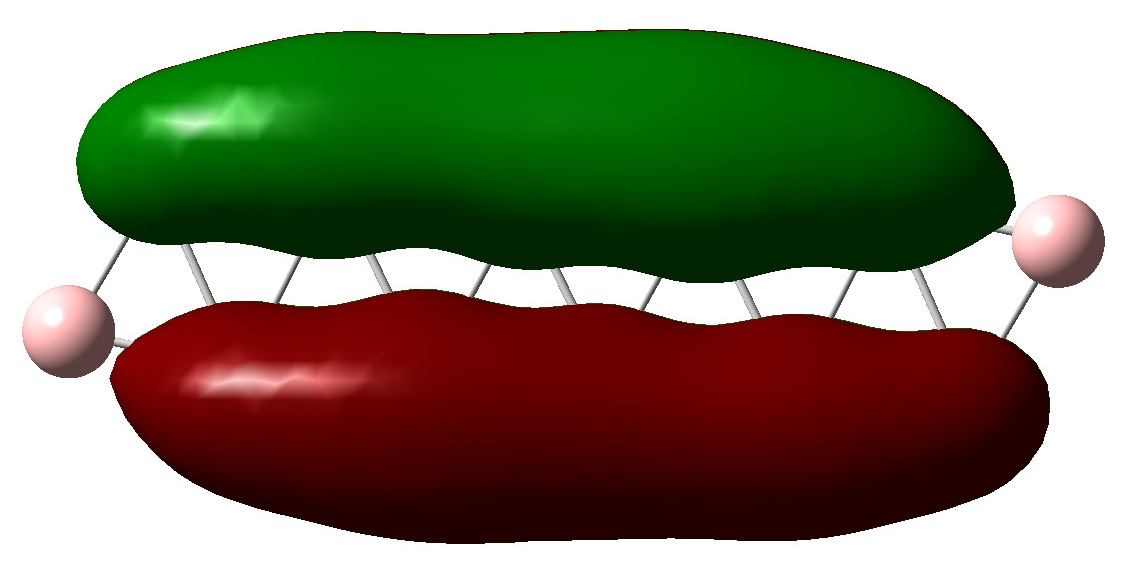}~~~\includegraphics[scale=0.037]{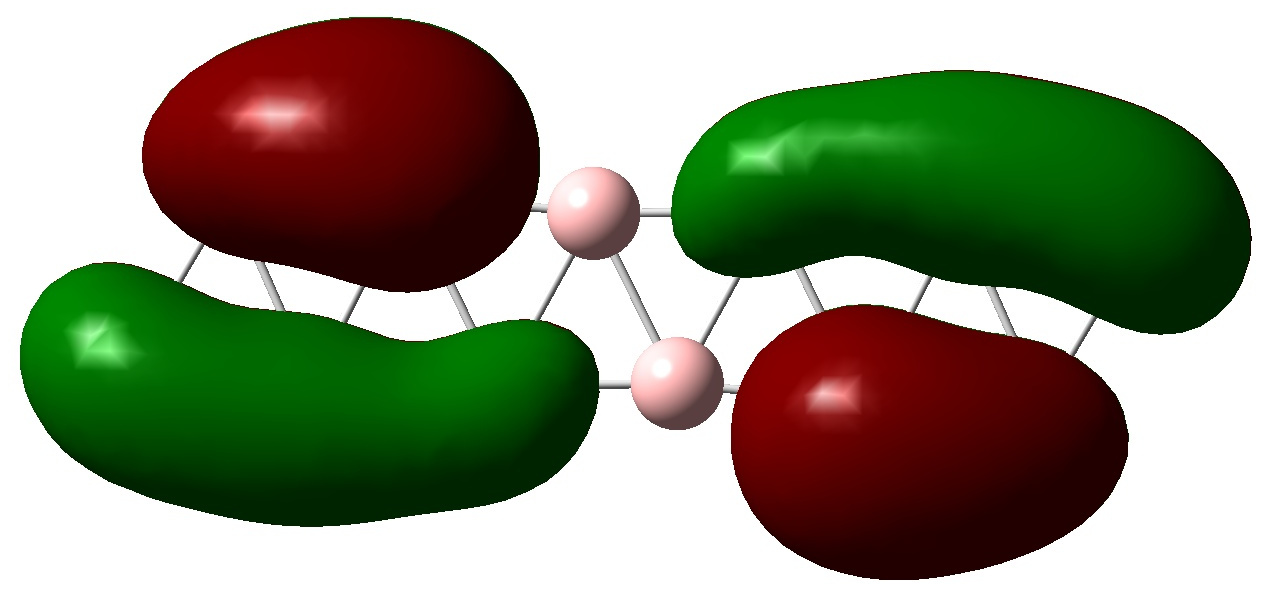}~~~\includegraphics[scale=0.036]{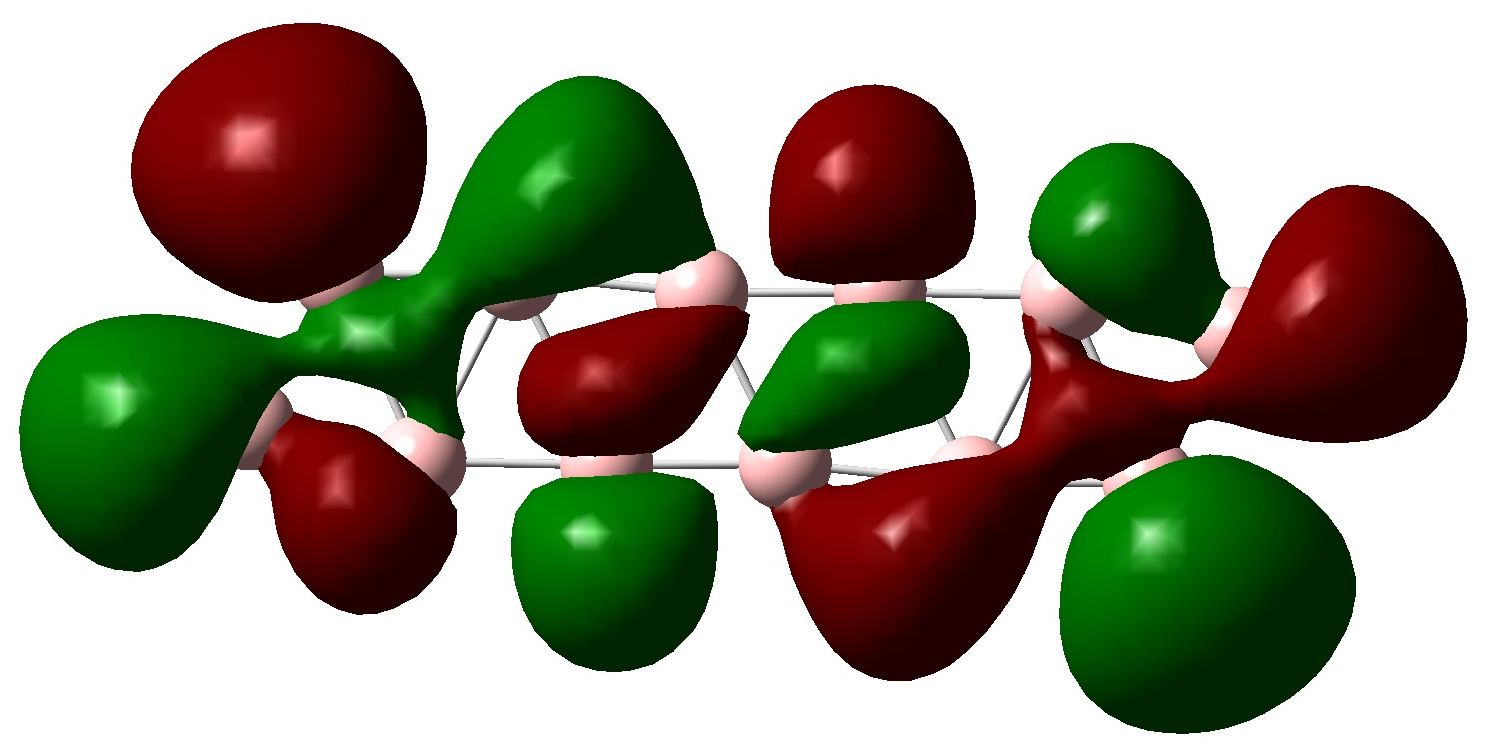}

\textbf{~~~~~~H-1}~\textbf{~~~~~~~~~~~H}~\textbf{~~~~~~~~~~~~~~L}~~~~~~~~~~~~~~\textbf{L+1}~~~~~~~~~~~~~\textbf{L+2}~~~~~~~~\textbf{~~~L+3}~~~~~~~~~\textbf{L+4~~~~~~~~~~L+5~~~}

\vspace{1cm}

\includegraphics[scale=0.037]{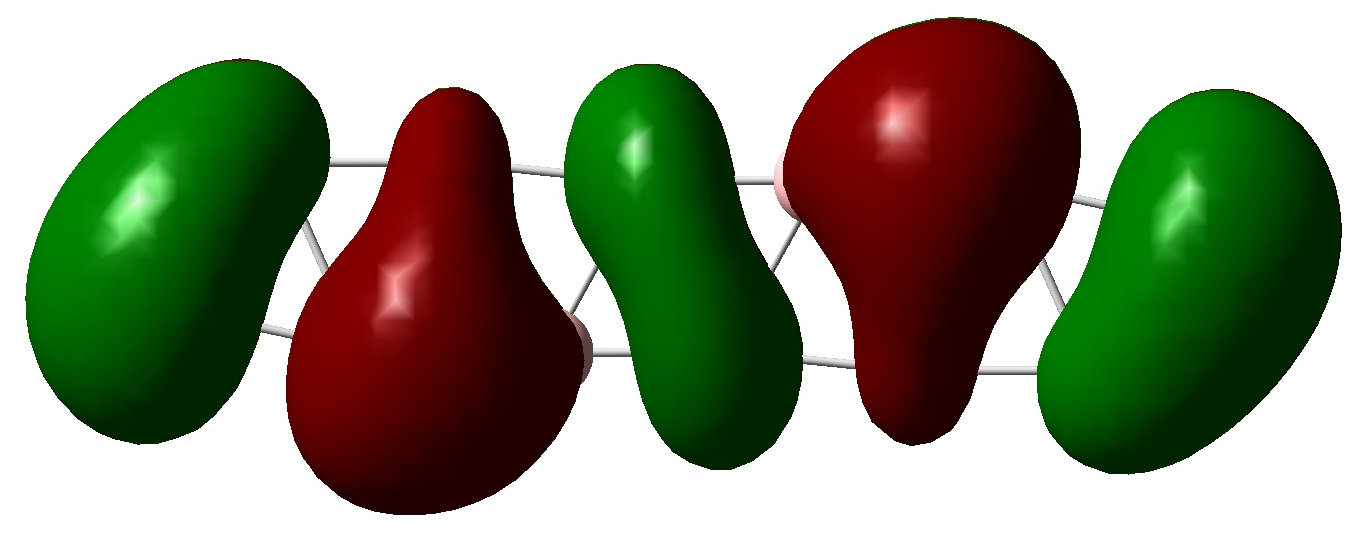}~~~\includegraphics[scale=0.037]{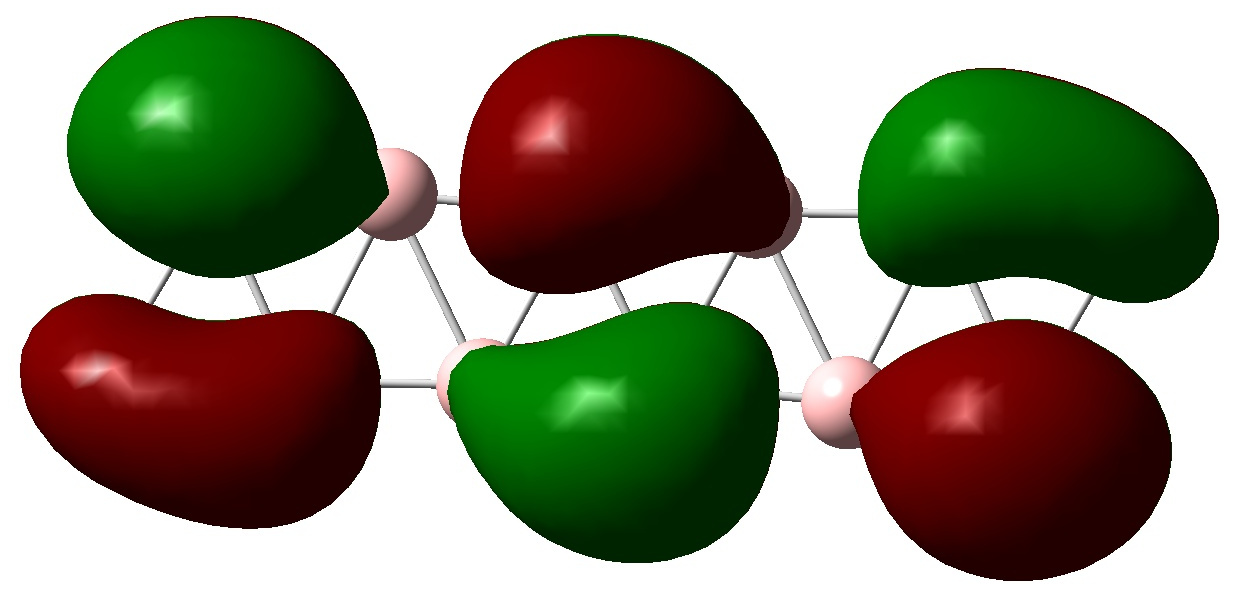}

\textbf{~~~L+6~~~~~~~~~~L+7}

\caption{Frontier molecular orbitals of chain ($C_{2h}$) isomer of B\protect\textsubscript{12}
(see Fig. 1c, of the main article), contributing to important peaks
in its optical absorption spectra. The symbols H and L corresponds
to HOMO and LUMO orbitals. }
\end{figure}

\begin{table}[H]
\caption{Many-particle wave functions of the excited states associated with
the peaks in the optical absorption spectrum of the chain ($C_{2h}$)
isomer of B\protect\textsubscript{12} {[}Fig. 1c, main article{]},
corresponding to excited states of $B_{u}$ symmetry. $E$ denotes
the excitation energy (in eV) of the state involved, and $f$ stands
for oscillator strength corresponding to the electric-dipole transition
to that state, from the ground state. In the ``Polarization'' column,
x, y, and z denote the the direction of polarization of the absorbed
light, with respect to the Cartesian coordinate axes defined in the
figure below. In the ``Wave function'' column, \textbf{$|HF\rangle$}
indicates the Hartree-Fock (HF) configuration, while other configurations
are defined as virtual excitations with respect to it. $H$ and $L$
denote HOMO and LUMO orbitals, while the degenerate orbitals are denoted
by additional subscripts 1, 2 etc. In parentheses next to each configuration,
its coefficient in the CI expansion is indicated. Below, GS does not
indicate a peak, rather it denotes the ground states wave function
of the isomer. All transitions correspond to photons polarized in
plane of the chain. }

\begin{tabular}{ccccccc}
\hline 
Peak & \hspace{1cm} & E (eV) & \hspace{1cm} & $f$ & \hspace{1cm} & Wave function\tabularnewline
\hline 
\hline 
GS\textsuperscript{} &  &  &  &  &  & $\arrowvert HF\rangle(0.8367)$\tabularnewline
 &  &  &  &  &  & $\arrowvert H\rightarrow L;H\rightarrow L\rangle(0.2095)$\tabularnewline
 &  &  &  &  &  & \tabularnewline
I &  & 1.64 &  & 2.3438 &  & $\arrowvert H\rightarrow L\rangle(0.6916)$\tabularnewline
 &  &  &  &  &  & $\arrowvert H-1\rightarrow L+1\rangle(0.3994)$\tabularnewline
 &  &  &  &  &  & \tabularnewline
II &  & 3.16 &  & 3.1597 &  & $\arrowvert H\rightarrow L+5\rangle(0.3966)$\tabularnewline
 &  &  &  &  &  & $\arrowvert H-1\rightarrow L+1\rangle(0.3600)$\tabularnewline
 &  & 3.24 &  & 2.5110 &  & $\arrowvert H-1\rightarrow L+3\rangle(0.6241)$\tabularnewline
 &  &  &  &  &  & $\arrowvert H-7\rightarrow L\rangle(0.3219)$\tabularnewline
 &  &  &  &  &  & \tabularnewline
III &  & 3.63 &  & 9.6596 &  & $\arrowvert H-4\rightarrow L\rangle(0.5134)$\tabularnewline
 &  &  &  &  &  & $\arrowvert H-1\rightarrow L+1\rangle(0.3374)$\tabularnewline
 &  &  &  &  &  & \tabularnewline
IV &  & 3.94 &  & 1.2794 &  & $\arrowvert H-5\rightarrow L+2\rangle(0.4566)$\tabularnewline
 &  &  &  &  &  & $\arrowvert H\rightarrow L;H-5\rightarrow L\rangle(0.3762)$\tabularnewline
 &  &  &  &  &  & \tabularnewline
V &  & 4.49 &  & 48.6542 &  & $\arrowvert H-1\rightarrow L+1\rangle(0.4369)$\tabularnewline
 &  &  &  &  &  & $\arrowvert H\rightarrow L\rangle(0.3285)$\tabularnewline
 &  &  &  &  &  & \tabularnewline
VI &  & 5.63 &  & 10.3584 &  & $\arrowvert H-8\rightarrow L\rangle(0.5171)$\tabularnewline
 &  &  &  &  &  & $\arrowvert H-6\rightarrow L+1\rangle(0.2364)$\tabularnewline
 &  &  &  &  &  & \tabularnewline
VII &  & 6.20 &  & 1.7806 &  & $\arrowvert H-3\rightarrow L+4\rangle(0.3575)$\tabularnewline
 &  &  &  &  &  & $\arrowvert H\rightarrow L+2;H-1\rightarrow L+1\rangle(0.2426)$\tabularnewline
 &  &  &  &  &  & \tabularnewline
VIII &  & 6.46 &  & 1.3157 &  & $\arrowvert H-6\rightarrow L+3\rangle(0.5479)$\tabularnewline
 &  &  &  &  &  & $\arrowvert H-1\rightarrow L+7\rangle(0.2820)$\tabularnewline
 &  &  &  &  &  & \tabularnewline
IX &  & 6.74 &  & 0.8341 &  & $\arrowvert H-9\rightarrow L+2\rangle(0.2964)$\tabularnewline
 &  &  &  &  &  & $\arrowvert H-1\rightarrow L+7\rangle(0.2485)$\tabularnewline
\hline 
\end{tabular}
\end{table}

\begin{table}[H]
\caption{Many-particle wave functions of the excited states corresponding to
the peaks in the optical absorption spectrum of the chain ($C_{2h}$)
isomer of B\protect\textsubscript{12} {[}Fig. 1c, main article{]},
corresponding to excited states of $A_{u}$ symmetry. $E$ denotes
the excitation energy (in eV) of the state involved, and $f$ stands
for oscillator strength corresponding to the electric-dipole transition
to that state, from the ground state. In the ``Polarization'' column,
x, y, and z denote the the direction of polarization of the absorbed
light, with respect to the Cartesian coordinate axes defined in the
figure below. In the ``Wave function'' column, \textbf{$|HF\rangle$}
indicates the Hartree-Fock (HF) configuration, while other configurations
are defined as virtual excitations with respect to it. $H$ and $L$
denote HOMO and LUMO orbitals, while the degenerate orbitals are denoted
by additional subscripts 1, 2 etc. In parentheses next to each configuration,
its coefficient in the CI expansion is indicated. Below, GS does not
indicate a peak, rather it denotes the ground states wave function
of the isomer. All transitions correspond to photons polarized perpendicular
to the plane of the chain.}

\begin{tabular}{ccccccc}
\hline 
Peak & \hspace{1cm} & E (eV) & \hspace{1cm} & $f$ & \hspace{1cm} & Wave function\tabularnewline
\hline 
\hline 
GS\textsuperscript{} &  &  &  &  &  & $\arrowvert HF\rangle(0.8367)$\tabularnewline
 &  &  &  &  &  & $\arrowvert H\rightarrow L;H\rightarrow L\rangle(0.2095)$\tabularnewline
 &  &  &  &  &  & \tabularnewline
I &  & 2.42 &  & 0.0170 &  & $\arrowvert H-1\rightarrow L+2\rangle(0.6840)$\tabularnewline
 &  &  &  &  &  & $\arrowvert H\rightarrow L;H-1\rightarrow L\rangle(0.4107)$\tabularnewline
 &  &  &  &  &  & \tabularnewline
II &  & 3.59 &  & 0.0932 &  & $\arrowvert H\rightarrow L+4\rangle(0.6707)$\tabularnewline
 &  &  &  &  &  & $\arrowvert H-2\rightarrow L+1\rangle(0.3854)$\tabularnewline
 &  &  &  &  &  & \tabularnewline
III &  & 4.07 &  & 0.1111 &  & $\arrowvert H-5\rightarrow L+1\rangle(0.3510)$\tabularnewline
 &  &  &  &  &  & $\arrowvert H\rightarrow L;H\rightarrow L+1\rangle(0.3505)$\tabularnewline
 &  & 4.14 &  & 0.0387 &  & $\arrowvert H-5\rightarrow L+1\rangle(0.6436)$\tabularnewline
 &  &  &  &  &  & $\arrowvert H\rightarrow L;H-7\rightarrow L+1\rangle(0.2097)$\tabularnewline
 &  &  &  &  &  & \tabularnewline
IV &  & 4.80 &  & 0.1128 &  & $\arrowvert H\rightarrow L+1;H-1\rightarrow L+1\rangle(0.6009)$\tabularnewline
 &  &  &  &  &  & $\arrowvert H\rightarrow L+6\rangle(0.5196)$\tabularnewline
 &  & 4.93 &  & 0.0627 &  & $\arrowvert H-5\rightarrow L+3\rangle(0.7141)$\tabularnewline
 &  &  &  &  &  & $\arrowvert H\rightarrow L+2;H-5\rightarrow L+3\rangle(0.2418)$\tabularnewline
 &  &  &  &  &  & \tabularnewline
V &  & 5.88 &  & 0.0077 &  & $\arrowvert H\rightarrow L+1;H-1\rightarrow L+3\rangle(0.4496)$\tabularnewline
 &  &  &  &  &  & $\arrowvert H\rightarrow L+5;H\rightarrow L+1\rangle(0.2934)$\tabularnewline
 &  &  &  &  &  & \tabularnewline
VI &  & 6.22 &  & 0.0153 &  & $\arrowvert H\rightarrow L+2;H-2\rightarrow L+1\rangle(0.2964)$\tabularnewline
 &  &  &  &  &  & $\arrowvert H-7\rightarrow L+4\rangle(0.2713)$\tabularnewline
 &  & 6.45 &  & 0.0548 &  & $\arrowvert H\rightarrow L+1;H-2\rightarrow L+2\rangle(0.3201)$\tabularnewline
 &  &  &  &  &  & $\arrowvert H\rightarrow L;H-7\rightarrow L+3\rangle(0.2747)$\tabularnewline
 &  & 6.68 &  & 0.0381 &  & $\arrowvert H\rightarrow L+1;H-2\rightarrow L+2\rangle(0.2930)$\tabularnewline
 &  &  &  &  &  & $\arrowvert H\rightarrow L+2;H\rightarrow L+4\rangle(0.2453)$\tabularnewline
\hline 
\end{tabular}
\end{table}

\newpage{}

\section{Atomic Coordinates }

In this section we present tables containing the atomic coordinates,
in Å units, of various isomers studied in this work.

\subsection{Quasi-planar}

\begin{table}[H]
\begin{tabular}{cccc}
\hline 
Atom & x & y & z\tabularnewline
\hline 
\hline 
B & 0.854596 & 0.493401 & 0.396007\tabularnewline
B & -0.854596 & 0.493401 & 0.396007\tabularnewline
B & 0.000000 & -0.986802 & 0.396007\tabularnewline
B & 0.000000 & 2.063378 & -0.045471\tabularnewline
B & 1.786937 & -1.031689 & -0.045471\tabularnewline
B & -1.786937 & -1.031689 & -0.045471\tabularnewline
B & 2.425806 & 0.485458 & -0.175268\tabularnewline
B & 1.633322 & 1.858081 & -0.175268\tabularnewline
B & -1.633322 & 1.858081 & -0.175268\tabularnewline
B & -2.425806  &  0.485458 & -0.175268\tabularnewline
B & -0.792484 & -2.343538 & -0.175268\tabularnewline
B & 0.792484 & -2.343538 & -0.175268\tabularnewline
\hline 
\end{tabular}

\caption{Optimized coordinates of Quasi-planar (C\protect\textsubscript{3v})
isomer}

\end{table}

\subsection{Double ring }

\begin{table}[H]
\begin{tabular}{cccc}
\hline 
Atom & x & y & z\tabularnewline
\hline 
\hline 
B & 0.000000 & 1.645803 & 0.755665\tabularnewline
B & -1.425308 & 0.822902 & 0.755665\tabularnewline
B & -1.425308 & -0.822902 & 0.755665\tabularnewline
B & 0.000000 & -1.645803 &  0.755665\tabularnewline
B & 1.425308 & -0.822902 & 0.755665\tabularnewline
B & 1.425308 & 0.822902 & 0.755665\tabularnewline
B & 0.822902 & 1.425308 & -0.755665\tabularnewline
B & 1.645803 & 0.000000 & -0.755665\tabularnewline
B & 0.822902 & -1.425308 & -0.755665\tabularnewline
B & -0.822902 & -1.425308 & -0.755665\tabularnewline
B & -1.645803 & 0.000000 & -0.755665\tabularnewline
B & -0.822902 & 1.425308 & -0.755665\tabularnewline
\hline 
\end{tabular}

\caption{Optimized coordinates of Double ring (D\protect\textsubscript{6d}
) isomer}
\end{table}

\subsection{Chain}

\begin{table}[H]
\begin{tabular}{cccc}
\hline 
Atom & x & y & z\tabularnewline
\hline 
\hline 
B & -0.430275 & 4.400542 & 0.000000\tabularnewline
B & 0.905913 & 3.618456 & 0.000000\tabularnewline
B & -0.755353 & 2.873999 & 0.000000\tabularnewline
B & 0.914514 & 2.038590 & 0.000000\tabularnewline
B & -0.777694 & 1.210978 & 0.000000\tabularnewline
B & 0.755353 & 0.376209 & 0.000000\tabularnewline
B & -0.755353 & -0.376209 & 0.000000\tabularnewline
B & 0.777694 & -1.210978 & 0.000000\tabularnewline
B & -0.914514 & -2.038590 & 0.000000\tabularnewline
B & 0.755353 & -2.873999 & 0.000000\tabularnewline
B & -0.905913 & -3.618456 & 0.000000\tabularnewline
B & 0.430275  & -4.400542 & 0.000000\tabularnewline
\hline 
\end{tabular}

\caption{Optimized coordinates of Chain (C\protect\textsubscript{2h}) isomer}
\end{table}

\subsection{Hexagons}

\begin{table}[H]
\begin{tabular}{cccc}
\hline 
Atom & x & y & z\tabularnewline
\hline 
\hline 
B & 0.000000 & 0.000000 & 0.844921\tabularnewline
B & 0.000000 & 1.517663 & 0.000000\tabularnewline
B & 0.000000 & -1.517663 & 0.000000\tabularnewline
B & 0.000000 & -1.664249 & 1.586458\tabularnewline
B & 0.000000 & -0.768140 & 2.882657\tabularnewline
B & 0.000000 & 0.768140 & 2.882657\tabularnewline
B & 0.000000 & 1.664249 & 1.586458\tabularnewline
B & 0.000000 & 0.000000 & -0.844921\tabularnewline
B & 0.000000 & -0.768140 & -2.882657\tabularnewline
B & 0.000000 & 0.768140 & -2.882657\tabularnewline
B & 0.000000 & 1.664249 & -1.586458\tabularnewline
B & 0.000000 & -1.664249 & -1.586458\tabularnewline
\hline 
\end{tabular}

\caption{Optimized coordinates of Hexagons (D\protect\textsubscript{2h}) isomer}
\end{table}

\subsection{Quad}

\begin{table}[H]
\begin{tabular}{cccc}
\hline 
Atom & x & y & z\tabularnewline
\hline 
\hline 
B & 0.000000 & 2.015258 & 0.000000\tabularnewline
B & 0.000000 & 0.000000 & 1.268281\tabularnewline
B & 0.000000 & 0.000000 & -1.268281\tabularnewline
B & 0.000000 & -2.015258 & 0.000000\tabularnewline
B & -1.682909 & 0.000000 & -0.912871\tabularnewline
B & -1.490298 & -1.389700 & 0.000000\tabularnewline
B & -1.490298 & 1.389700 & 0.000000\tabularnewline
B & -1.682909 & 0.000000 & 0.912871\tabularnewline
B & 1.682909 & 0.000000 & 0.912871\tabularnewline
B & 1.490298 & -1.389700 & 0.000000\tabularnewline
B & 1.490298 & 1.389700 & 0.000000\tabularnewline
B & 1.682909 & 0.000000 & -0.912871\tabularnewline
\hline 
\end{tabular}

\caption{Optimized coordinates of Quad (D\protect\textsubscript{2h}) isomer}
\end{table}

\subsection{Pris-3-square-1}

\begin{table}[H]
\begin{tabular}{cccc}
\hline 
Atom & x & y & z\tabularnewline
\hline 
\hline 
B & 0.000000 & 1.574917 & 0.000000\tabularnewline
B & 1.574917 & 0.000000 & 0.000000\tabularnewline
B & -1.574917 & 0.000000 & 0.000000\tabularnewline
B & 0.000000 & -1.574917 & 0.000000\tabularnewline
B & -0.862562 & 0.862562 & 1.264540\tabularnewline
B & -0.862562 & -0.862562 & 1.264540\tabularnewline
B & 0.862562 & 0.862562 & 1.264540\tabularnewline
B & 0.862562 & -0.862562 & 1.264540\tabularnewline
B & 0.862562 & 0.862562 & -1.264540\tabularnewline
B & 0.862562 & -0.862562 & -1.264540\tabularnewline
B & -0.862562 & 0.862562 & -1.264540\tabularnewline
B & -0.862562 & -0.862562 & -1.264540\tabularnewline
\hline 
\end{tabular}

\caption{Optimized coordinates of Pris-3-square-1 (D\protect\textsubscript{4h})
isomer}
\end{table}

\subsection{Pris-3-square-2}

\begin{table}[H]
\begin{tabular}{cccc}
\hline 
Atom & x & y & z\tabularnewline
\hline 
\hline 
B & 0.000000 & 1.870473 & 0.000000\tabularnewline
B & 1.870473 & 0.000000 & 0.000000\tabularnewline
B & -1.870473 & 0.000000 & 0.000000\tabularnewline
B & 0.000000 & -1.870473 & 0.000000\tabularnewline
B & -0.896666 & 0.896666 & 0.994197\tabularnewline
B & -0.896666 & -0.896666 & 0.994197\tabularnewline
B & 0.896666 & 0.896666 & 0.994197\tabularnewline
B & 0.896666 & -0.896666 & 0.994197\tabularnewline
B & 0.896666 & 0.896666 & -0.994197\tabularnewline
B & 0.896666 & -0.896666 & -0.994197\tabularnewline
B & -0.896666 & 0.896666 & -0.994197\tabularnewline
B & -0.896666 & -0.896666 & -0.994197\tabularnewline
\hline 
\end{tabular}

\caption{Optimized coordinates of Pris-3-square-2 (D\protect\textsubscript{4h})
isomer}
\end{table}

\subsection{Parallel triangular }

\begin{table}[H]
\begin{tabular}{cccc}
\hline 
Atom & x & y & z\tabularnewline
\hline 
\hline 
B & 0.000000 & 1.069852 & 0.924715\tabularnewline
B & -0.926519 & -0.534926 & 0.924715\tabularnewline
B & 0.926519 & -0.534926 & 0.924715\tabularnewline
B & 0.000000 & 1.069852 & -0.924715\tabularnewline
B & 0.926519 & -0.534926 & -0.924715\tabularnewline
B & -0.926519 & -0.534926 & -0.924715\tabularnewline
B & 0.000000 & 0.989584 & 2.529660\tabularnewline
B & -0.857005 & -0.494792 & 2.529660\tabularnewline
B & 0.857005 & -0.494792 & 2.529660\tabularnewline
B & 0.000000 & 0.989584 & -2.529660\tabularnewline
B & 0.857005 & -0.494792 & -2.529660\tabularnewline
B & -0.857005 & -0.494792 & -2.529660\tabularnewline
\hline 
\end{tabular}

\caption{Optimized coordinates of Parallel triangular (D\protect\textsubscript{3h})
isomer}
\end{table}

\subsection{Planar square }

\begin{table}[H]
\begin{tabular}{cccc}
\hline 
Atom & x & y & z\tabularnewline
\hline 
\hline 
B & 0.000000 & 1.237919 & 0.000000\tabularnewline
B & 1.237919 & 0.000000 & 0.000000\tabularnewline
B & -1.237919 & 0.000000 & 0.000000\tabularnewline
B & 0.000000 & -1.237919 & 0.000000\tabularnewline
B & -1.237919 & 2.340594 & 0.000000\tabularnewline
B & 1.237919 & 2.340594 & 0.000000\tabularnewline
B & 2.340594 & 1.238828 & 0.000000\tabularnewline
B & 2.340594 & -1.238828 & 0.000000\tabularnewline
B & -2.340594 & -1.238828 & 0.000000\tabularnewline
B & -2.340594 & 1.238828 & 0.000000\tabularnewline
B & 1.238828 & -2.340594 & 0.000000\tabularnewline
B & -1.238828 & -2.340594 & 0.000000\tabularnewline
\hline 
\end{tabular}

\caption{Optimized coordinates of Planar square (D\protect\textsubscript{4h})
isomer}
\end{table}

\subsection{Perfect Icosahedron}

\begin{table}[H]
\begin{tabular}{cccc}
\hline 
Atom & x & y & z\tabularnewline
\hline 
\hline 
B & 0.000000 & 0.865000 & 1.399599\tabularnewline
B & 0.000000 & 0.865000 & -1.399599\tabularnewline
B & 0.000000 & -0.865000 & 1.399599\tabularnewline
B & 0.000000 & -0.865000 & -1.399599\tabularnewline
B & 1.399599 & 0.000000 & 0.865000\tabularnewline
B & -1.399599 & 0.000000 & 0.865000\tabularnewline
B & 1.399599 & 0.000000 & -0.865000\tabularnewline
B & -1.399599 & 0.000000 & -0.865000\tabularnewline
B & 0.865000 & 1.399599 & 0.000000\tabularnewline
B & 0.865000 & -1.399599 & 0.000000\tabularnewline
B & -0.865000 & 1.399599 & 0.000000\tabularnewline
B & -0.865000 & -1.399599 & 0.000000\tabularnewline
\hline 
\end{tabular}

\caption{Optimized coordinates of perfect Icosahedron (I\protect\textsubscript{h})
isomer}
\end{table}

\subsection{Distorted Icosahedron}

\begin{table}[H]
\begin{tabular}{cccc}
\hline 
Atom & x & y & z\tabularnewline
\hline 
\hline 
B & 0.000000 & 0.852920 & -1.337059\tabularnewline
B & 0.000000 & 0.852920 & 1.337059\tabularnewline
B & 0.000000 & -0.852920 & -1.337059\tabularnewline
B & 0.000000 & -0.852920 & 1.337059\tabularnewline
B & -1.394554 & 0.000000 & -0.843230\tabularnewline
B & 1.394554 & 0.000000 & -0.843230\tabularnewline
B & -1.394554 & 0.000000 & 0.843230\tabularnewline
B & 1.394554 & 0.000000 & 0.843230\tabularnewline
B & -0.897896 & 1.466072 & 0.000000\tabularnewline
B & -0.897896 & -1.466072 & 0.000000\tabularnewline
B & 0.897896 & 1.466072 & 0.000000\tabularnewline
B & 0.897896 & -1.466072 & 0.000000\tabularnewline
\hline 
\end{tabular}

\caption{Optimized coordinates of distorted Icosahedron (D\protect\textsubscript{2h})
isomer}
\end{table}

\section{Nature of the Wave Function of the Lowest Singlet State of B\protect\textsubscript{12}
Icosahedron }

We performed calculations on the lowest singlet state of the B$_{12}$
isomer with perfect icosahedral symmetry ($I_{h}$), by employing
multi-reference singles-doubles CI (MRSDCI) level of theory, and cc-pVDZ
basis set, to understand the nature of its wave function. For the
purpose we used the HF orbitals, and the seven reference configurations
listed in the table below. From the magnitudes of the coefficients
of the reference configurations in the CI expansion of the wave function,
it is obvious that the closed-shell HF reference state dominates the
wave function. Therefore, any single-reference approach such as CCSD,
using the HF state as the reference, will give reasonable results
for this state. 

\begin{table}[H]
\caption{Many-particle wave function of the lowest singlet state of isomer
of B$_{12}$ with perfect icosahedral symmetry ($I_{h}$) (Fig. 1e
of the main article). \textbf{$|HF\rangle$} indicates the closed-shell
Hartree-Fock (HF) configuration, while other configurations are defined
as virtual excitations with respect to it. $H$ and $L$ denote HOMO
and LUMO orbitals. In the parentheses next to each configuration,
its coefficient in the CI expansion is indicated. \label{tab:Icos}}

\begin{tabular}{c}
\hline 
Dominant configurations in the MRSDCI wave function\tabularnewline
\hline 
\hline 
$\arrowvert HF\rangle(0.8097)$\tabularnewline
$\arrowvert H\rightarrow L+3;H\rightarrow L+3\rangle(0.2697)$\tabularnewline
$\arrowvert H\rightarrow L+3;H-1\rightarrow L+6\rangle(0.1444)$\tabularnewline
$\arrowvert H\rightarrow L+2;H-1\rightarrow L+3\rangle(0.1339)$\tabularnewline
$\arrowvert H\rightarrow L+3;H-2\rightarrow L+3\rangle(0.1049)$\tabularnewline
$\arrowvert H\rightarrow L+3;H-8\rightarrow L+1\rangle(0.0922)$\tabularnewline
$\arrowvert H-1\rightarrow L+4;H-3\rightarrow L+6\rangle(0.0901)$\tabularnewline
\hline 
\end{tabular}
\end{table}